\documentclass[journal,twoside]{IEEEtran}

\usepackage[font=small]{caption}
\usepackage{subcaption}

\usepackage{array,amsfonts,amssymb,bm,paralist}
\usepackage[cmex10]{amsmath}
\usepackage{graphicx}
\graphicspath{{./Figures/}{./Figures/CRB/}}

\usepackage{cite}
\usepackage[normalem]{ulem} 
\usepackage{textcomp}
\usepackage[colorlinks=true, allcolors=blue]{hyperref}
\usepackage[inline]{enumitem}
\usepackage{cleveref}
\usepackage{mathtools}
\usepackage[usenames,dvipsnames]{xcolor}
\usepackage{rotating}
\usepackage{dsfont}

\def\del{\partial}

\newcommand{\CR}{\ensuremath{\mathbb{C}^{\R}}}

\def\T{{\top}}

\newcommand{\norm}[1]{\lVert #1 \rVert}

\DeclareFontFamily{U}{mathx}{\hyphenchar\font45}
\DeclareFontShape{U}{mathx}{m}{n}{
      <5> <6> <7> <8> <9> <10>
      <10.95> <12> <14.4> <17.28> <20.74> <24.88>
      mathx10
      }{}
\DeclareSymbolFont{mathx}{U}{mathx}{m}{n}
\DeclareFontSubstitution{U}{mathx}{m}{n}
\DeclareMathAccent{\widecheck}{0}{mathx}{"71}

\newcommand{\Frac}[2]{{{#1}/{#2}}}  

\newcommand{\beq}{\begin{equation}}
\newcommand{\eeq}{\end{equation}}
\newcommand{\beqan}{\begin{eqnarray*}}
\newcommand{\eeqan}{\end{eqnarray*}}

\newcommand{\eqlabel}[1]{ \stackrel{(#1)}{=} }

\newcommand{\openCase}  {\left\{ \begin{array}{@{\,}ll}}
\newcommand{\openCasell}{\left\{ \begin{array}{@{\,}ll}}
\newcommand{\openCasecl}{\left\{ \begin{array}{@{\,}cl}}
\newcommand{\openCaserl}{\left\{ \begin{array}{@{\,}rl}}
\newcommand{\openCaseTablell}{\left\{ \begin{array}{@{}ll}}
\newcommand{\openCaseTablecl}{\left\{ \begin{array}{@{}cl}}
\newcommand{\openCaseTablerl}{\left\{ \begin{array}{@{}rl}}
\newcommand{\closeCase} {\end{array} \right.}

\def\mtilde{\widetilde{m}}
\def\Mtilde{\widetilde{M}}

\def\ttilde{\widetilde{t}}
\def\Ttilde{\widetilde{T}}

\def\xtilde{\widetilde{x}}
\def\Xtilde{\widetilde{X}}

\def\lmax{\lambda_{\rm max}}

\DeclareMathOperator*{\argmin}{arg\,min}
\DeclareMathOperator*{\argmax}{arg\,max}

\newcommand{\iP}[1]{\mathrm{P}({#1})} 
\def\smid{\,|\,}  
\def\sMid{\,;\,}  
\newcommand{\E}[1]{\mathrm{E}\!\left[\,{#1}\,\right]}   

\usepackage{algorithm, algorithmic}
\usepackage{multirow}

\definecolor{ao(english)}{rgb}{0.0, 0.5, 0.0}

\newcommand{\lambf}{\boldsymbol{\lambda}}
\newcommand{\etabf}{\boldsymbol{\eta}}
\newcommand{\ybf}{\mathbf{y}}
\def\CR{Cram{\'e}r--Rao}
\def\CRB{Cram{\'e}r--Rao bound}

\newcommand{\lamChatBold}{{\widehat{\lambf}{}^{\rm C}}}

\newcommand{\lamCTVhat}{\widehat{{\lambda}}^{\rm C TV}}
\newcommand{\lamCTVhatBold}{\widehat{{\lambf}}{}^{\rm C TV}}
\newcommand{\lamNCTVhat}{\widehat{\lambf}{}^{\rm NC TV}}
\newcommand{\lamChat}{\widehat{\lambda}^{\rm C}}
\newcommand{\lamNChat}{\widehat{\lambf}{}^{\rm NC}}
\newcommand{\etaChat}{\widehat{{\eta}}^{\rm C}}
\newcommand{\etaChatBold}{\widehat{{\etabf}}{}^{\rm C}}
\newcommand{\etaCTVhat}{\widehat{{\eta}}^{\rm C TV}}
\newcommand{\etaCTVhatBold}{\widehat{{\etabf}}{}^{\rm C TV}}
\newcommand{\etaNCTVhat}{\widehat{\etabf}{}^{\rm NCTV}}
\newcommand{\etaNChat}{\widehat{\etabf}{}^{\rm NC}}

\newcommand{\etaup}{\eta_{\rm v}}
\newcommand{\betaup}{\beta_{\rm v}}
\newcommand{\betactv}{\beta_{\rm CTV}}
\newcommand{\betanctv}{\beta_{\rm NCTV}}
\def\etaHmmBold{\widehat{\etabf}{}^{\rm HMM\, C} }
\def\lamHmmBold{\widehat{\lambf}{}^{\rm HMM\, C}}
\def\etaHmmNcBold{\widehat{\etabf}{}^{\rm HMM\, NC} }
\def\lamHmmNcBold{\widehat{\lambf}{}^{\rm HMM\, NC}}

\newcommand{\betaleft}{\beta_{\rm h}}
\newcommand{\etaleft}{\eta_{\rm h}}
\def\lambar{\bar{\lambda}}
\def\lamtilde{\tilde{\lambda}}
\def\betac{\beta_{\rm C}}
\def\betanc{\beta_{\rm NC}}
\def\sigmax{\sigma_x}

\def\etaFTbold{\widehat{\etabf}{}^{\rm FT}}

\def\etaBaseline{\widehat{\eta}^{\rm baseline}}
\def\etaBaselineBold{\widehat{\etabf}{}^{\rm baseline}}

\def\etaOracle{\widehat{\eta}^{\rm DT|\lambda}}
\def\etaOracleBold{\widehat{\etabf}{}^{\rm DT|\lambda}}

\def\etaCTML{\widehat{\eta}^{\rm CT}}
\def\etaCTMLgiven{\widehat{\eta}^{\rm CT|\lamtilde}}

\def\etaLFbold{\widehat{\etabf}{}^{\rm LF}}

\def\lambdaLFbold{\widehat{\lambf}{}^{\rm LF}}

\def\lambdaCTML{\widehat{\lambda}^{\rm CT}}

\def\etaDTML{\widehat{\eta}^{\rm DT}}
\def\etaDTMLgiven{\widehat{\eta}^{\rm DT|\tilde{\lambda}}}
\def\etaDTMLgivenBold{\widehat{\etabf}{}^{\rm DT|\tilde{\lambda}}}

\def\lamDTML{\widehat{\lambda}^{\rm DT}}
\def\gTV{g_{\rm TV}}

\def\yvec{\mathbf{Y}}
\def\mtilde{\widetilde{m}}
\def\ttildevec{\widetilde{\mathbf{T}}}
\def\xtildevec{\widetilde{\mathbf{X}}}

\def\fmat{\mathcal{I}}

\DeclareMathOperator{\Poisson}{Poisson}

\begin{document}

\title{
Online Beam Current Estimation in \\ Particle Beam Microscopy
}

        
\author{Sheila~W.~Seidel,
        Luisa Watkins,
        Minxu Peng,\\
        Akshay Agarwal,
        Christopher Yu,
        and~Vivek K Goyal
\thanks{This work was supported in part by
a Draper Fellowship,
a Boston University Clare Boothe Luce Scholar Award,
and by the US National Science Foundation under Grant No.~1815896.}
\thanks{S. W. Seidel is with the Department of Electrical and Computer Engineering, Boston University, Boston, MA 02215 USA and is a Draper Scholar with Charles Stark Draper Laboratory, Cambridge, MA 02139 USA (sseidel@bu.edu).}
\thanks{L. Watkins, M. Peng, and V. K. Goyal are with the Department of Electrical and Computer Engineering, Boston University, Boston, MA 02215 USA (luisaw@bu.edu; mxpeng@bu.edu; v.goyal@ieee.org).}
\thanks{A. Agarwal was with the Department of Electrical and Computer Engineering, Massachusetts Institute of Technology, Cambridge, MA 02139 USA (akshayag@mit.edu).}
\thanks{C. Yu is with Charles Stark Draper Laboratory, Cambridge, MA 02139 USA (cyu@draper.com).}
}

\markboth{Online Beam Current Estimation in Particle Beam Microscopy}%
{Seidel \MakeLowercase{\textit{et al.}}: }

\maketitle

\begin{abstract}
In conventional particle beam microscopy,
knowledge of the beam current is essential for accurate micrograph formation and sample milling.
This generally necessitates offline calibration of the instrument.
In this work, we establish that beam current can be estimated online, from the same secondary electron count data that is used to form micrographs.
Our methods depend on the
recently introduced time-resolved measurement concept, which combines multiple short measurements at a single pixel and has previously been shown to partially mitigate the effect of  beam current variation on micrograph accuracy.
We analyze the problem of jointly estimating beam current and secondary electron yield using the \CRB.
Joint estimators operating at a single pixel and estimators that exploit models for
inter-pixel correlation
and Markov beam current variation
are proposed and tested on synthetic microscopy data.
Our estimates of secondary electron yield that incorporate explicit beam current estimation beat state-of-the-art methods, resulting in micrograph accuracy nearly indistinguishable from what is obtained with perfect beam current knowledge. 
Our novel beam current estimation could help improve milling outcomes, prevent sample damage, and enable online instrument diagnostics.
\end{abstract}


\IEEEpeerreviewmaketitle

\begin{figure}
\centering
        \begin{subfigure}{0.475\linewidth}
        \centering
        \includegraphics[width=1\linewidth]{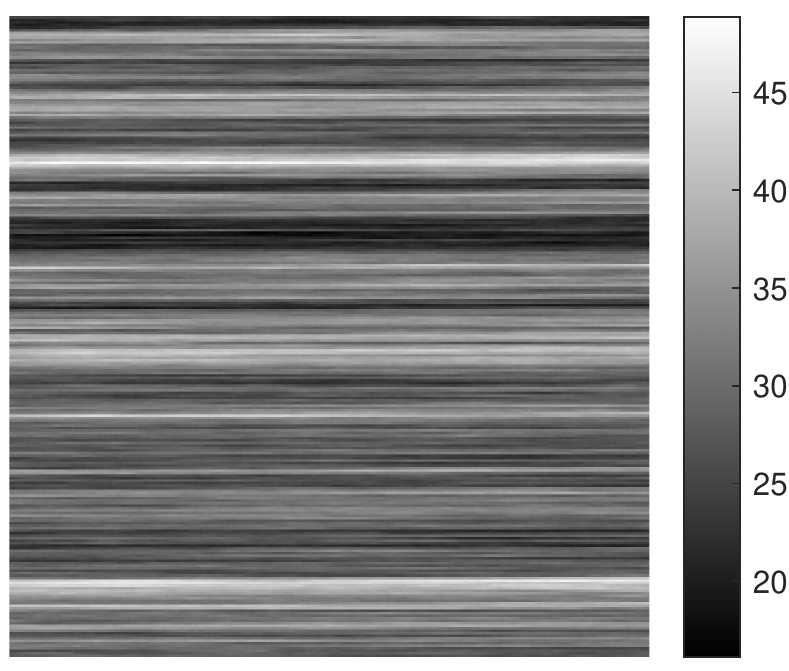}
        \caption{\centering Beam current incident on raster scanned sample in HIM }
        \label{fig:fib_beamCurrent}
    \end{subfigure}
    \begin{subfigure}{0.4\linewidth}
        \centering
        \includegraphics[width=1\linewidth]{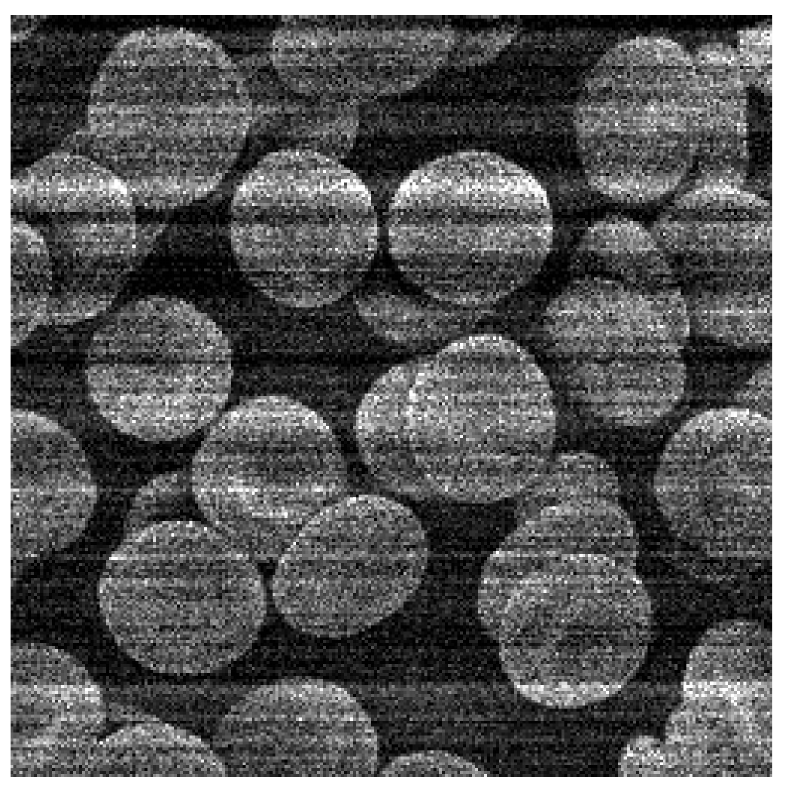}
        \caption{\centering Micrograph resulting from beam current in (a)}
        \label{fig:fib_stripes}
    \end{subfigure}\\
    \centering
        \begin{subfigure}{0.475\linewidth}
        \centering
        \includegraphics[width=1\linewidth]{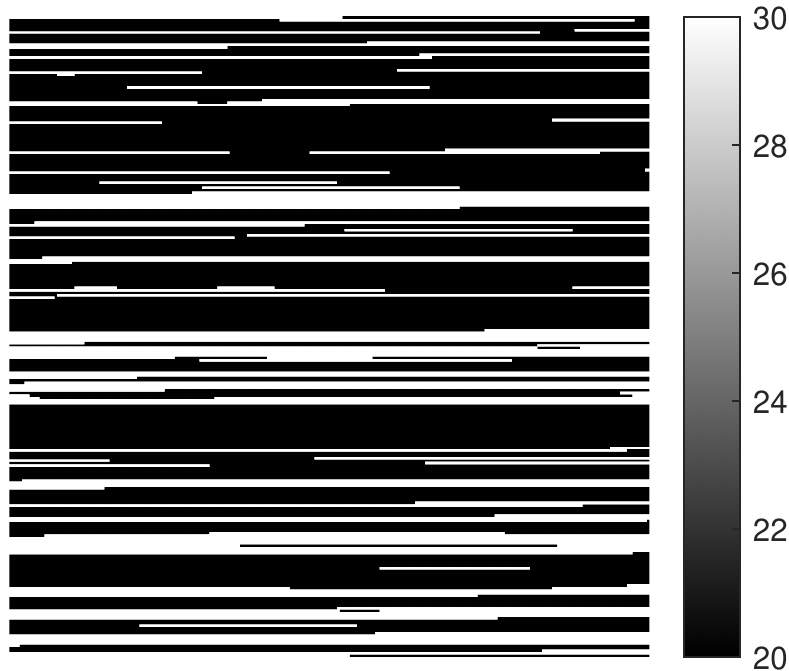}
        \caption{\centering Beam current incident on sample in Neon beam system}
        \label{fig:neon_beamCurrent}
    \end{subfigure}
    \begin{subfigure}{0.4\linewidth}
        \centering
        \includegraphics[width=1\linewidth]{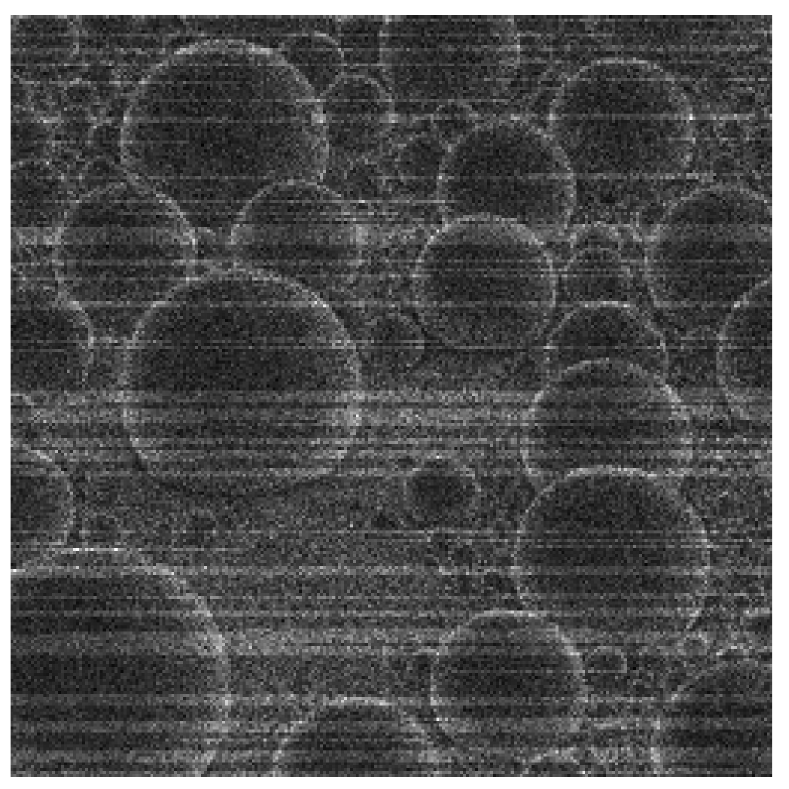}
        \caption{\centering Micrograph resulting from beam current in (c)}
        \label{fig:neon_stripes}
    \end{subfigure}

    \caption{Synthetic examples of stripe artifacts in simple models for helium and neon beam microscopes.
     A helium beam current is continuous-valued and slowly varying;
     a neon beam current has discrete jumps between known values.
    In (a) and (c), the displayed beam current is the mean number of incident ions per pixel over the pixel dwell time.
    }
    \label{fig:stripe_artifacts}
\end{figure}

\section{Introduction}
\IEEEPARstart{P}{article} beam microscopes image samples by detecting the secondary electrons (SEs) expelled from the sample by an incident beam of charged particles.
Scanning electron microscopes (SEM) employ an electron beam while newer focused ion beam (FIB) microscopes use a beam composed of heavier ions, such as helium, neon, or gallium.
FIB microscopes tend to have higher SE yield, larger depth of field, and finer resolution~\cite{WardNE:06,Notte2016,morgan2006introduction}.
Due to the heavier incident particles, these instruments are also often used to perform milling.
Image quality and milling accuracy depend heavily on the ability to maintain a stable beam current.
However, contamination within the instrument~\cite{Rahman2013}, or age of the source tip, may cause the beam intensity to fluctuate away from the desired setting. When the beam is raster scanned across a sample, these fluctuations give rise to horizontal stripe artifacts in the micrograph.
\Cref{fig:fib_beamCurrent} shows the beam current incident on a horizontally raster-scanned sample when the beam has slow, smooth intensity variation as often seen in SEM and helium ion microscope (HIM) systems~\cite{Barlow2016}.
\Cref{fig:fib_stripes} depicts the corresponding stripe artifacts that arise when a micrograph is formed under the incorrect assumption of constant beam current.
Neon beam microscopes have been less widely adopted because of difficulties in maintaining a stable beam current.
Here, the beam current can be modeled as toggling between known values~\cite{NotteNeon2010,rahmanNeon2012}, as shown in \Cref{fig:neon_beamCurrent}.
The resulting artifacts are shown in \Cref{fig:neon_stripes}.
Real-time knowledge of the beam current could prevent these artifacts, provide the operator an indicator of instrument fitness, and enable the instrument to adjust dwell time to improve milling outcomes and avoid sample damage.

Existing particle beam microscopes do not measure beam current, however algorithms have been developed to remove stripe content from micrographs post facto.
For example, in~\cite{Barlow2016,Barlow2018,schwartz2019}, low-frequency image content is removed, with~\cite{schwartz2019} also including total variation denoising.
Several algorithms~\cite{Khalilian2019,liu2018,ott2019} have also been developed to reduce the `curtaining' effect that arises due to variations in an ion beam's milling rate.
Unlike the stripe artifacts shown in \Cref{fig:fib_stripes,fig:neon_stripes}, which arise when the image formation algorithm incorrectly assumes a constant beam current, curtaining stripes accurately reflect grooves made in the underlying sample during milling.

In this paper, we establish that beam current can be accurately estimated \emph{online}, from the same SE count data used to form a micrograph, without the use of a calibrated sample.
This may seem to be like estimating two quantities from a single noisy measurement.
While that would be impossible without additional assumptions,
here we take advantage of the
 time-resolved (TR) measurement concept introduced in~\cite{PENG2020}.
TR measurement divides the dwell time into $n$ shorter sub-acquisitions and can be implemented without changes to the instrument.
It enables estimation of the number of incident particles despite the variability in the numbers of SEs generated by the incident particles.
This was previously shown to improve micrograph accuracy~\cite{PENG2020,peng2020analysis}
and to provide a
 natural robustness to imperfectly known beam current
~\cite{Watkins2021a,Watkins2021b}.
In this work, we go beyond robustness to explicitly estimate the  beam current. In addition to improving the accuracy of the estimated SE yield (i.e., producing a better micrograph),
this is of interest for control of sample damage, milling accuracy, and instrument diagnostics.

\subsection{Main Contributions}
\begin{itemize}
    \item \emph{An analysis of the joint estimation problem.}
    The \CRB{} (CRB) for joint estimation of SE yield and beam current from TR measurement is derived and used to show that under certain conditions,
    joint estimation is vanishingly more challenging than estimation of SE yield alone.
    \item \emph{Demonstration of joint estimation of SE yield and beam current at a single pixel.}
    Proposed estimators are evaluated on synthetic data and compared to the CRB.
    \item \emph{Joint estimation algorithms that exploit models for beam current variation and inter-pixel correlation.}
    Causal and non-causal joint estimators are proposed
    to exploit continuous and discrete Markov models for beam current,
    and for the continuous case total variation regularization of SE yield is also incorporated.
    Tested on synthetic microscopy data, the SE yield estimates are shown to beat state of the art methods, and beam current estimates are shown to closely match the ground truth.
    \item \emph{Numerical methods for Neyman Type A distributions.}
    Expressions and approximations for derivatives of the Neyman Type A negative log likelihood in terms of Touchard polynomials are given.
\end{itemize}

\subsection{Outline}
In \Cref{sec:measurement_models}, we summarize microscope abstractions, measurement models, and basic analyses from \cite[Sect.~II]{peng2020analysis}.
In \Cref{sec:feasibility_of_joint_estimation}, we use the \CRB{} to explore the feasibility of joint estimation.
Then, joint estimation at a single pixel is demonstrated in \Cref{sec:single_pixel_estimation}.
At the time scale of pixel-to-pixel scanning, beam current does not vary arbitrarily.
Thus, we develop methods to exploit simple Markov models for the beam current.
In \Cref{sec:multi_pixel_estimation},
motivated by electron and helium ion beams,
we consider joint estimation for continuous-valued smoothly varying beam current.
In \Cref{sec:neon_beam_algorithms},
motivated by neon ion beams~\cite{rahmanNeon2012,NotteNeon2010},
we consider joint estimation
where beam current is known to flip back and forth between two known values.
\Cref{sec:conclusion} provides concluding comments on the promise of joint estimation in particle beam microscopy.

\section{Measurement Models and Baseline Estimators}
\label{sec:measurement_models}
In this section, we describe our measurement model for the particle beam microscope, assuming direct SE detection;
see \cite[Sect.~II]{peng2020analysis} for additional details.
Though not yet common in commercial hardware, direct SE detection provides higher signal-to-noise ratio than employing scintillators and photomultiplier tubes~\cite{yamada1991electron,jin2008applications,mcmullan2009detective,Agarwal:2021arXiv}.
The advantages of time-resolved sensing do not depend on direct SE detection, as demonstrated experimentally in~\cite{PENG2020}.
Although our model describes both electron- and ion-beam microscopy, we will refer to incident particles as ions.

\subsection{Conventional Measurement}

The incident ion beam may be accurately modeled as a Poisson process~\cite{sim2004effect}.
So, at a single pixel with dwell time $t$, the number of incident ions $M$ is a Poisson random variable with mean $\lambda = \Lambda t$, where $\Lambda$ is rate of incident ions per unit time.
Dose is conventionally defined as the number of incident ions per unit area.
Since the absolute spatial and temporal scales are not important in our abstractions or methods,
we will refer to $\lambda$ as the \emph{dose} and, especially when it is unknown and potentially varying, as the \emph{beam current}.
The number of SEs expelled by the $i$th incident ion is also modeled as a Poisson random variable:
$X_i \sim \Poisson(\eta)$, where $\eta$ is the sample \emph{SE yield} at that pixel.
The goal in forming a micrograph is to show how $\eta$ varies from pixel to pixel.
The number of incident ions $M$ is not directly observed; it is at best inferred from the detected SEs.

At each pixel, a conventional microscope measures the sum of all SEs over the dwell time: $Y = \sum_{i=1}^M X_i$.
Modeled in this way, $Y$ is a \emph{Neyman Type A} compound Poisson random variable with probability mass function (PMF) given by
\begin{equation}
  \label{eq:neyman}
    \mathrm{P}_{Y}(y \sMid \eta, \lambda) = \frac{e^{-\lambda} \eta^y}{y!} \sum_{m = 0}^{\infty} \frac{(\lambda e^{-\eta}) ^m m^y}{m!}
    \quad
    y = 0,\,1,\,\ldots,
\end{equation}
with mean
\begin{equation}
  \label{eq:Y-mean}
    \E{Y} = \lambda \eta.
\end{equation}
The conventional estimate for $\eta$ assumes $\lambda$ is known and operates independently at each pixel:
\begin{equation}
  \etaBaseline(Y,\,\lambda) = \frac{Y}{\lambda}.
  \label{eq:eta-baseline}
\end{equation}
When $\lambda$ is assumed to be constant, but actually varies in time, the conventional estimate \eqref{eq:eta-baseline} gives rise to prominent stripe artifacts like the ones shown in \Cref{fig:fib_stripes,fig:neon_stripes}.

\subsection{Continuous-Time Time-Resolved Measurement}
The continuous-time (CT) measurement model is an idealization of an FIB microscope first introduced in~\cite{peng2020analysis}.
It allows us to study the limits of TR measurement.
Here, we imagine that we have direct detection of SEs with perfect temporal precision of each SE burst.
In our model, some incident ions result in no detected SEs because $\mathrm{P}(X_i = 0)$ is nonzero.
There is no observed difference between the lack of an incident ion and an incident ion resulting in no detected SEs.
Thus, the CT measurement is
\begin{equation}
      \label{eq:CTTR-observation}
      \left\{\Mtilde,\,(\Ttilde_1,\Xtilde_1),\,(\Ttilde_2,\Xtilde_2),\,\ldots\,\,(\Ttilde_{\Mtilde},\Xtilde_{\Mtilde})\right\},
\end{equation}
where $\Mtilde$ is the number of incident ions that result in at least one SE,
 $\Ttilde_i$ is the time of the $i$th such incident ion,
and $\Xtilde_i$ is the corresponding SE counts.
The probability of an incident ion yielding more than one SE is $1-e^{-\eta}$, so we have $\Mtilde \sim \Poisson(\lambda(1-e^{-\eta}))$ with PMF
\begin{equation}
    \label{eq:Mtilde-PMF}
    \mathrm{P}_{\Mtilde}(\mtilde \sMid \eta,\lambda) = \exp(-\lambda(1 - e^{-\eta}))\frac{(\lambda(1 - e^{-\eta}))^{\mtilde}}{{\mtilde}!}.
\end{equation}
The $X_i$s are independent and identically distributed with PMF
\begin{equation}
    \label{eq:PMF_Xtilde}
    \mathrm{P}_{\Xtilde_i}(j \sMid \eta) = \frac{e^{-\eta}}{1 - e^{-\eta}} \cdot \frac{\eta^{j}}{j!},
    \qquad j = 1,\,2,\,\ldots,
\end{equation}
which is the zero-truncation of the $\Poisson(\eta)$ distribution.

Several estimators of $\eta$ from a CT measurement when $\lambda$ is known were studied in~\cite{peng2020analysis}.
Here, we are interested in the \emph{assumed dose} $\lamtilde$ not necessarily being equal to the dose $\lambda$.
The CT maximum likelihood (ML) estimate
evaluated using the assumed dose $\lamtilde$
 is the unique root of the following equation:
\begin{align}\label{eq:eta_trml_conti}
\etaCTMLgiven = \frac{Y}{\Mtilde + \lamtilde e^{-\etaCTMLgiven}}.
\end{align}

\subsection{Discrete-Time Time-Resolved Measurement}
The discrete-time (DT) measurement model assumes
the per-pixel dwell time $t$ is split into $n$ sub-acquisitions of equal duration.  Each sub-acquisition then has the same distribution as a conventional measurement with dose $\lambda/n$.
A key observation of this paper is that the  $n$-length DT measurement vector contains rich information about both the dose $\lambda$ and SE yield $\eta$.
Imagine short sub-acquisitions with dose $\lambda/n$ small enough that observing more than one incident ion per sub-acquisition is unlikely.
In this case, with large enough $\eta$, the number of sub-acquisitions where the number of observed SEs is strictly positive is roughly equal to the number of incident ions.

In this work, we will use subscript $k$ for pixel index,
so pixel $k$ has beam current $\lambda_k$ and SE yield $\eta_k$; vectors $\etabf$ and $\lambf$ contain the values of $\eta$ and $\lambda$ for each of the $p$ pixels in a sample. The vector $\ybf\in\mathbb{R}^{pn}$ gathers measurements across all pixels, with the vector of $n$ time-resolved measurements at the $k$th pixel given by $\ybf_k = [y^{(1)}_k,y^{(2)}_k,\dots,y^{(n)}_k]$. 
Since the entries in $\ybf_k$ are independent, their joint PMF is 
\begin{equation}
    \label{eq:DTTR-distribution}
    \mathrm{P}_{\yvec_k}(\ybf_k \sMid \eta_k,\lambda_k)
    = \prod_{i=1}^n \mathrm{P}_Y\!\left( y^{(i)}_k \sMid \eta_k,\lambda_k/n \right),
\end{equation}
where $\mathrm{P}_Y(\cdot \sMid \cdot, \cdot)$ is given by \eqref{eq:neyman}.
Under assumed dose $\lamtilde$, the DT ML estimator
finds the value of $\eta_k$,
separately at each sample pixel,
that maximizes the likelihood in \eqref{eq:DTTR-distribution}:
\begin{align}
\label{eq:eta_DTML}
\etaDTMLgiven_k(\ybf_k,\,\lamtilde) = \argmax_{\eta_k} \, \prod_{i=1}^{n} \mathrm{P}_{Y}\!\left(y^{(i)}_k \sMid \eta_k, \lamtilde/n \right).
\end{align}
In this work, we use TR data to estimate \emph{both} $\lambda$ and $\eta$.
When oracle knowledge of the dose $\lambda$ is assumed while estimating $\eta$,
the estimate is denoted $\etaOracle_k$.

\section{Feasibility of Joint Estimation of SE Yield and Beam Current}
\label{sec:feasibility_of_joint_estimation}
A single conventional measurement $Y$ combines SE yield $\eta$ and beam current $\lambda$ inseparably,
as suggested by \eqref{eq:Y-mean}.
However, in this section, we show that using time-resolved data, joint estimation of $\eta$ and $\lambda$ becomes possible.
In \Cref{subsec:crb} we derive CT and DT CRBs for the joint estimation problem. In \Cref{sec:low_eta_challenge}, we discuss the challenge of joint estimation at low-$\eta$ pixels.

\subsection{\CR~Bound }
\label{subsec:crb}
The CRB provides a lower bound for the variance of an unbiased estimator of our unknown parameter $\theta = [\eta,\,\lambda]$. In this section we derive the CRB for joint estimation of $\eta$ and $\lambda$ under both CT and DT measurement models and use them to explore the feasibility and challenge of joint estimation.

\subsubsection{Continuous-Time \CR~Bound}
\label{subsubsec:ct_crb}
\begin{figure}
    \centering
    \includegraphics[width=.8\linewidth]{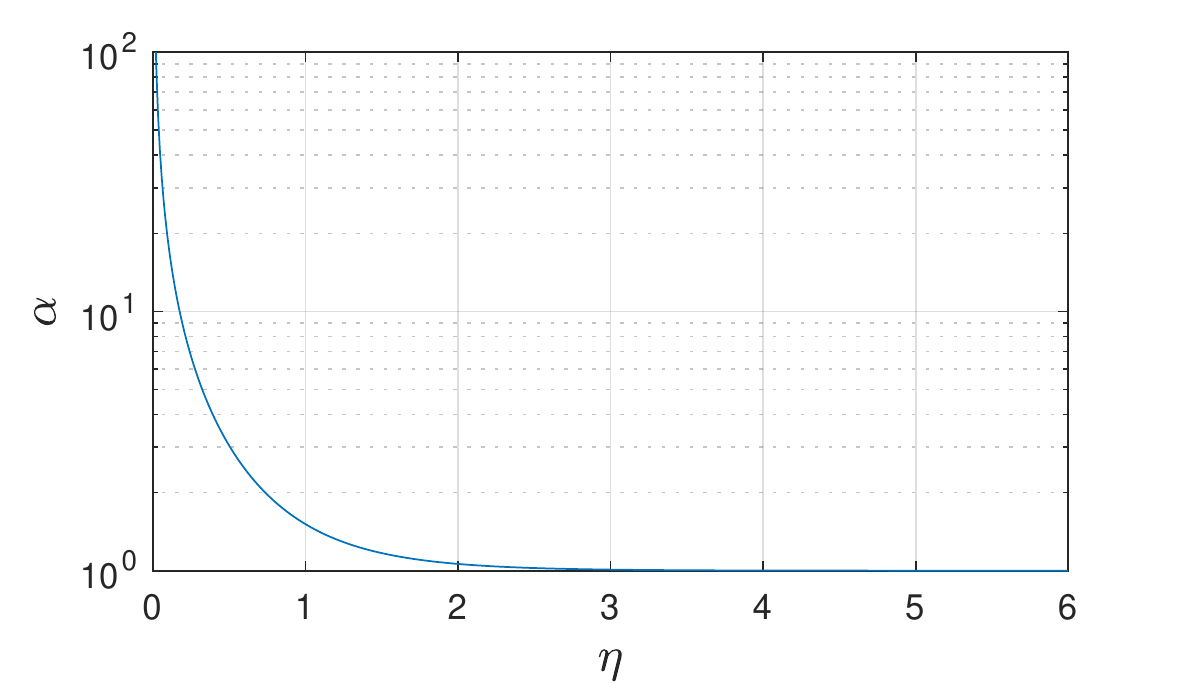}
    \caption{Factor from \eqref{eq:alpha} by which the CRB for joint estimation of $\eta$ and $\lambda$ exceeds that of estimating one parameter with the other given.
   }
    \label{fig:prodTerm}
\end{figure}

With a CT measurement \eqref{eq:CTTR-observation},
the entries of the Fisher information (FI) matrix are given by
\begin{align}
\label{eq:fisherInfoCT}
    [\fmat^{\rm CT}]_{i,j} = \mathrm{E}\Bigg[ &\Bigg(\frac{\partial{\rm log}\,\mathrm{P}_{\mtilde,\ttildevec,\xtildevec}(\mtilde,\ttilde,\xtilde \sMid \eta,\lambda) }{\partial \theta_i}\Bigg)\nonumber\\
    &\Bigg(\frac{\partial{\rm log}\,\mathrm{P}_{\mtilde,\ttildevec,\xtildevec}(\mtilde,\ttilde,\xtilde \sMid \eta,\lambda) }{\partial \theta_j}\Bigg) \, \Bigg\lvert \, \eta,\lambda\Bigg].
\end{align}
As shown in~\cite[Sect.~III-B]{peng2020analysis},
\begin{equation}
    \label{eq:fisher_ct_eta}
    [\fmat^{\rm CT}(\eta,\lambda)]_{1,1} = \lambda\left(\frac{1}{\eta}-e^{-\eta}\right).
\end{equation}
The FI about $\lambda$ in the CTTR measurement is
\begin{align}
    [\fmat^{\rm CT}]_{2,2} &\eqlabel{a} \E{\Mtilde}\fmat_{\Xtilde_i}(\lambda;\,\eta) + \fmat_{\Mtilde}(\lambda;\,\eta)\nonumber\\
    &\eqlabel{b} \fmat_{\Mtilde}(\lambda;\,\eta),
\end{align}
where (a) follows from the application of the chain rule for FI, the fact that the conditional distribution of $\Ttilde_i$ given $\Mtilde$ has no dependence on $\eta$, and the independence of $\{\Mtilde, \Xtilde_1, \Xtilde_2, \ldots, \Xtilde_{\Mtilde}\}$;
and (b) from the fact that $P_{\Xtilde_i}(j;\,\eta)$ is not a function of $\lambda$ and thus $I_{\Xtilde_i}(\lambda;\,\eta) = 0$. It follows that
\begin{align}
[\fmat^{\rm CT}&]_{2,2} = \E{\left(\frac{\partial {\rm log}\mathrm{P}_{\Mtilde}(\Mtilde;\lambda,\eta)}{\partial \lambda}\right)^{\!2};\,\lambda}\nonumber\\
&=\sum_{j=0}^\infty\left(-(1-e^{-\eta})+\frac{j}{\lambda}\right)^{\!2}\frac{e^{-\lambda(1-e^{-\eta})}[\lambda(1-e^{-\eta})]^j}{j!}\nonumber\\
&=\frac{1-e^{-\eta}}{\lambda}.
\end{align}
The cross terms in the the FI matrix are
\begin{align}
    [\fmat^{\rm CT}]_{1,2} &=  [\fmat^{\rm CT}]_{2,1}\nonumber\\
    =& \, \E{\left(\frac{\partial {\rm log}\mathrm{P}_{\Mtilde}(\Mtilde;\lambda,\eta)}{\partial \lambda}\right)\left(\frac{\partial {\rm log}\mathrm{P}_{\Mtilde}(\Mtilde;\lambda,\eta)}{\partial \eta}\right)}\nonumber\\
    =& \, \sum_{j=0}^\infty\left( -(1-e^{-\eta})+\frac{j}{\lambda}\right)\left(\frac{e^{-\eta}}{1-e^{-\eta}}j-\lambda e^{-\eta}\right)\nonumber\\
     & \qquad \cdot \frac{e^{-\lambda(1-e^{-\eta})}[\lambda(1-e^{-\eta})]^j}{j!}\nonumber\\
    =& \, e^{-\eta}.
    \label{eq:crossTerms}
\end{align}
Thus, the full CT FI matrix is
\begin{equation}
\label{eq:ct_fi_mat}
{\fmat}^{\rm CT} = 
\begin{bmatrix}
{\lambda\left(\frac{1}{\eta}-e^{-\eta}\right)} & {e^{-\eta}} \\
{e^{-\eta}} & \frac{1}{\lambda}(1-e^{-\eta})
\end{bmatrix}.
\end{equation}
When one parameter is to be estimated and the other is given, the CT CRB is given by
\begin{subequations}
    \label{eq:crb_ct_oneUnknown}
\begin{align}
    \sigma_{\eta|\lambda}^2 \geq {\rm CRB}^{\rm CT}(\eta|\lambda) &= \left[[\fmat^{\rm CT}]_{1,1}\right]^{-1} = \frac{1}{\lambda}\left(\frac{1}{\eta}-e^{-\eta}\right)^{\!-1},\label{eq:ct_crb_eta_given_lambda} \\
    \sigma_{\lambda|\eta}^2 \geq {\rm CRB}^{\rm CT}(\lambda|\eta) &= \left[[\fmat_{\rm CT}]_{2,2}\right]^{-1}  = \frac{\lambda}{1-e^{-\eta}}.
\end{align}
\end{subequations}
When both $\eta$ and $\lambda$ are unknown, the CT CRB for each parameter is computed by inverting $\fmat_{\rm CT}$:
\begin{subequations}
    \label{eq:crb_ct_bothUnknown}
\begin{align}
       \sigma_{\eta}^2 &\geq {\rm CRB}^{\rm CT}(\eta) = \left[[\fmat^{\rm CT}]^{-1}\right]_{1,1} = \frac{1-e^{-\eta}}{\frac{\lambda}{\eta}(1-e^{-\eta})-\lambda e^{-\eta}}, \\
    \sigma_{\lambda}^2 &\geq {\rm CRB}^{\rm CT}(\lambda) = \left[[\fmat^{\rm CT}]^{-1}\right]_{2,2}  = \frac{\lambda(1-\eta e^{-\eta})}{(1-e^{-\eta})-\eta e^{-\eta}}.
\end{align}
\end{subequations}
Note that the CRBs when both parameters are unknown \eqref{eq:crb_ct_bothUnknown} may be written in terms of the CRBs for that same parameter when the other parameter is known \eqref{eq:crb_ct_oneUnknown}:
\begin{subequations}
  \label{eq:crb_ct_alpha}
\begin{align}
    {\rm CRB}_{\rm CT}(\eta) &= \alpha(\eta) \,{\rm CRB}_{\rm CT}(\eta|\lambda), \\
    {\rm CRB}_{\rm CT}(\lambda) &= \alpha(\eta) \,{\rm CRB}_{\rm CT}(\lambda|\eta),
\end{align}
\end{subequations}
where
\begin{equation}
    \label{eq:alpha}
    \alpha(\eta) = \frac{1-(1+\eta)e^{-\eta}+\eta e^{-2\eta}}{1-(1+\eta)e^{-\eta}}.
\end{equation}
The factor $\alpha(\eta)$, which is plotted in \Cref{fig:prodTerm}, represents the added challenge of the joint estimation problem compared to estimating one parameter given the other.
When $\eta \geq 2$, as is typical for FIB, $\alpha(\eta) \approx 1$, and furthermore
 $\lim_{\eta\rightarrow\infty}\alpha(\eta)=1$;
i.e., asymptotically, jointly estimating both parameters is no more challenging.
When $\eta$ is low, $\Mtilde$ becomes a less suitable proxy for, and contains less information about, the number of incident ions $M$.

\subsubsection{Discrete-Time \CR~Bound}
\label{subsubsec:dt_crb}
The Fisher information matrix about unknown parameter $\theta$ in $n$ time resolved measurements, each with a per sub-acquisition dose of $\lambda/n$ is given by
\begin{align}
     &[\fmat^{\rm DT}(\eta,\lambda;\,n)]_{i,j} \nonumber\\
     &= n \, \E{\left(\frac{\partial {\rm log}\mathrm{P}_{Y}\big(y;\, \eta,\frac{\lambda}{n}\big)}{\partial\theta_i}\right) \!
             \left(\frac{\partial {\rm log}\mathrm{P}_{Y}\big(y;\, \eta,\frac{\lambda}{n}\big)}{\partial\theta_j}\right) \Bigg| \, \eta,\lambda,n} \nonumber\\
     &=n\sum_{y = 0}^\infty\left(\frac{\partial {\rm log}\mathrm{P}_{Y}(y;\, \eta,\Frac{\lambda}{n})}{\partial\theta_i}\right) \nonumber \\
     & \qquad \qquad        \left(\frac{\partial {\rm log}\mathrm{P}_{Y}(y;\, \eta,\Frac{\lambda}{n})}{\partial\theta_j}\right)
                           \mathrm{P}_Y(y;\,\eta,\Frac{\lambda}{n}).
  \label{eq:IDT}
\end{align}
As derived in~\cite{PENG2020}, 
\begin{align}
  \label{eq:dlogPy_deta}
    \frac{\partial {\rm log}\mathrm{P}_{Y}(y;\, \eta,\Frac{\lambda}{n})}{\partial\eta}
    = \frac{y}{\eta}-\frac{\mathrm{P}_Y(y+1;\eta,\Frac{\lambda}{n})}{\mathrm{P}_Y(y;\eta,\Frac{\lambda}{n})}\frac{y+1}{\eta},
\end{align}
and similarly, the derivative with respect to $\lambda$ is
\begin{align}
  \label{eq:dlogPy_dlambda}
    \frac{\partial {\rm log}\mathrm{P}_{Y}(y;\, \eta,\Frac{\lambda}{n})}{\partial\lambda}
    = -\frac{1}{n}+\frac{\mathrm{P}_Y(y+1;\eta,\Frac{\lambda}{n})}{\mathrm{P}_Y(y;\eta,\Frac{\lambda}{n})}\frac{y+1}{\eta}\frac{1}{\lambda}.
\end{align}
Substituting \eqref{eq:dlogPy_deta} and \eqref{eq:dlogPy_dlambda} into \eqref{eq:IDT} and truncating the series appropriately enables numerical approximation of the FI matrix.
CRBs are then analogous to \eqref{eq:crb_ct_oneUnknown} and \eqref{eq:crb_ct_bothUnknown}.

\subsection{The Challenge of Low SE Yield Samples}
\label{sec:low_eta_challenge}
\begin{figure}
    \centering
    \begin{subfigure}{0.49\linewidth}
        \centering
        \includegraphics[width=1\linewidth]{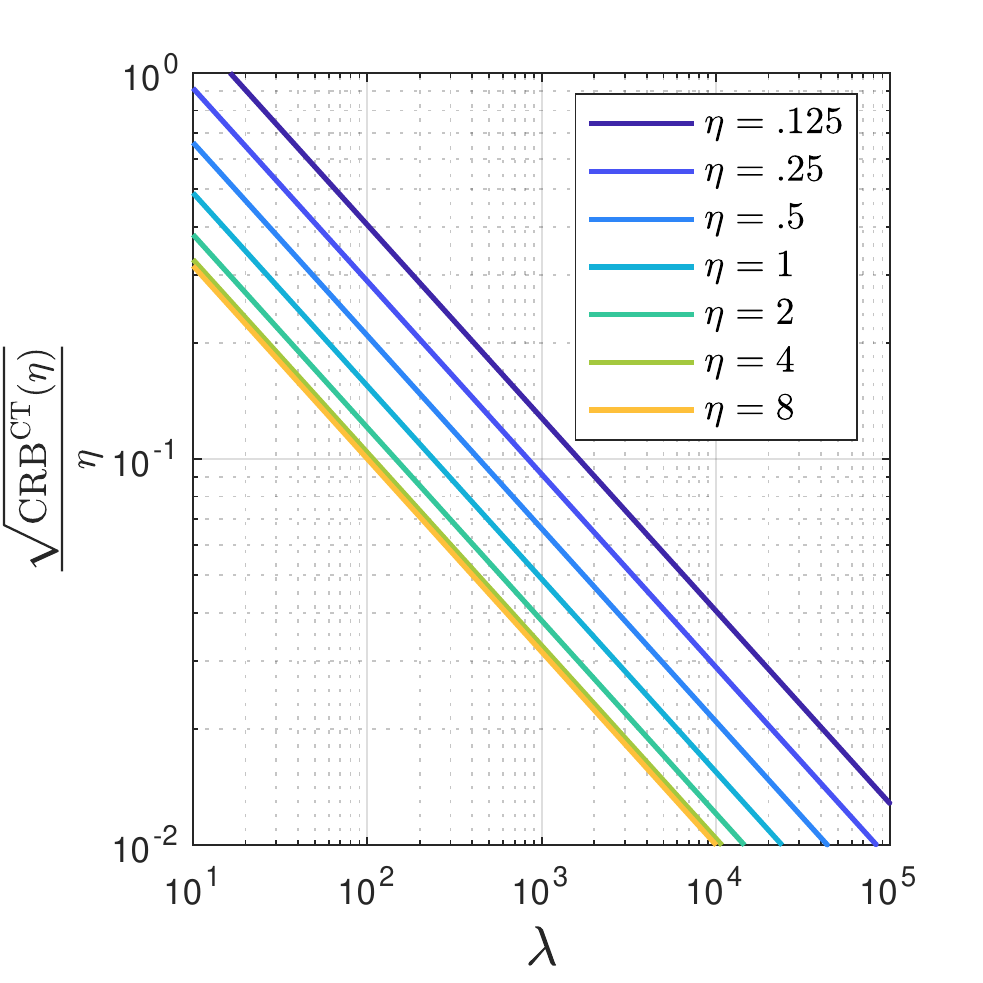}
        \caption{}
        \label{fig:crbCT_vs_dose_eta}
    \end{subfigure}
    \begin{subfigure}{0.49\linewidth}
        \centering
        \includegraphics[width=1\linewidth]{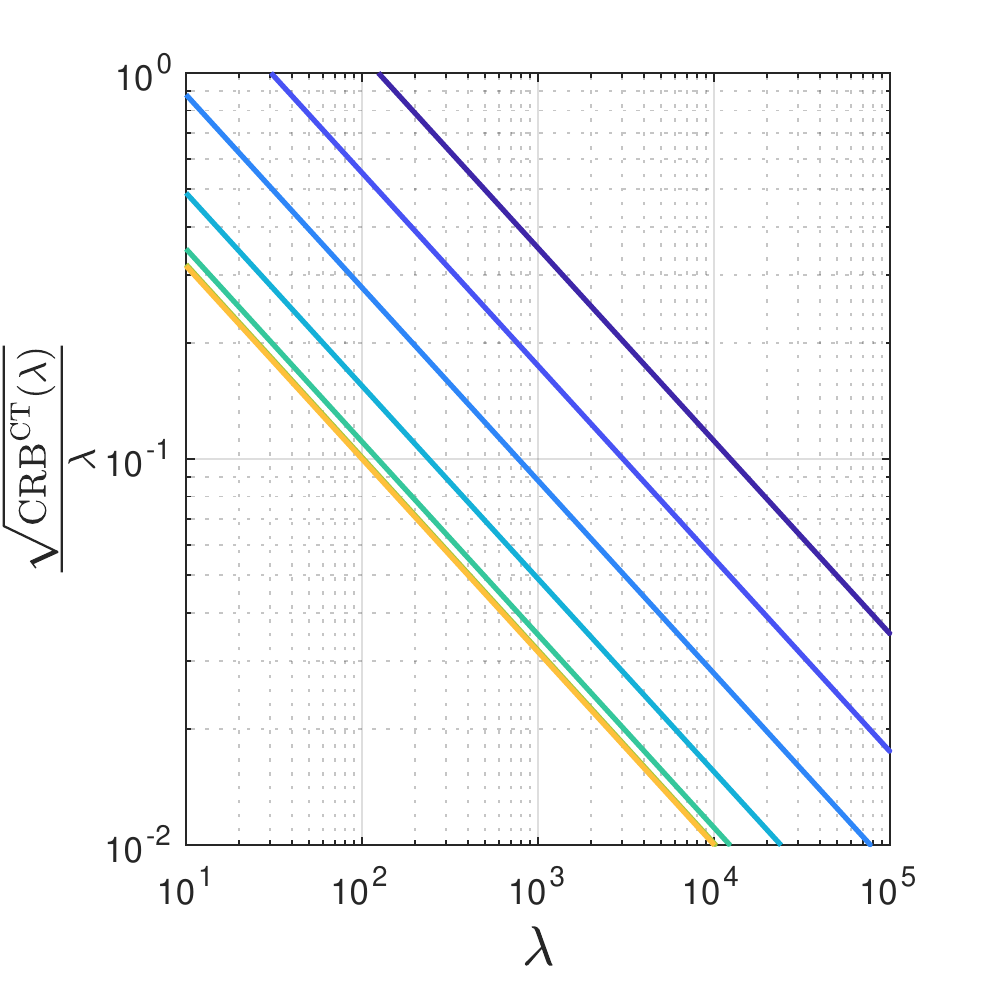}
        \caption{}
    \end{subfigure}
    \caption{
    Normalized CRBs as functions of $\lambda$ for several values of $\eta$.
    As $\eta$ decreases, the dose $\lambda$ required to achieve any fixed desired relative error increases.
    }
    \label{fig:crbCT_vs_dose}
\end{figure}

In \Cref{subsec:crb}, we showed that as $\eta$ gets smaller, jointly estimating $\eta$ and $\lambda$ becomes increasingly more difficult than estimation of $\eta$ or $\lambda$ alone.
\Cref{fig:crbCT_vs_dose} further illustrates the challenge of lower $\eta$ samples.
Here we plot the normalized continuous-time CRBs for $\hat{\eta}$ and $\hat{\lambda}$ as  functions of the total dose $\lambda$.
Different colored curves denote different $\eta$ values. Note that the intersection of each curve with a horizontal line indicates the dose $\lambda$ required to achieve the corresponding normalized CRB\@.
For example, in order for the normalized standard deviation to be lower bounded by $10^{-1}$, a dose on the order of $10^2$ is required when $\eta = 4$; when $\eta = 0.25$, a dose of nearly $10^3$ is required.

In \Cref{sec:single_pixel_estimation}, we study joint estimation of $\eta$ and $\lambda$ at a $single$ pixel. Although we show that this is possible, as demonstrated in \Cref{fig:crbCT_vs_dose}, even pixels with large $\eta$ require a large dose to form high-quality estimates.
In \Cref{sec:multi_pixel_estimation,sec:neon_beam_algorithms}, we exploit \emph{inter-pixel} correlations of both $\eta$ and $\lambda$ to enable high-quality joint estimation at moderate
doses.

\section{Joint Estimation at a Single Pixel}
\label{sec:single_pixel_estimation}
In this section, we derive CT and DT ML estimates for $\eta$ and $\lambda$ using only measurements acquired at a \emph{single} pixel. In \Cref{sec:singlePixelEstimatorPerformance}, we evaluate the performance of these estimators and compare to the CRBs derived in  \Cref{subsec:crb}
\subsection{Continuous-Time Time-Resolved ML Estimation}
\label{sec:CTTR_ML}

Recall the continuous-time measurement model in \eqref{eq:CTTR-observation}. The $\Xtilde_i$ variables are independent, so using \eqref{eq:PMF_Xtilde},
the joint distribution of the SE count vector
(conditioned on $\Mtilde = \mtilde$)
is
\begin{align}
    \label{eq:Xtilde-likelihood}
  \prod_{i=1}^{\mtilde}
    \mathrm{P}_{\Xtilde_i}(j_i \sMid \eta)
    &= \left(\frac{e^{-\eta}}{1 - e^{-\eta}}\right)^{\!\mtilde} \frac{\eta^{j_1+j_2+\cdots+j_{\mtilde}}}{j_1! \, j_2! \, \cdots \, j_{\mtilde}!}\nonumber\\
    &= c \left(\frac{e^{-\eta}}{1 - e^{-\eta}}\right)^{\!\mtilde} \eta^{y},
\end{align}
where the simplification comes from identifying the sum as the total SE count $y$
and replacing the product of factorials with an unspecified constant
because this is immaterial to estimation of $\eta$ and $\lambda$.
Combining \eqref{eq:Mtilde-PMF} and \eqref{eq:Xtilde-likelihood},
the relevant likelihood is
\begin{align}
    \mathrm{P}_{\Mtilde,Y}&(\mtilde,y \smid \eta,\lambda) \nonumber \\
    &= c \exp(-\lambda(1 - e^{-\eta}))\frac{(\lambda(1 - e^{-\eta}))^{\mtilde}}{{\mtilde}!}
        \!\left(\frac{e^{-\eta}}{1 - e^{-\eta}}\right)^{\!\mtilde} \eta^{y} \nonumber \\
    &= c \exp(-\lambda(1 - e^{-\eta}))\lambda^{\mtilde} {e^{-\eta \mtilde}} \eta^{y},
    \label{eq:MtildeY-likelihood}
\end{align}
where $\mtilde!$ has been absorbed into the constant $c$
because this is immaterial to estimation of $\eta$ and $\lambda$.
Omitting the constant,
\begin{equation}
    \label{eq:NLL}
    - \log \mathrm{P}_{\Mtilde,Y}(\mtilde,y \smid \eta,\lambda)
    = \lambda(1 - e^{-\eta}) - \mtilde \log \lambda + \eta \mtilde - y \log \eta.
\end{equation}
Taking derivatives
of ${- \log \mathrm{P}_{\Mtilde,Y}(\mtilde,y \smid \eta,\lambda)}$
gives
\begin{eqnarray}
  \frac{\del}{\del\lambda}[\sim] & = &  (1 - e^{-\eta}) - \frac{\mtilde}{\lambda}, \\
  \frac{\del}{\del\eta}[\sim]    & = &  \lambda e^{-\eta} + \mtilde - \frac{y}{\eta}.
\end{eqnarray}
Setting these to zero to find the joint ML estimate gives that $\etaCTML$ is the root of 
\begin{equation}
\label{eq:ctml_eta}
    \frac{\eta}{1-\exp(-\eta)} = \frac{y}{\mtilde},
\end{equation}
which then can be substituted to give
\begin{equation}
\label{eq:ctml_lambda}
    \lambdaCTML = \frac{\mtilde}{1 - \exp(-\etaCTML)}.
\end{equation}
These values can be justified heuristically without the ML property.
Since $\mtilde/(1-\exp(-\eta)$ is a good proxy for the number of incident ions,
\eqref{eq:ctml_eta} sets $\etaCTML$ to be the number of detected SEs $y$ divided by this estimate for the number of incident ions.
In~\cite{peng2020analysis},
this was called the \emph{continuous-time Lambert quotient mode estimator}, and
it was shown to differ from the ML estimate of $\eta$ with $\lambda$ known.

Note that when at most a single SE is observed in response to each of the incident ions
(i.e., $\Xtilde_i = 1$ for all $i$),
we have $y = \mtilde$, so the right side of \eqref{eq:ctml_eta} equals 1.
The left side of \eqref{eq:ctml_eta} approaches 1 as $\eta$ approaches 0;
thus, we assign $\etaCTML = 0$, and
substituting in
\eqref{eq:ctml_lambda} gives $\lambdaCTML =\infty$.
We address this singularity by placing a reasonable upper bound $\lmax$ on our estimate for $\lambda$.
The smallest nonzero estimate we can obtain for $\eta$ is then $1/\lmax$.
Requiring a large dose to be able to accurately estimate a small value of $\eta$ at a single pixel is consistent with the normalized CRBs in \Cref{fig:crbCT_vs_dose_eta}.
This limitation is one motivation for
our use of inter-pixel correlations in

\Cref{sec:multi_pixel_estimation,sec:neon_beam_algorithms}.

\subsection{Discrete-Time Time-Resolved ML Estimation}
\label{sec:DTTR_measurement}
At the $k$th pixel, we acquire a vector $\ybf_k$ of $n$ time-resolved measurements with joint PMF given in \eqref{eq:DTTR-distribution}.
The corresponding joint ML estimate is
\begin{align}
  (\etaDTML_k ,\lamDTML_k)
    &= \argmax_{\eta_k, \lambda_k} \mathrm{P}_{\yvec_k}(\ybf_k \sMid \eta_k,\lambda_k )\nonumber\\
    &= \argmin_{\eta_k, \lambda_k} \left[-\log \mathrm{P}_{\yvec_k}(\ybf_k \sMid \eta_k,\lambda_k )\right].
    \label{eq:dtml}
\end{align}
The objective function in \eqref{eq:dtml} is a sum of $n$ terms, each a logarithm of the Neyman Type A PMF in \eqref{eq:neyman}.
While difficult to work with analytically, since the decision variable is only two-dimensional and the objective function is smooth, numerical evaluation of \eqref{eq:dtml} is not difficult.
The numerical experiments in following sections use gradient descent methods based on derivatives
\eqref{eq:loglike-first-derivative-approx2} and \eqref{eq:loglike-first-derivative-eta-approx2}
derived in the appendix.

Similar to the CT case in \Cref{sec:CTTR_ML},
observing at most a single SE  per sub-acquisition creates a singularity
whereby
\eqref{eq:dtml} gives $(\etaDTML_k ,\lamDTML_k) = (0,\infty)$.
In practice,
we can again place a reasonable upper bound $\lmax$ on our $\lambda$ estimate and
then estimate $\eta$ accordingly.

\begin{figure*}
    \centering
    \begin{subfigure}{0.245\linewidth}
        \centering
        \includegraphics[width=1\linewidth]{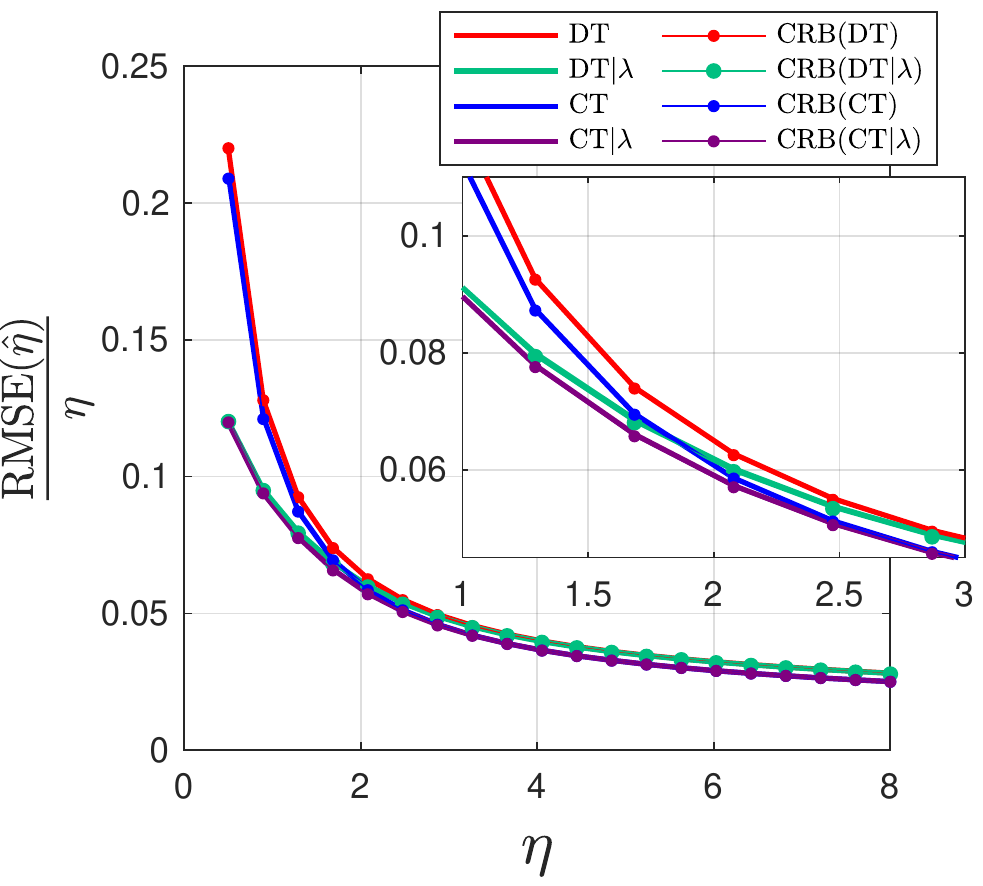}
        \caption{$\frac{{{\rm RMSE}(\widehat{\eta})}}{\eta}$ vs.\ $\eta$}
        \label{fig:mseEta_vs_eta}
    \end{subfigure}
    \begin{subfigure}{0.245\linewidth}
        \centering
        \includegraphics[width=1\linewidth]{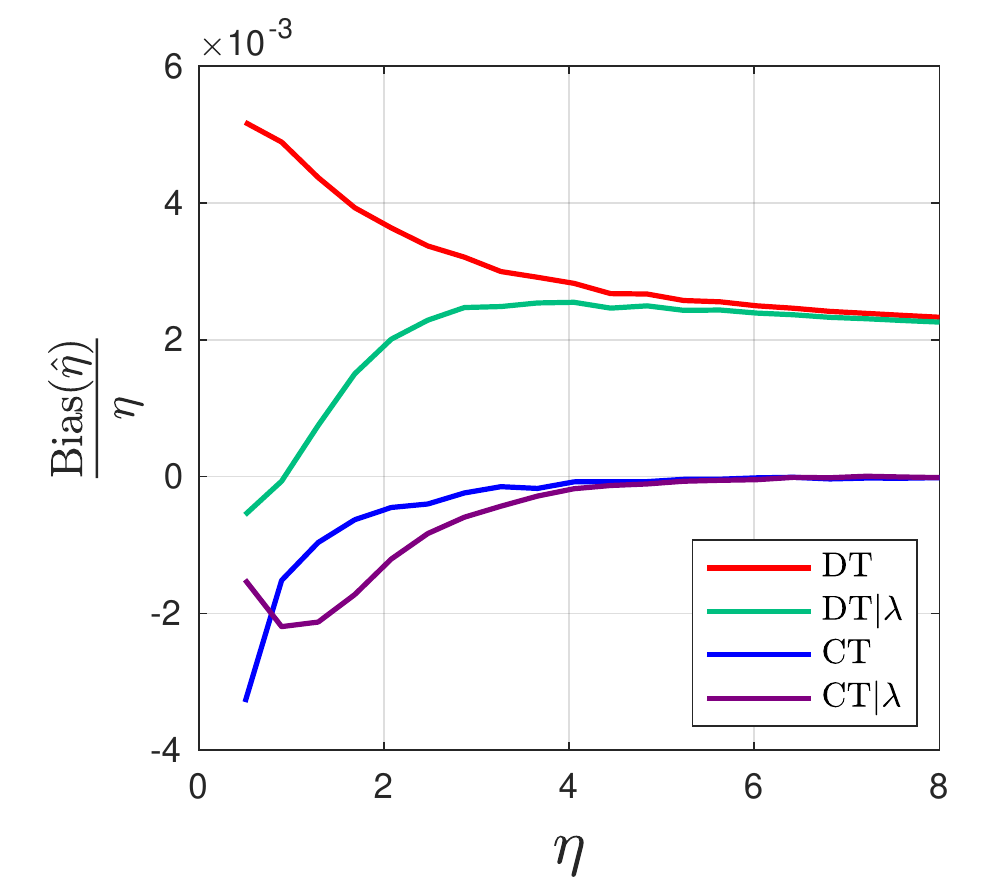}
        \caption{$\frac{{\rm Bias}(\widehat{\eta})}{\eta}$ vs.\ $\eta$}
        \label{fig:biasEta_vs_eta}
    \end{subfigure}
    \begin{subfigure}{0.245\linewidth}
        \centering
        \includegraphics[width=1\linewidth]{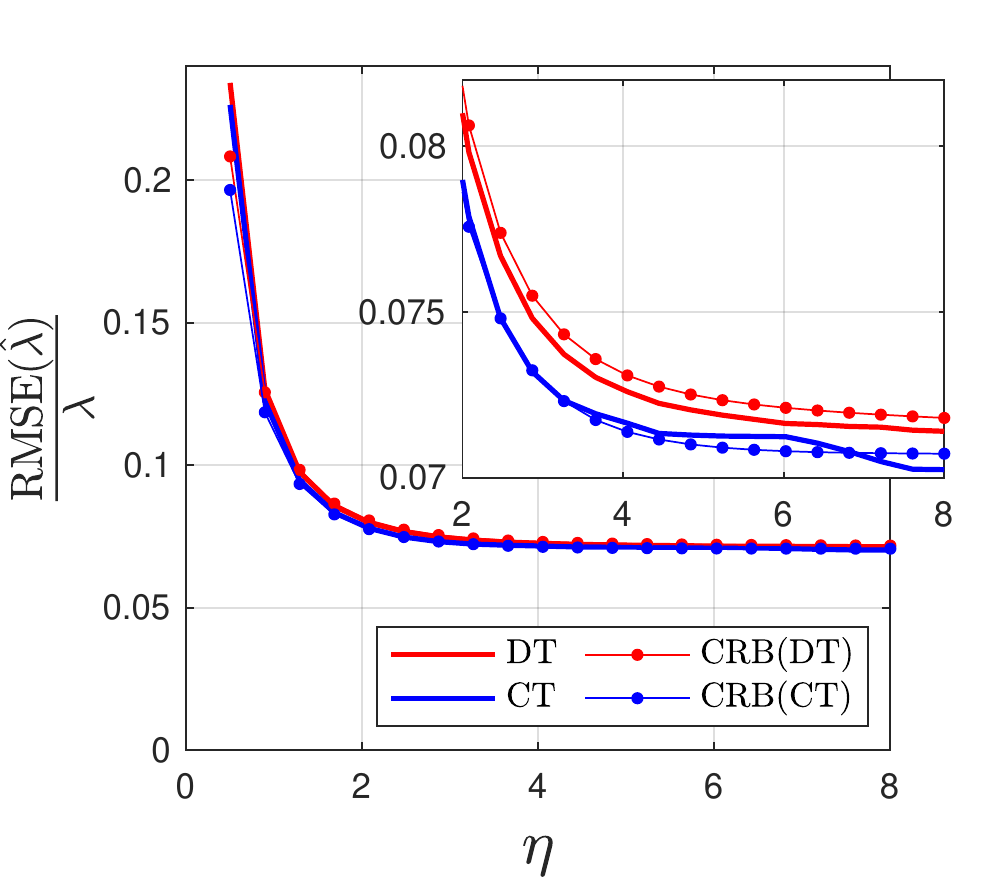}
        \caption{$\frac{{{\rm RMSE}(\widehat{\lambda})}}{\lambda}$ vs.\ $\eta$}
        \label{fig:mseLam_vs_eta}
    \end{subfigure}
    \begin{subfigure}{0.245\linewidth}
        \centering
        \includegraphics[width=1\linewidth]{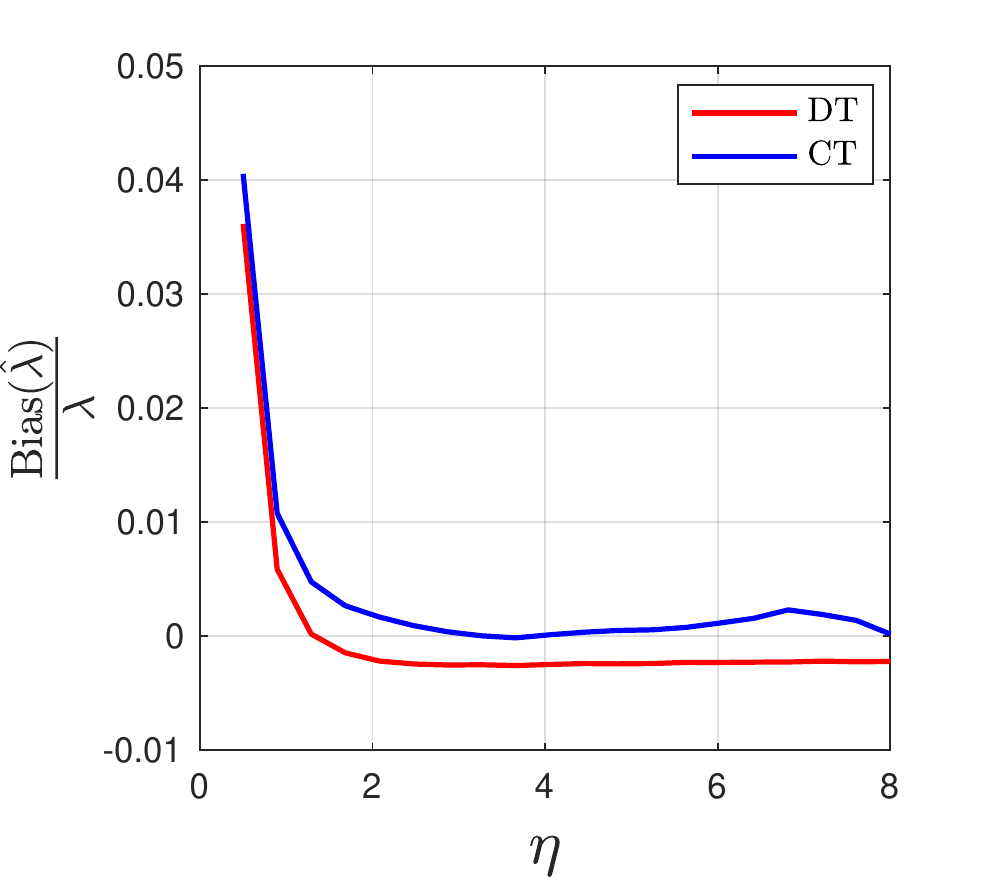}
        \caption{$\frac{{\rm Bias}(\widehat{\lambda})}{\lambda}$ vs.\ $\eta$}
        \label{fig:biasLam_vs_eta}
    \end{subfigure}
    \caption{Normalized RMSE and bias as functions of $\eta$ for single-pixel estimators with $\lambda = 200$ and, in discrete-time cases, $\Frac{\lambda}{n} = 0.1$.
    In panels (a) and (c), the normalized square roots of the \CRB s are plotted for reference.
    } 
    \label{fig:singlePixelPerformance_vs_eta}
\end{figure*}

\begin{figure*}
    \centering
    \begin{subfigure}{0.245\linewidth}
        \centering
        \includegraphics[width=1\linewidth]{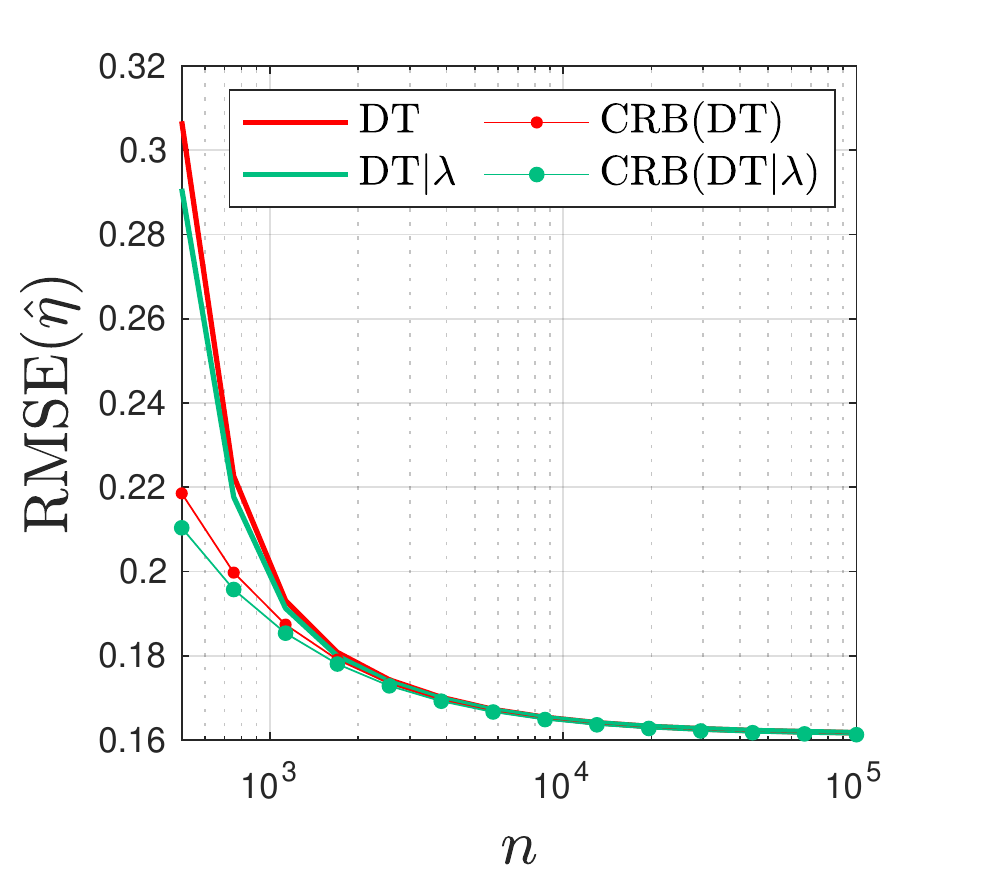}
        \caption{${\rm RMSE}(\widehat{\eta})$ vs.\ $n$}
        \label{fig:mseEta_vs_n}
    \end{subfigure}
    \begin{subfigure}{0.245\linewidth}
        \centering
        \includegraphics[width=1\linewidth]{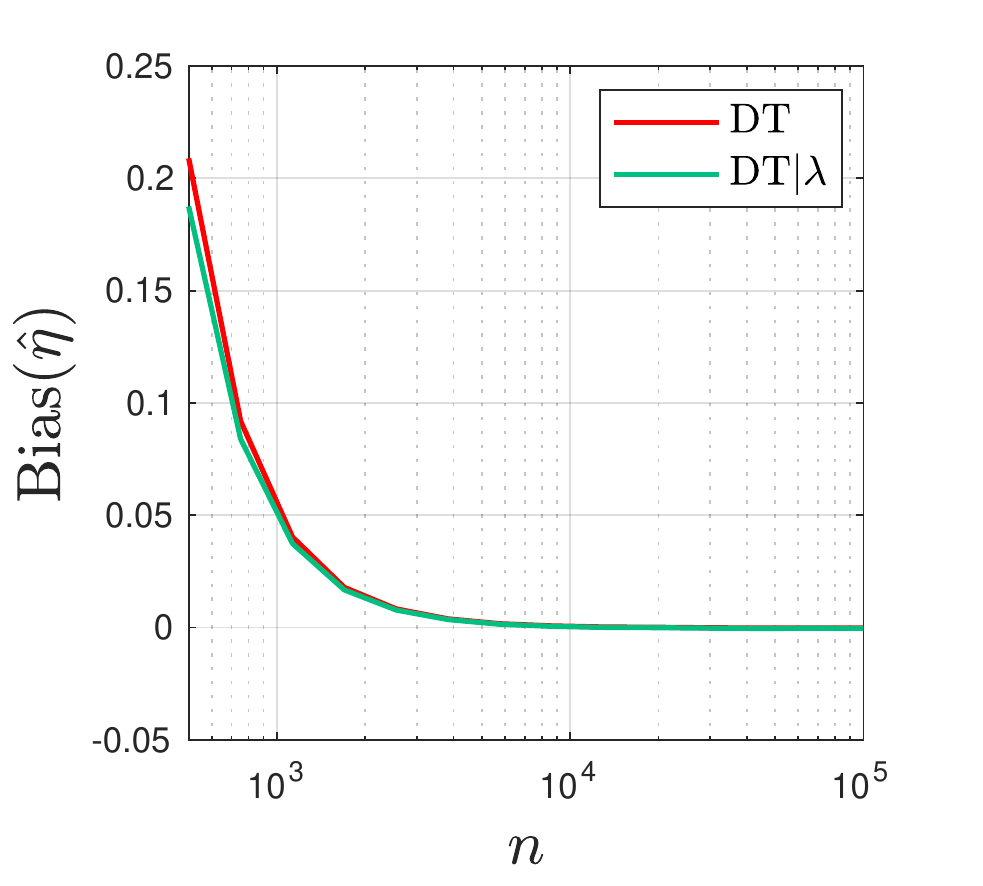}
        \caption{${\rm Bias}(\widehat{\eta})$ vs.\ $n$}
        \label{fig:biasEta_vs_n}
    \end{subfigure}
    \begin{subfigure}{0.245\linewidth}
        \centering
        \includegraphics[width=1\linewidth]{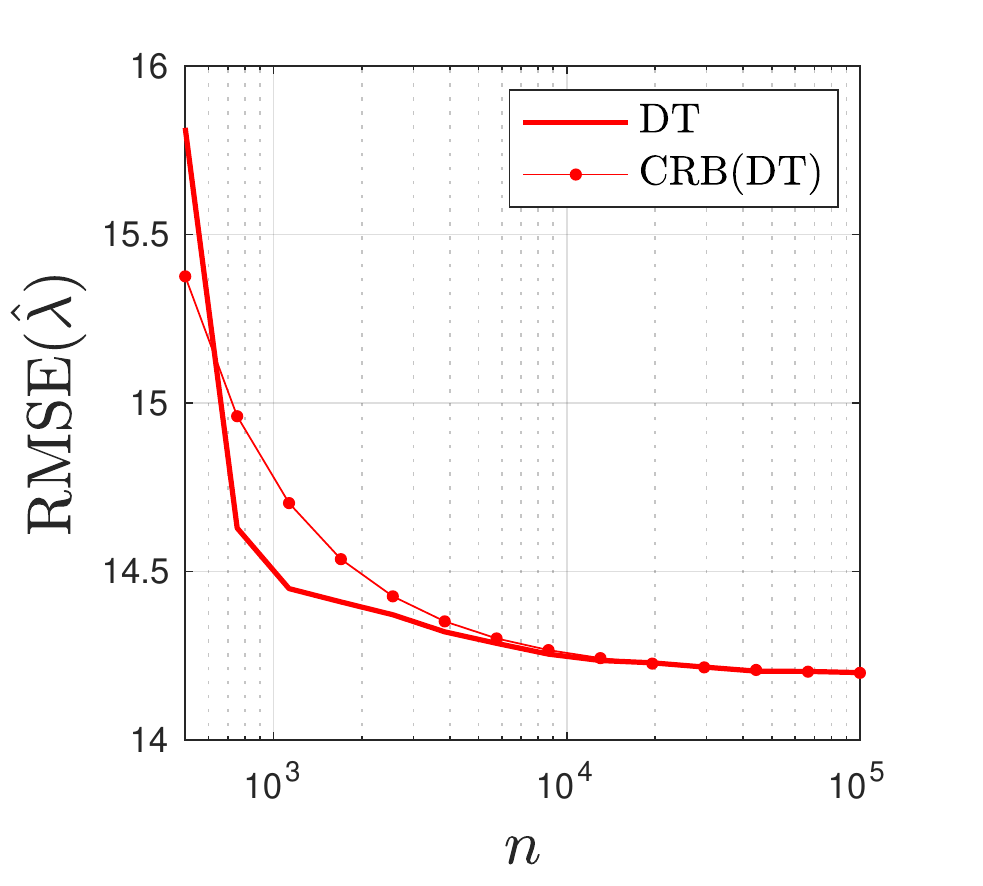}
        \caption{${\rm RMSE}(\widehat{\lambda})$ vs.\ $n$}
        \label{fig:mseLam_vs_n}
    \end{subfigure}
    \begin{subfigure}{0.245\linewidth}
        \centering
        \includegraphics[width=1\linewidth]{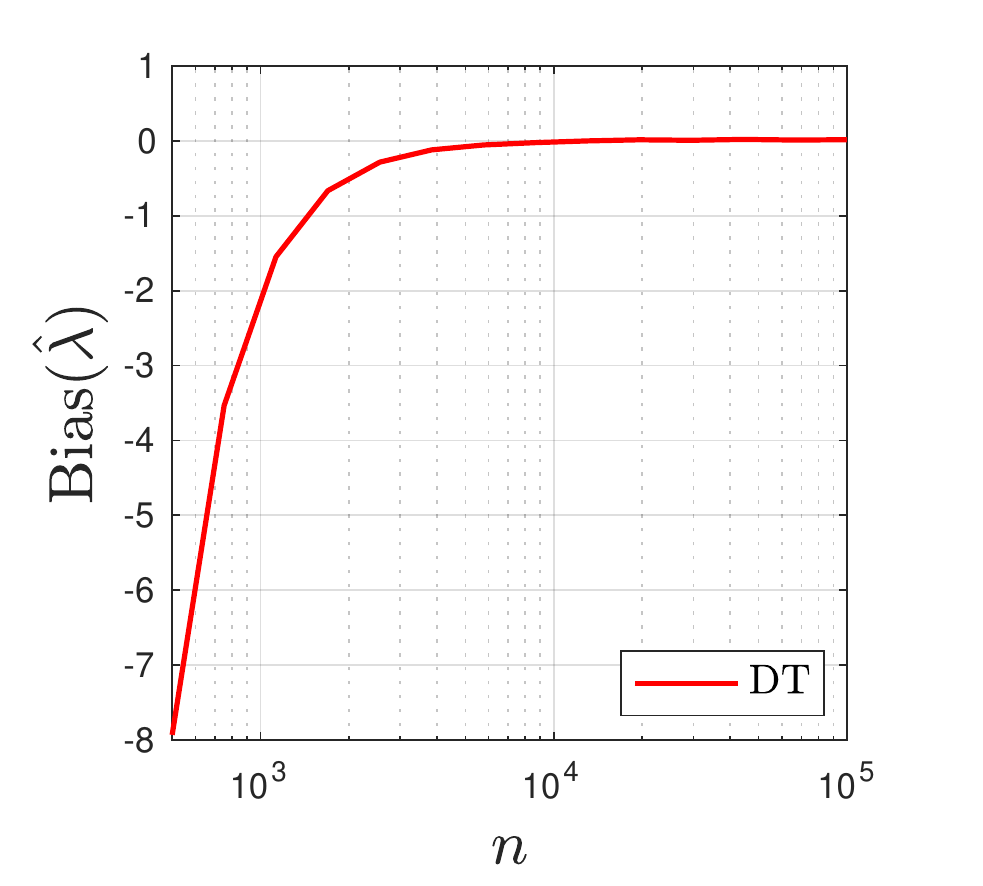}
        \caption{${\rm Bias}(\widehat{\lambda})$ vs.\ $n$}
        \label{fig:biasLam_vs_n}
    \end{subfigure}
    \caption{RMSE and bias as functions of $n$ for single-pixel estimators with $\lambda = 200$ and $\eta = 5$.
    } 
    \label{fig:singlePixelPerformance_vs_n}
\end{figure*}

\subsection{Estimator Performance}
\label{sec:singlePixelEstimatorPerformance}

In \Cref{fig:singlePixelPerformance_vs_eta}, we plot the root mean-squared error (RMSE) and bias for our estimators, both normalized by the true parameter value, as functions of $\eta$.
We compare existing methods for estimating $\eta$, which require knowledge of the dose, with our new methods for jointly estimating $\eta$ and $\lambda$:
\begin{itemize}
    \item CT$|\lambda$: $\widehat{\eta}^{\rm CT|\lambda}_k$ is the CT estimate in \eqref{eq:eta_trml_conti}, evaluated using oracular knowledge of the dose ($\lamtilde = \lambda$).
    \item DT$|\lambda$: $\etaOracle_k$ is the DT estimate in \eqref{eq:eta_DTML}, evaluated using oracular knowledge of the dose ($\lamtilde = \lambda$).
    \item CT: $(\etaCTML_k,\lambdaCTML_k)$ is the CT joint estimate from \eqref{eq:ctml_eta} and \eqref{eq:ctml_lambda} 
    \item DT: $(\etaDTML_k ,\lamDTML_k)$ is the DT joint estimate from \eqref{eq:dtml}.
\end{itemize}
We also plot the normalized square roots of the corresponding CRBs: ${\rm CRB (CT)}$ of \eqref{eq:crb_ct_bothUnknown}, ${\rm CRB(CT|}\lambda{\rm )}$ of \eqref{eq:ct_crb_eta_given_lambda}, as well as ${\rm CRB (DT)}$ and ${\rm CRB (DT|\lambda)}$ computed as described in \Cref{subsubsec:dt_crb}.
As predicted by the CRBs derived in \Cref{subsec:crb}, our joint estimators and the estimators of $\eta$ when $\lambda$ is given achieve similar performance at higher $\eta$.
In fact, \Cref{fig:mseEta_vs_eta,fig:mseLam_vs_eta} both show a very close match of all estimators to their corresponding CRBs, with the (biased) DT estimate of $\lambda$ slightly outperforming the CRB.
With $\lambda = 200$ as used here,
the normalized RMSEs of both $\widehat{\eta}$ and $\widehat{\lambda}$ get very large as $\eta$ gets smaller, especially when $\lambda$ is not given, motivating our use of inter-pixel correlations in \Cref{sec:multi_pixel_estimation,sec:neon_beam_algorithms}.
The RMSE and bias of $\widehat{\eta}$ and $\widehat{\lambda}$ are generally smaller for CT methods, with very small bias of $\widehat{\eta}$ and $\widehat{\lambda}$ for all estimators.

\Cref{fig:singlePixelPerformance_vs_n} shows estimator RMSE (along with the square root of the CRBs) and bias as functions of the number of sub-acquisitions $n$ for fixed total dose $\lambda = 200$ and SE yield $\eta = 5$.
Note that when $n$ gets large, DT performance converges to the CT asymptote.
At $\Frac{\lambda}{n} = 0.1$, a value attainable by current hardware and used in the DT experiments that follow in \Cref{sec:performance_eval}, DT estimator RMSE is close to the CT limit.
\Cref{fig:mseEta_vs_n,fig:biasEta_vs_n} show that joint estimation and estimation of $\eta$ given $\lambda$ are similarly difficult when $n$ is sufficiently large.
In \Cref{fig:mseEta_vs_n} we observe that at larger $n$, RMSE($\etaDTML$) approaches the CRB\@.
The RMSE of $\lamDTML$ dips below the CRB (\Cref{fig:mseLam_vs_n}) at certain lower values of $n$, which may be explained by the non-negligible bias shown in \Cref{fig:biasLam_vs_n}.

\section{Exploiting a Smoothly Varying Beam Current}
\label{sec:multi_pixel_estimation}
\Cref{sec:single_pixel_estimation} demonstrated that joint estimation of $\eta$ and $\lambda$ is possible at a single pixel through time-resolved measurement. However, as shown in \Cref{fig:crbCT_vs_dose} and discussed in \Cref{sec:CTTR_ML}, high fidelity estimates may require a large dose, especially at low-$\eta$ pixels.  
In this section, we use a simple model for smoothly varying beam  current---meant to be representative of electron and helium ion beams---to form high quality estimates of both $\etabf$ and $\lambf$ at moderate doses.
To meet a variety of use cases, we propose both causal and non-causal algorithms, each with and without total variation (TV) regularization on $\etabf$.
In the interests of brevity and relevance to contemporary instruments, we consider only discrete-time measurements.

\subsection{An Autoregressive Model for Beam Current}
We model beam current as a first-order Gaussian autoregressive process:
\begin{equation}
    \lambda_k = x_k+a\lambda_{k-1}+c,
    \label{eq:ar1}
\end{equation}
where $a$ is the correlation coefficient for neighboring pixels in a row
and all $x_k \sim \mathcal{N}(0,\sigmax^2)$ variables are independent.
The mean and variance of the beam current are
\begin{equation}
    \lambar
    = \E{\lambda}
    = \frac{c}{1-a}
\qquad
\mbox{and}
\qquad
    \sigma_\lambda^2 = \frac{\sigmax^2}{1-a^2}.
\end{equation}
This describes an incident beam with slow, unknown variations about $\lambar$, which may be the intended beam current setting.

\subsection{Causal Estimation}
\label{sec:causal_est}
We seek \emph{causal} estimates, $\etaChat_k$ and $\lamChat_k$, of  $\eta_{k}$ and $\lambda_k$ at the $k$th pixel, given only the measurement vector at that pixel $\yvec_k = \ybf_k$ and $\lambda_{k-1}$.
Although $\lambda_{k-1}$ is not perfectly known, we assume it is approximately equal to our estimate of the beam current formed at the last pixel:
$\lambda_{k-1} \approx \lamChat_{k-1}$.
Inspired by \eqref{eq:ar1}, we use the prior
\begin{equation}
    f(\lambda_k \smid \lambda_{k-1}) \sim \mathcal{N}(a\lamChat_{k-1} + c,\sigmax^2)
    \label{eq:causal_prior}
\end{equation}
and formulate a MAP estimate:
\begin{align}
 (\etaChat_k&, \lamChat_k)
   = \argmax_{\eta_k, \lambda_k} f(\eta_k,\lambda_k \smid \ybf_k) \nonumber\\
  &= \argmax_{\eta_k, \lambda_k} \mathrm{P}_{\yvec_k}(\ybf_k \sMid \eta_k,\lambda_k) f(\lambda_k \smid \lambda_{k-1})\nonumber\\
  &=\argmin_{\eta_k, \lambda_k} \Big[ {-\log \mathrm{P}_{\yvec_k}(\ybf_k \sMid \eta_k,\lambda_k )} \nonumber\\
  & \qquad \qquad \quad \; + \frac{1}{2\sigmax^2}(\lambda_k-(a\lamChat_{k-1}+c))^2 \Big] ,
 \label{eq:causal_map}
\end{align}
where $\mathrm{P}_{\yvec_k}(\cdot \sMid \cdot, \cdot)$ is the joint PMF given in \eqref{eq:DTTR-distribution}. In practice, we introduce a tuning parameter $\betac$:
\begin{align}
  (\etaChat_k, \lamChat_k)
    =\argmin_{\eta_k, \lambda_k}
      \big[& {-\log \mathrm{P}_{\yvec_k}(\ybf_k \sMid \eta_k,\lambda_k )} \nonumber\\
           &+ \betac(\lambda_k-(a\lamChat_{k-1}+c))^2 \big].
    \label{eq:causal_tune}
\end{align}
At each new pixel, \eqref{eq:causal_tune} is solved using gradient descent. At the first pixel, indexed by $k = 1$, we solve \eqref{eq:causal_tune} using $\lamChat_{0} = \lambar$.

\subsection{Causal Estimation with Total Variation Regularization}
\label{sec:causal_estimation_tv}

Total variation regularization on $\etabf$ may be added to the cost function in \eqref{eq:causal_tune} to exploit the fact that microscopy images are often piecewise smooth.
Our new TV-regularized causal estimate of $\etabf$ (and the corresponding estimate of $\lambf$) minimize the following cost function:
\begin{align}
\label{eq:causal_tune_tv}
 (\etaCTVhat_k, \lamCTVhat_k)
   &= \argmin_{\eta_k, \lambda_k} \big[ {-\log \mathrm{P}_{\yvec_k}(\ybf_k \sMid \eta_k,\lambda_k )} \nonumber\\
   & \qquad + \betac(\lambda_k-(a\lambda_{k-1}+c))^2 + \gTV(\eta_k) \big],
\end{align}
where $\gTV(\eta_k)$ is a TV cost term.
In the causal, raster-scanned scenario, the neighboring pixels that have already been visited are those to the left and above the current pixel. Thus, our TV cost term is given by
\begin{equation}
\label{eq:tv_cost_term}
    g_{\rm TV}(\eta) = \betaleft |\eta -\etaleft| + \betaup |\eta -\etaup|,
\end{equation}
where $\etaup$ and $\etaleft$ are $\eta$ values already estimated (and assumed known) at the vertically and horizontally adjacent pixels.
Parameters $\betaleft$ and $\betaup$ may be tuned to promote more horizontal or vertical smoothness.

We solve \eqref{eq:causal_tune_tv} using proximal gradient methods.
The proximal operator for the term in \eqref{eq:tv_cost_term} is
\begin{align}
\label{eq:proxTV}
  {\rm prox}_{{g_{\rm TV}}}(x)
    = \argmin_{\alpha}
       \frac{1}{2}\norm{x-\alpha}_2^2
       + \betaleft |\alpha -\etaleft| 
       + \betaup |\alpha -\etaup| .
\end{align}
When $\etaleft<\etaup$ holds, the minimization in
\eqref{eq:proxTV} gives 
\begin{align}
    {\rm prox}_{{g_{\rm TV}}}(x) =\begin{dcases}
    x+\betaleft+\betaup, & \text{if } x< \etaleft-\betaleft-\betaup;\\
    x-\betaleft-\betaup, & \text{if } x> \etaup+\betaleft+\betaup;\\
    x-\betaleft+\betaup, & \text{if } \etaleft + \betaleft - \betaup < x \\
                         & \qquad < \etaup + \betaleft - \betaup;\\
    \etaleft, & \text{if } |x-\etaleft+\betaup|\leq\betaleft;\\
    \etaup, & \text{if } |x-\etaup-\betaup|\leq\betaup,\\
    \end{dcases}
    \label{eq:prox_tv_twoBeta}
\end{align}
which, when $\betaleft = \betaup = \betactv$,\footnote{In this paper, we evaluate our algorithm with $\betaleft = \betaup = \betactv$. However promoting more similarity between pixels that are vertically adjacent (i.e., $\betaup> \betaleft$) might be useful to mitigate horizontal stripe artifacts.} reduces to
\begin{equation}
    {\rm prox}_{{g_{\rm TV}}}(x) =\begin{dcases}
    x+2\betactv, & \text{if } x< \etaleft-2\betactv;\\
    x-2\betactv, & \text{if } x> \etaup+2\betactv;\\
    x, & \text{if } \etaleft < x< \etaup; \\
    \etaleft, & \text{if } \etaleft-2\betactv<x<\etaleft;\\
    \etaup, & \text{if } \etaup\leq x\leq\etaup +2\betactv.\\
    \end{dcases}
    \label{eq:prox_tv_equalBeta}
\end{equation}
The case of $\etaleft > \etaup$ is similar.
Equation \eqref{eq:causal_tune_tv} is solved at each pixel, with $\betaup=0$ at all pixels in the first row of the image and $\betaleft=0$ for the first column of the image.

\subsection{Non-Causal Estimation}
\label{sec:noncausal_est}
The non-causal estimation algorithm operates on the \emph{entire} measurement vector $\ybf$ and estimates $\etabf$ and $\lambf$ simultaneously at all pixels.
This formulation allows us to leverage stronger priors on $\lambf$, as well as on $\etabf$ as we show later in \Cref{sec:joint_estimation_tv}.
Given \eqref{eq:ar1}, $\lambf$ is jointly Gaussian,
\begin{equation}
   \lambf \sim\mathcal{N}(\bar{\lambda}\mathds{1},{\bf\Sigma}),
\qquad
    {\bf \Sigma}_{i,j} = \frac{\sigma_x^2 }{1-a^2}a^{|i-j|},
\end{equation}
where $\mathds{1}\in\mathbb{R}^p$ is a vector of ones.
Since all sub-acquisitions and pixels are conditionally independent, the joint PMF of the \emph{entire} measurement vector $\ybf$ is
\begin{equation}
   \mathrm{P}_{\yvec}(\ybf \smid \etabf,\lambf) = \prod_{k = 1}^p\prod_{i=1}^n \mathrm{P}_Y\big( y^{(i)}_k \sMid \eta_k,\lambda_k/n \big),
\end{equation}
where $\mathrm{P}_{\rm Y}(\cdot \sMid \cdot, \cdot)$ is the PMF in \eqref{eq:neyman}.
The MAP estimate for
$(\etabf,\,\lambf,\,\lambar)$
is given by
\begin{align}
(&\etaNChat, \lamNChat, \widehat{\lambar})
  = \argmax_{\etabf, \lambf, \bar{\lambda}} \mathrm{P}_{\yvec}(\ybf \smid \etabf,\lambf) f(\lambf \smid \lambar)\nonumber\\
 &= \argmin_{\etabf, \lambf, \bar{\lambda}}
     \big[ {-\log\mathrm{P}_{\yvec}(\ybf|\etabf,\lambf)}
       + \frac{1}{2}(\lambf-\bar{\lambda}\mathds{1})^\T{\bf \Sigma}^{-1}(\lambf-\bar{\lambda}\mathds{1}) \big] \label{eq:MAP_non_causal}.
\end{align}
As in \eqref{eq:causal_tune}, we introduce a tuning parameter $\betanc$ to allow additional regularization:
\begin{align}
(\etaNChat ,\lamNChat,\widehat{\lambar})
  &= \argmin_{\etabf ,\lambf,\bar{\lambda}} \big[ {-\log\,\mathrm{P}_{\yvec}(\ybf \smid \etabf,\lambf)} \nonumber\\
  &\qquad \; + \betanc(\lambf-\bar{\lambda}\mathds{1})^\T{\bf \Sigma}^{-1}(\lambf-\bar{\lambda}\mathds{1}) \big] .\label{eq:cost_non_causal}
\end{align}
Note that this formulation does not require knowledge of the mean beam current but rather estimates $\lambar$ in addition to $\etabf$ and $\lambf$.
This cost function is differentiable and is thus we solve the minimization using gradient descent methods.
The derivatives of the first term in \eqref{eq:cost_non_causal} are derived in the appendix;
the derivative of the second term with respect to $\lambf$ is proportional to ${\bf\Sigma}^{-1}(\lambf-\bar{\lambda}\mathds{1})$.
To avoid storing a prohibitively large matrix, and because ${\bf \Sigma}$ is approximately circulant,
we perform multiplication by ${\bf \Sigma}^{-1}$ in the frequency domain using the fast Fourier transform.

\subsection{Non-Causal Estimation with Total Variation Regularization}
\label{sec:joint_estimation_tv}
As with our causal estimate, TV regularization may be added to \eqref{eq:cost_non_causal} to promote piecewise smooth estimates of $\etabf$:
\begin{align}
(\etaNCTVhat, &\,\lamNCTVhat,\,\widehat{\lambar})
  = \argmin_{\etabf ,\lambf,\bar{\lambda}} \Big[ {-\log\,\mathrm{P}_{\yvec}(\ybf \smid \etabf,\lambf)} \nonumber\\
&+ \betanc(\lambf-\bar{\lambda}\mathds{1})^\T{\rm {\bf \Sigma}}^{-1}(\lambf-\bar{\lambda}\mathds{1}) + \betanctv\norm{\etabf}_{\rm TV} \Big], \label{eq:cost_non_causal_tv}
\end{align}
where $\norm{\etabf}_{\rm TV}$ is given by
\begin{equation}
    \norm{\etabf}_{\rm TV} = \sum_{i,j}\sqrt{|\eta_{i+1,j}-\eta_{i,j}|^2 + |\eta_{i,j+1}-\eta_{i,j}|^2)}
    \label{eq:tv_cost}
\end{equation}
and $\betanctv$ is a tuning parameter.
Equation \eqref{eq:cost_non_causal_tv} is solved using proximal gradient methods; the proximal operator for \eqref{eq:tv_cost} is solved using \cite{beckTV}.

\subsection{Operational Considerations}
Although all of our proposed algorithms jointly estimate $\etabf$ and $\lambf$, different assumptions are made about the parameters $a$, $\sigma_x^2$ and $\lambar$ in  \eqref{eq:ar1}.
Note that both causal and non-causal algorithms assume knowledge of the correlation $a$ between pixels. Due to the tuning parameters $\betac$ and $\betanc$, no assumption is made about $\sigma_x^2$.
In practice, algorithm performance was found to not depend heavily on ideal choices
of $a$, $\betac$,  or $\betanc$, as we will show in \Cref{sec:performance_eval}, \Cref{fig:robustness_demonstration_c,fig:robustness_demonstration_nc}.
While our causal algorithms do require knowledge of the mean beam current $\lambar$, the non-causal algorithms estimate $\lambar$ in addition to $\lambf$ and $\etabf$. When causal operation is warranted but the mean beam current is not known, the non-causal algorithm could be run periodically to provide $\lambar$.

\subsection{Simulated Microscopy Results}
\label{sec:performance_eval}

\subsubsection{Data Generation}
\label{sec:him_data}
We evaluate the multi-pixel algorithms proposed in this section on synthetic HIM and SEM data.
Measurements for these two examples were generated using existing micrographs
as ground truth images.\footnote{All ground truth images in this work are from the ThermoFisher Scientific database: \url{https://www.fei.com/image-gallery/}}
Compared to SEM, HIM has higher SE yield and can thus produce high-quality images at lower doses.
To be representative of HIM, we scale the
ground truth to $\eta \in [2,\,8]$ and use mean dose $\lambar = 20$~\cite{notte2007introduction};
for SEM, we use $\eta \in [0.1,\,1]$ and $\lambar = 200$~\cite{LinJ:05}.
Beam current time series were produced according to the Gaussian first-order autoregressive model in \eqref{eq:ar1}.
In both test examples, the correlation coefficient for neighboring pixels in a row is $a = 0.999$ and the coefficient of variation is $\Frac{\sigma_{\lambda}}{\lambar} = 0.2$.
Data is generated pseudorandomly at each pixel following the separable joint PMF in \eqref{eq:DTTR-distribution},
where in each case the nominal sub-acquisition doses is 0.1 ($n = 200$ for HIM and $n = 2000$ for SEM).

\subsubsection{Methods}
\label{sec:him_methods}
We compare nine methods for estimating $\etabf$, some of which also generate an estimate of $\lambf$:
\begin{itemize}
    \item baseline: $\etaBaselineBold$ is the pixel-wise evaluation of \eqref{eq:eta-baseline} independently at each pixel using the nominal dose $\lambar$.
    \item frequency-domain filter (FDF)~\cite{Barlow2016}: Compute the 2D discrete Fourier transform of $\etaBaselineBold$.
          Let $q$ and $u$ be the horizontal and vertical frequency indexes.
          Coefficients that satisfy both $|q|\leq w$ and $|u|>h$ are nulled before applying the inverse transform to yield $\etaFTbold$.
    \item {DT$|\lambda$}: $\etaOracleBold$ is the pixel-wise ML estimate \eqref{eq:eta_DTML} computed with true beam current $\lambda$ (provided by an oracle).
    \item {DT$|\lamtilde$}: $\etaDTMLgivenBold$ is the pixel-wise ML estimate \eqref{eq:eta_DTML} computed with nominal beam current $\lamtilde$.
    \item linear filter: $(\etaLFbold,\,\lambdaLFbold)$ is the joint estimate computed with the method of~\cite{watkins2021}.
    \item causal: $(\etaChatBold,\,\lamChatBold)$ is the joint estimate from \eqref{eq:causal_tune}.
    \item non-causal: $(\etaNChat,\,\lamNChat)$ is the joint estimate from \eqref{eq:cost_non_causal}.
          (This also produces an estimate of the mean beam current $\lambar$.)
    \item causal with TV: $(\etaCTVhatBold,\,\lamCTVhatBold)$ is the joint estimate from \eqref{eq:causal_tune_tv}.
    \item non-causal with TV: $(\etaNCTVhat,\,\lamNCTVhat)$ is the joint estimate from \eqref{eq:cost_non_causal_tv}.
          (This also produces an estimate of the mean beam current $\lambar$.)
\end{itemize}
Parameters in the frequency-domain filter, linear filter, and methods from
\Cref{sec:causal_est,sec:causal_estimation_tv,sec:noncausal_est,sec:joint_estimation_tv}
are tuned to minimize RMSE\@.

\begin{figure*}
\centering
    \begin{subfigure}{0.195\linewidth}
    \centering
     \includegraphics[width=1\linewidth]{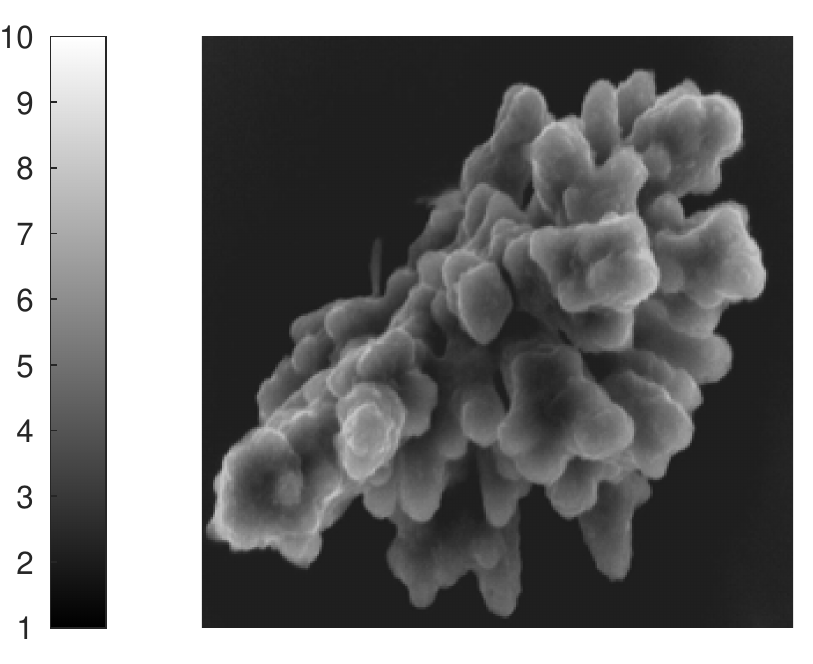}
     \caption{ground truth $\etabf$}
    \end{subfigure}
    \begin{subfigure}{0.19\linewidth}
    \centering
     \includegraphics[width=1\linewidth]{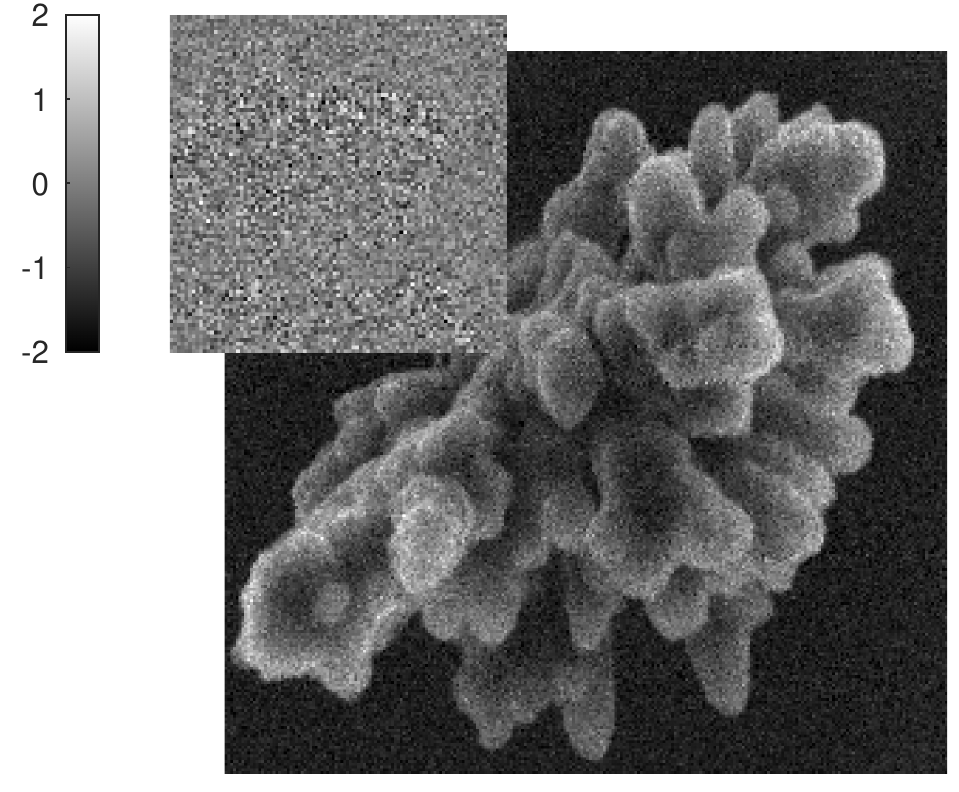}
     \caption{$\etaOracleBold$, RMSE = 0.4974}
     \label{fig:bouquet_oracle}
    \end{subfigure}
    \begin{subfigure}{0.19\linewidth}
    \centering
     \includegraphics[width=1\linewidth]{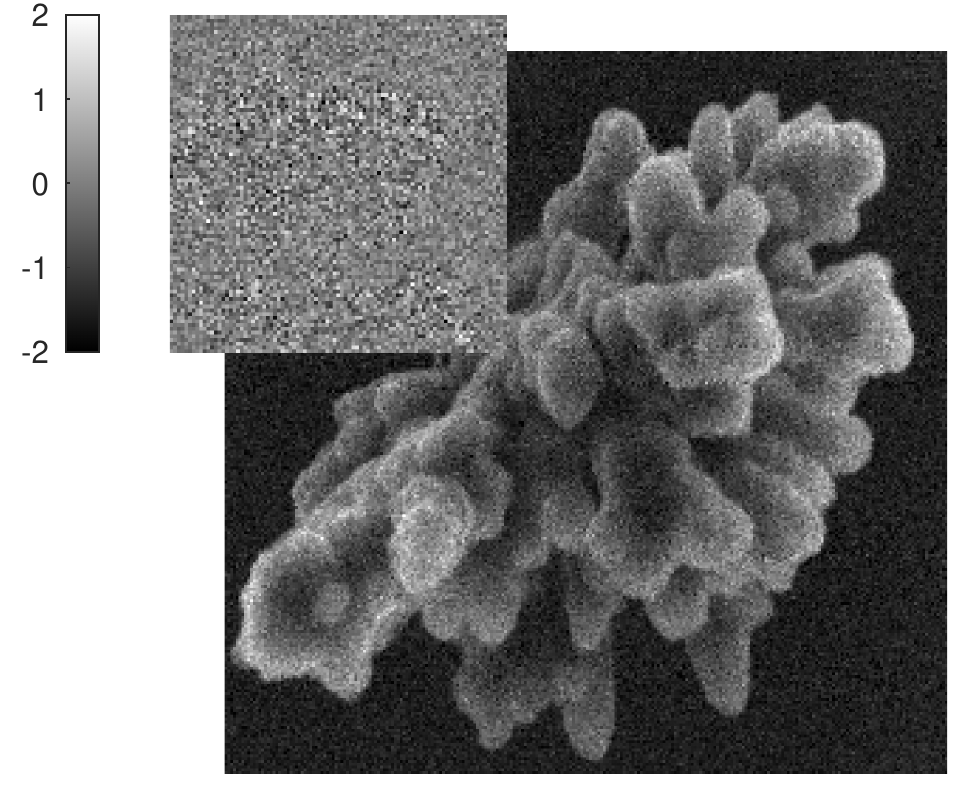}
     \caption{$\etaLFbold$, RMSE= 0.4985 }
     \label{fig:sticky_ft}
    \end{subfigure}
    \begin{subfigure}{0.19\linewidth}
    \centering
     \includegraphics[width=1\linewidth]{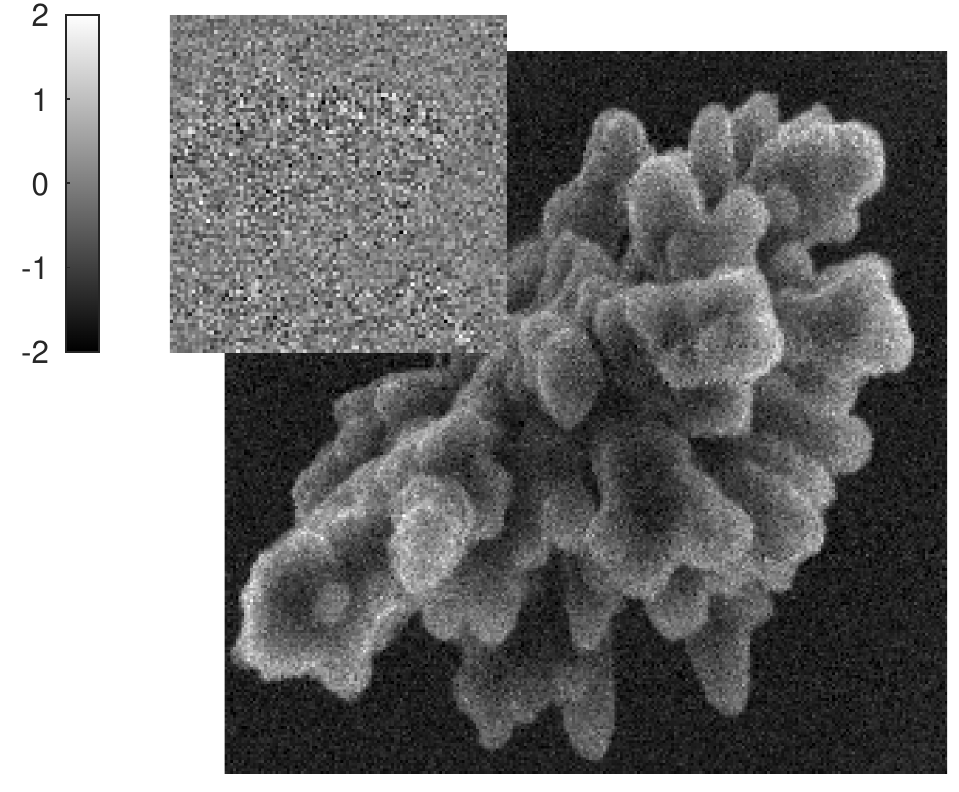}
     \caption{$\etaChatBold$, RMSE = 0.4981}
     \label{fig:bouquet_c}
    \end{subfigure}
    \begin{subfigure}{0.19\linewidth}
    \centering
     \includegraphics[width=1\linewidth]{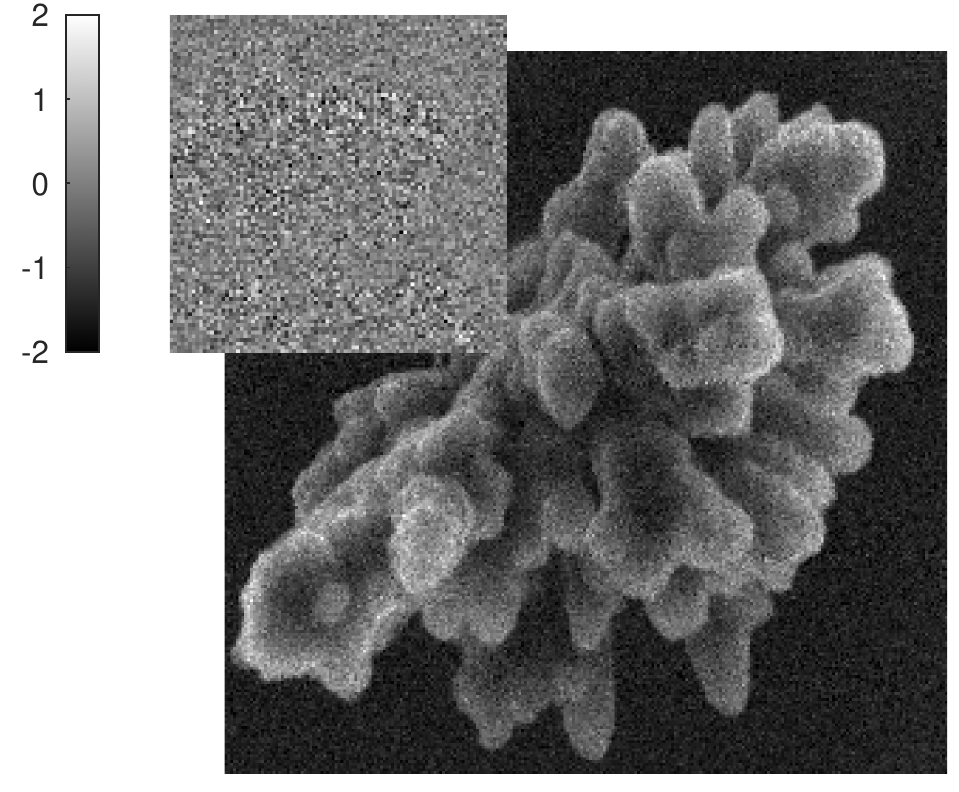}
     \caption{$\etaNChat$, RMSE = 0.4979}
     \label{fig:bouquet_nc}
    \end{subfigure}\\
     \begin{subfigure}{0.19\linewidth}
    \centering
     \includegraphics[width=1\linewidth]{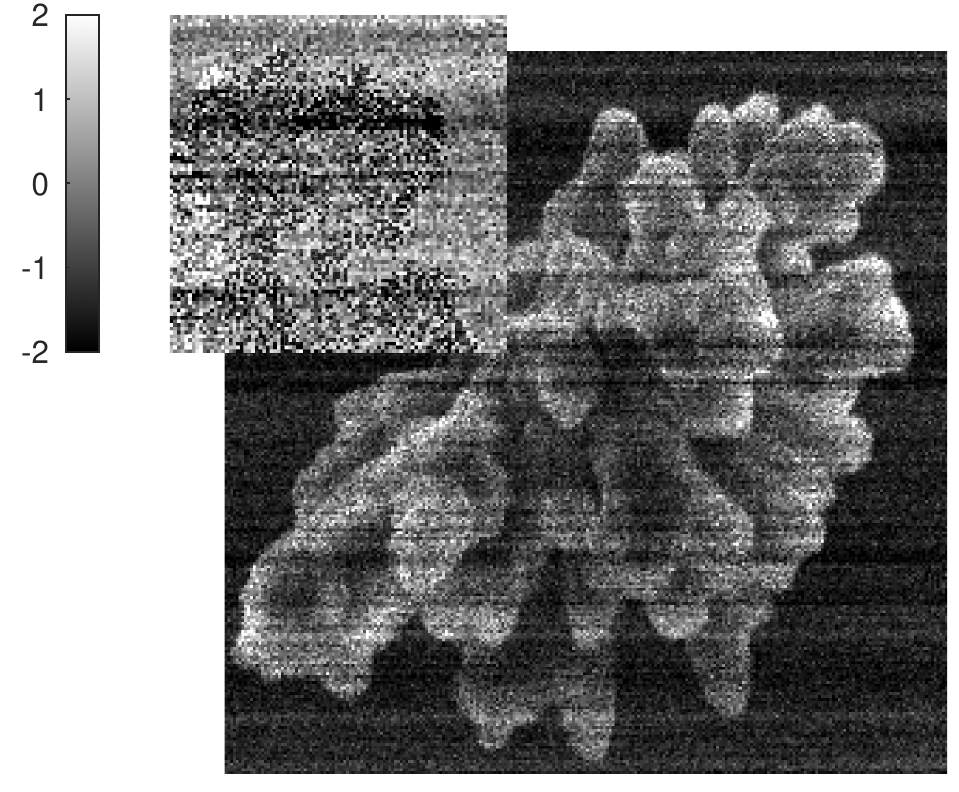}
     \caption{$\etaBaselineBold$, RMSE = 1.1129}
     \label{fig:bouquet_conv}
    \end{subfigure}
     \begin{subfigure}{0.19\linewidth}
    \centering
     \includegraphics[width=1\linewidth]{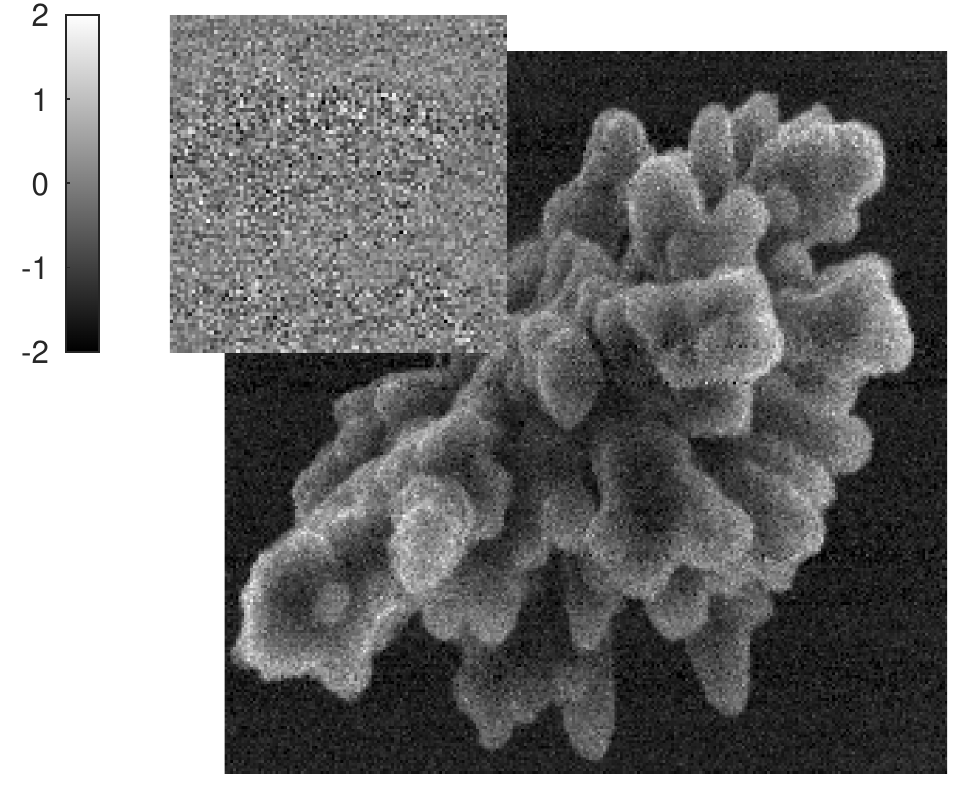}
     \caption{$\etaDTMLgivenBold$, RMSE = 0.5097}
     \label{fig:bouquet_trml}
    \end{subfigure}
        \begin{subfigure}{0.19\linewidth}
    \centering
     \includegraphics[width=1\linewidth]{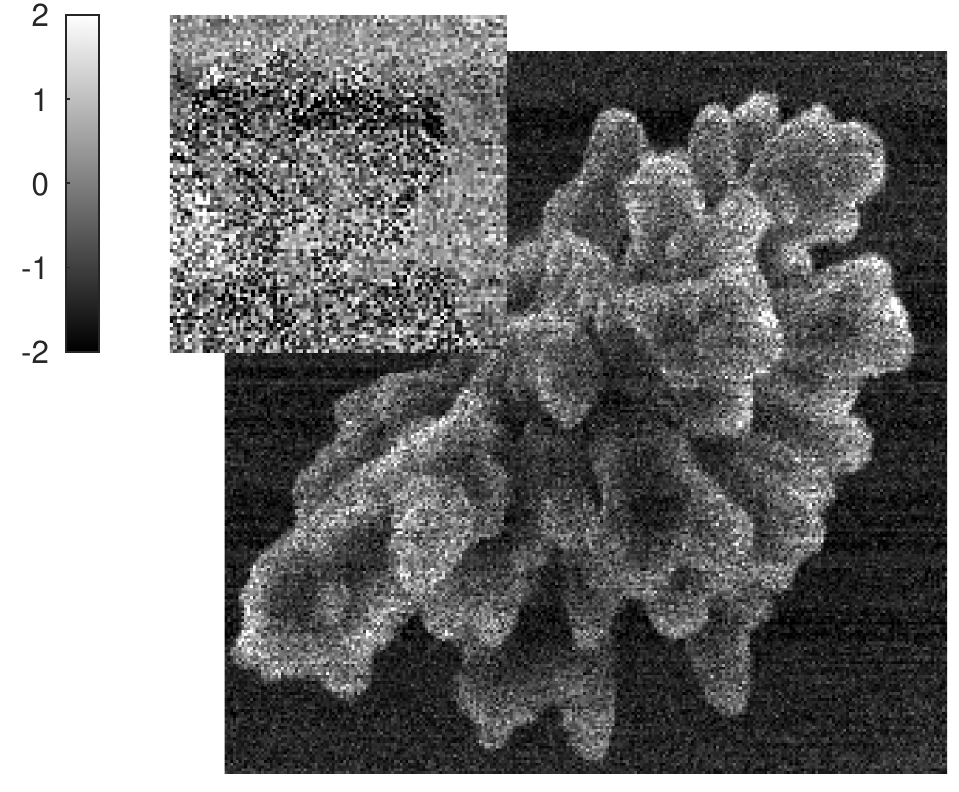}
     \caption{$\etaFTbold$, RMSE= 0.9817}
    \end{subfigure}
    \begin{subfigure}{0.19\linewidth}
    \centering
     \includegraphics[width=1\linewidth]{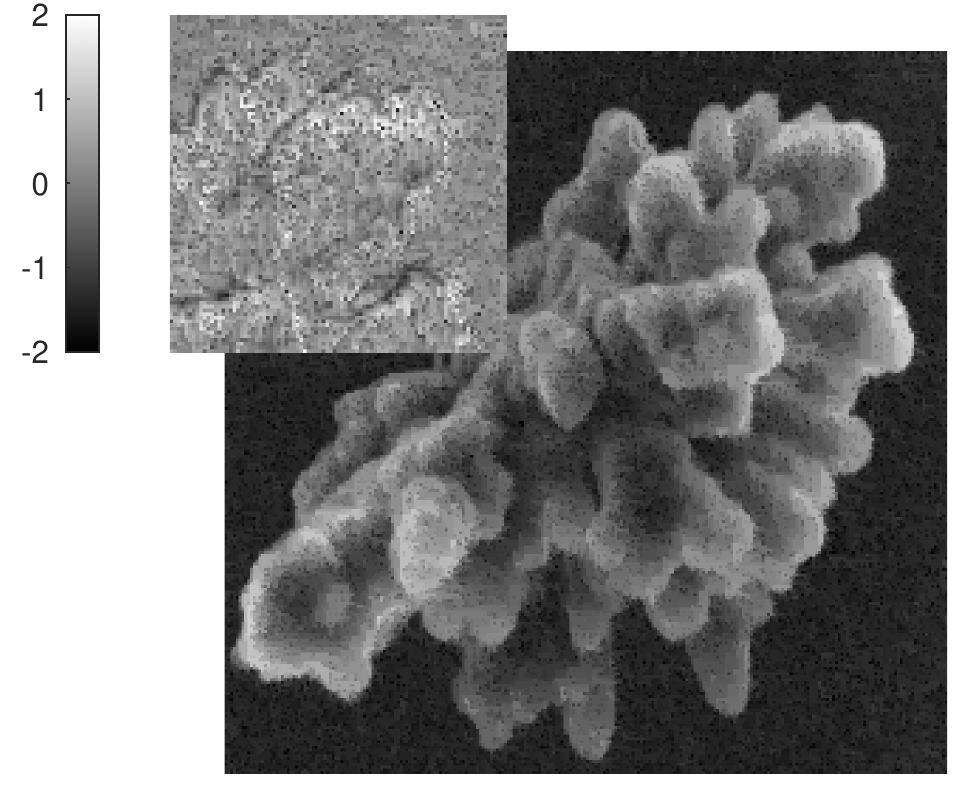}
     \caption{$\etaCTVhatBold$, RMSE = 0.4361}
     \label{fig:bouquet_ctv}
    \end{subfigure}
    \begin{subfigure}{0.19\linewidth}
    \centering
     \includegraphics[width=1\linewidth]{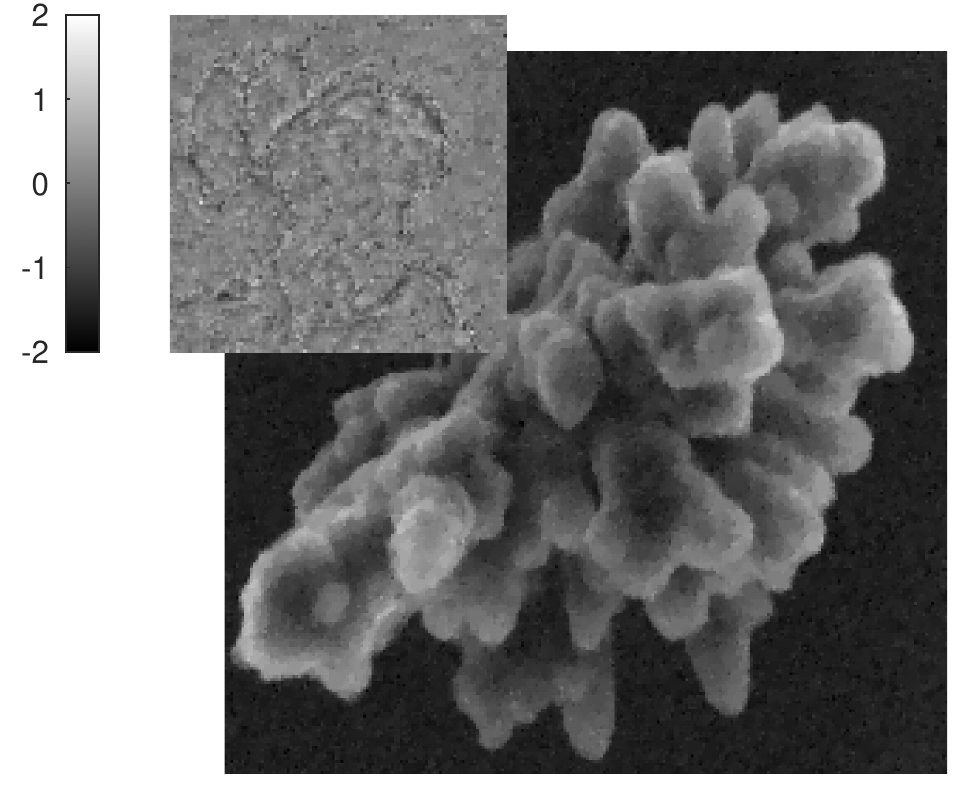}
     \caption{$\etaNCTVhat$, RMSE = 0.2296}
     \label{fig:bouquet_nctv}
    \end{subfigure}
    \caption{
    HIM example with
    ground truth $\eta \in [2,\,8]$, mean dose $\lambar = 20$ and
    nominal sub-acquisition dose $\Frac{\bar{\lambda}}{n} = 0.1$.
    The actual dose $\lambda$ is a Gaussian autoregressive process with correlation coefficent of $0.999$ for neighboring pixels in a row and coefficient of variation $\Frac{\sigma_{\lambda}}{\bar{\lambda}} = 0.2$.
    All micrograph images are on the same scale shown in (a), chosen so that no more than  2\% of pixels are saturated in any given image.
    Inset images show error $\widehat{\etabf}-\etabf$ for a subset of the image taken from the top right corner.
    Tuning parameters are: $\betanc = 200$, $\betanctv = 1 $, $\betac = 10$, and $\betactv = 1{\rm e}-4$. The non-causal estimator found $\widehat{\lambar} = 20.41$ and the non-causal estimator with TV regularization found $\widehat{\lambar} = 20.29$; the beam current empirical mean was $\frac{1}{p}\sum_{k=1}^p\lambda_k = 20.34$. \\
    }
    \label{fig:himEx}
\end{figure*}

\subsubsection{Results}
\label{sec:him_results}
\begin{figure*}
\centering
    \begin{subfigure}{0.21\linewidth}
    \centering
    \vspace{2mm}
     \includegraphics[width=1\linewidth]{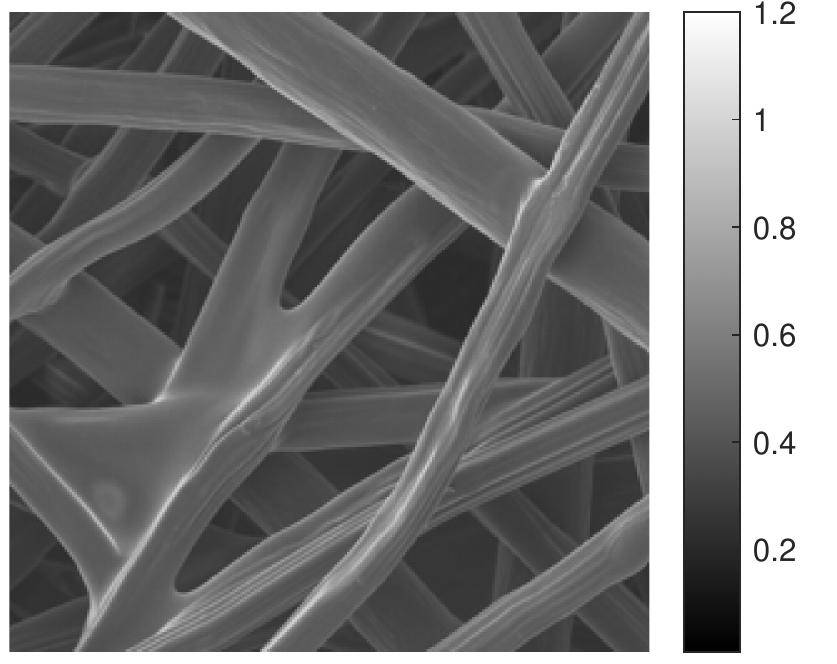}
     \caption{ground truth $\etabf$}
     \label{fig:sticky_truth}
    \end{subfigure}
    \begin{subfigure}{0.185\linewidth}
    \centering
     \includegraphics[width=1\linewidth]{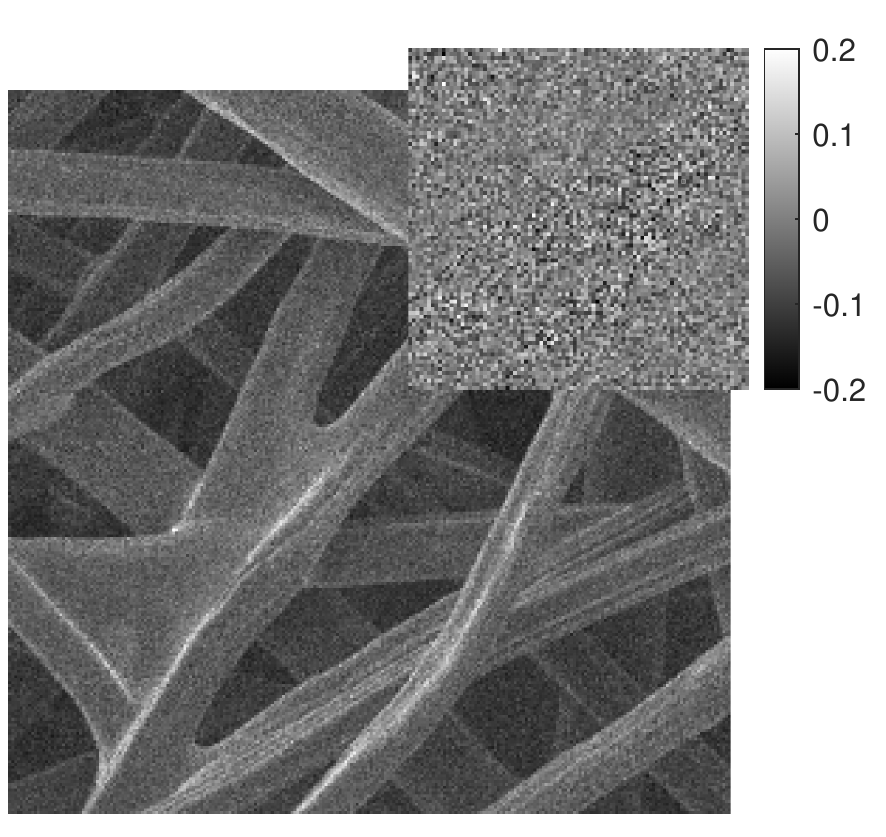}
     \caption{$\etaOracleBold$, RMSE= 5.54e-2 }
     \label{fig:sticky_oracle}
    \end{subfigure}
    \begin{subfigure}{0.185\linewidth}
    \centering
     \includegraphics[width=1\linewidth]{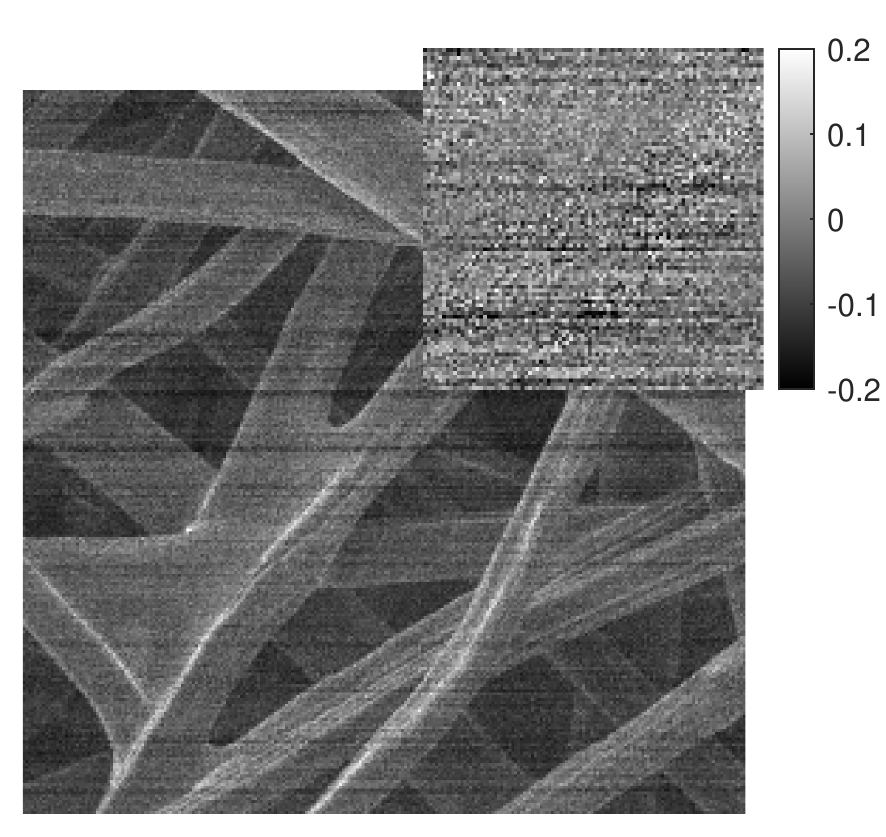}
     \caption{$\etaLFbold$, RMSE= 6.63e-2}
    \end{subfigure}
    \begin{subfigure}{0.185\linewidth}
    \centering
     \includegraphics[width=1\linewidth]{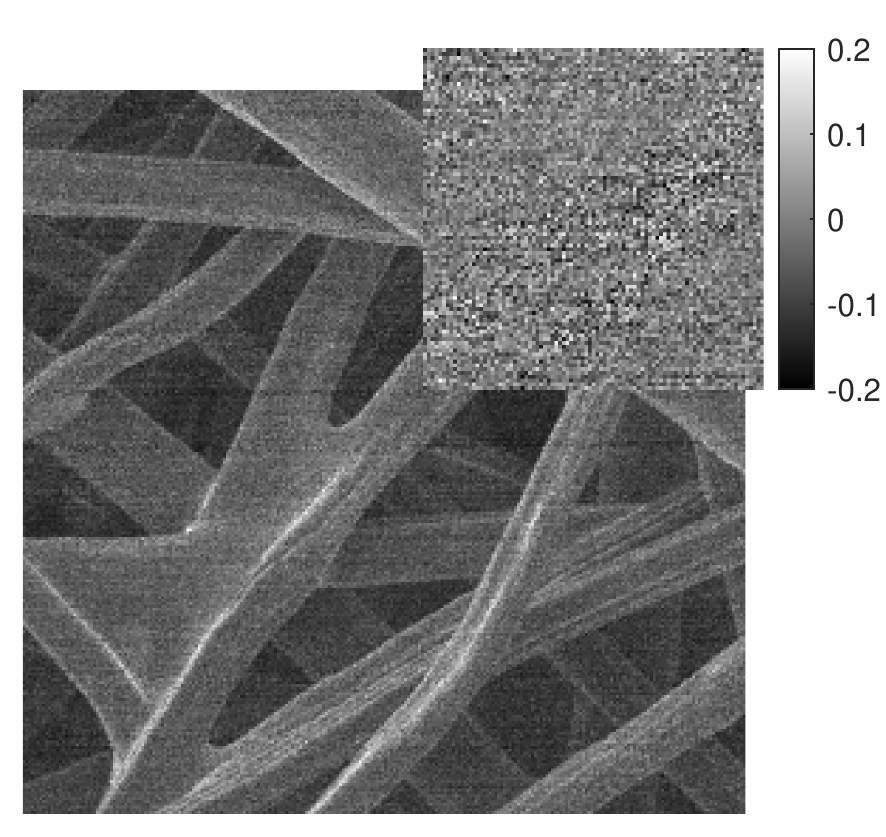}
     \caption{$\etaChatBold$, RMSE=5.83e-2}
     \label{fig:sticky_c}
    \end{subfigure}
    \begin{subfigure}{0.185\linewidth}
    \centering
     \includegraphics[width=1\linewidth]{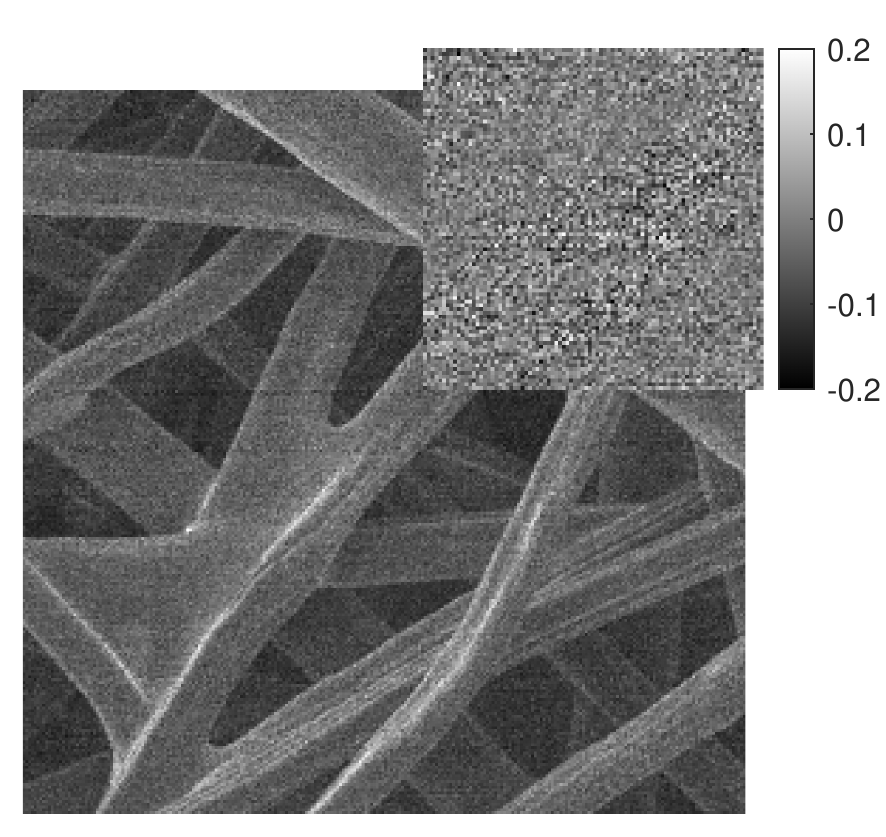}
     \caption{$\etaNChat$, RMSE=5.68e-2}
     \label{fig:sticky_nc}
    \end{subfigure}\\
    \hspace{1mm}
     \begin{subfigure}{0.187\linewidth}
    \centering
     \includegraphics[width=1\linewidth]{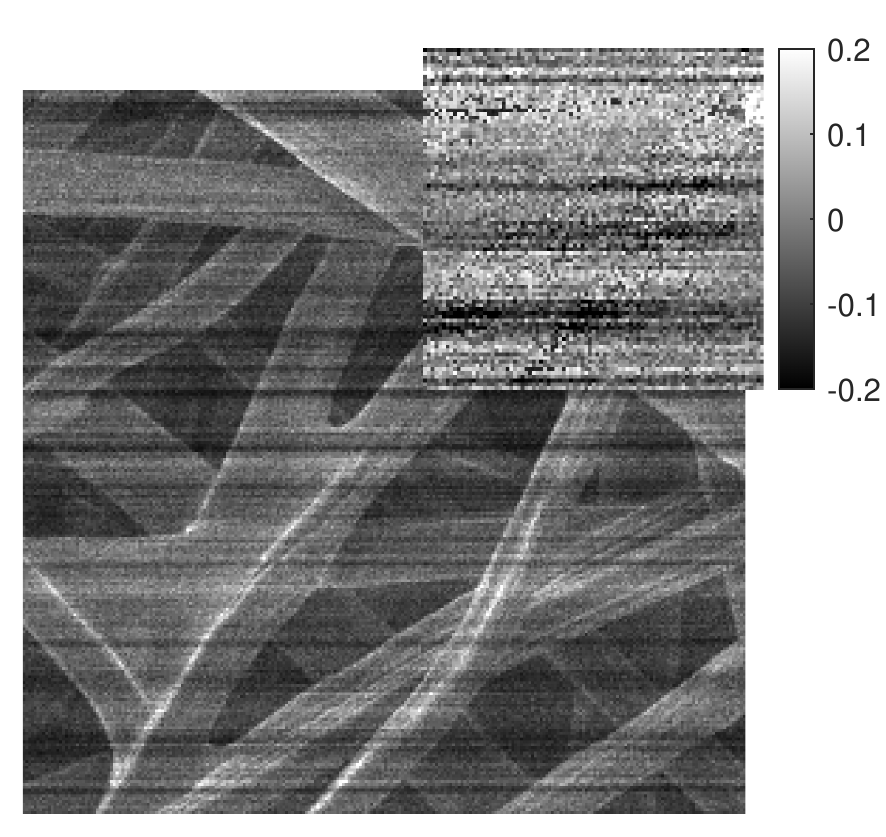}
     \caption{$\etaBaselineBold$, RMSE=9.72e-2}
     \label{fig:sticky_conv}
    \end{subfigure}
    \hspace{.5mm}
     \begin{subfigure}{0.185\linewidth}
    \centering
    \vspace{.25mm}
     \includegraphics[width=1\linewidth]{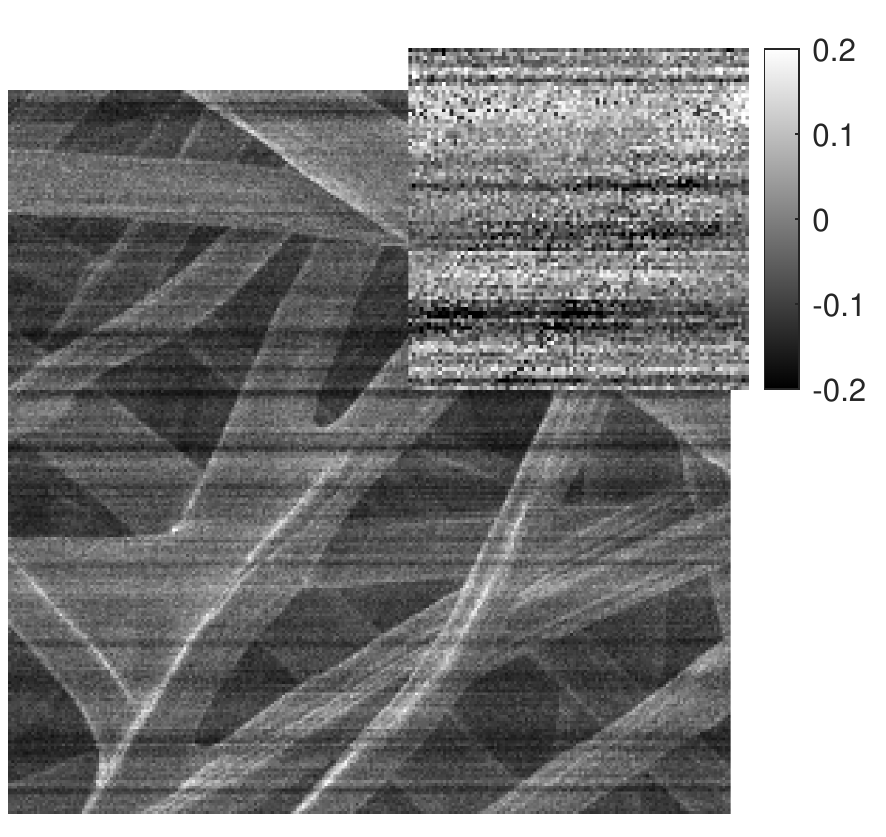}
     \caption{$\etaDTMLgivenBold$,
     RMSE=8.87e-2}
     \label{fig:sticky_trml}
    \end{subfigure}
    \begin{subfigure}{0.185\linewidth}
    \centering
     \includegraphics[width=1\linewidth]{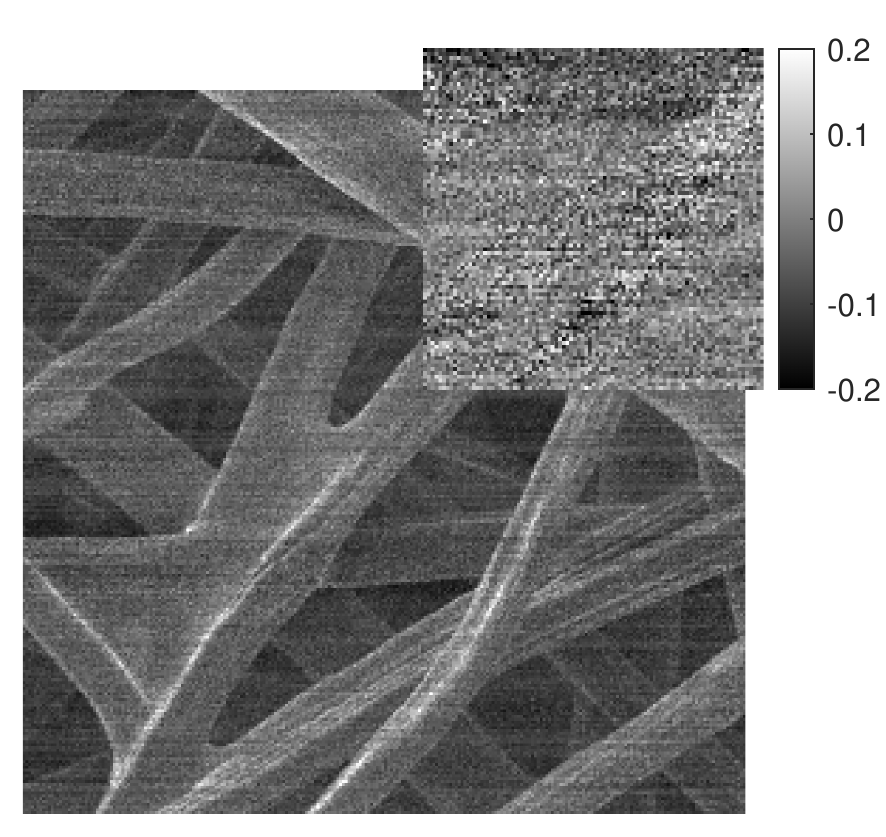}
     \caption{$\etaFTbold$, RMSE= 6.93e-2 }
    \end{subfigure}
    \begin{subfigure}{0.185\linewidth}
    \centering
     \includegraphics[width=1\linewidth]{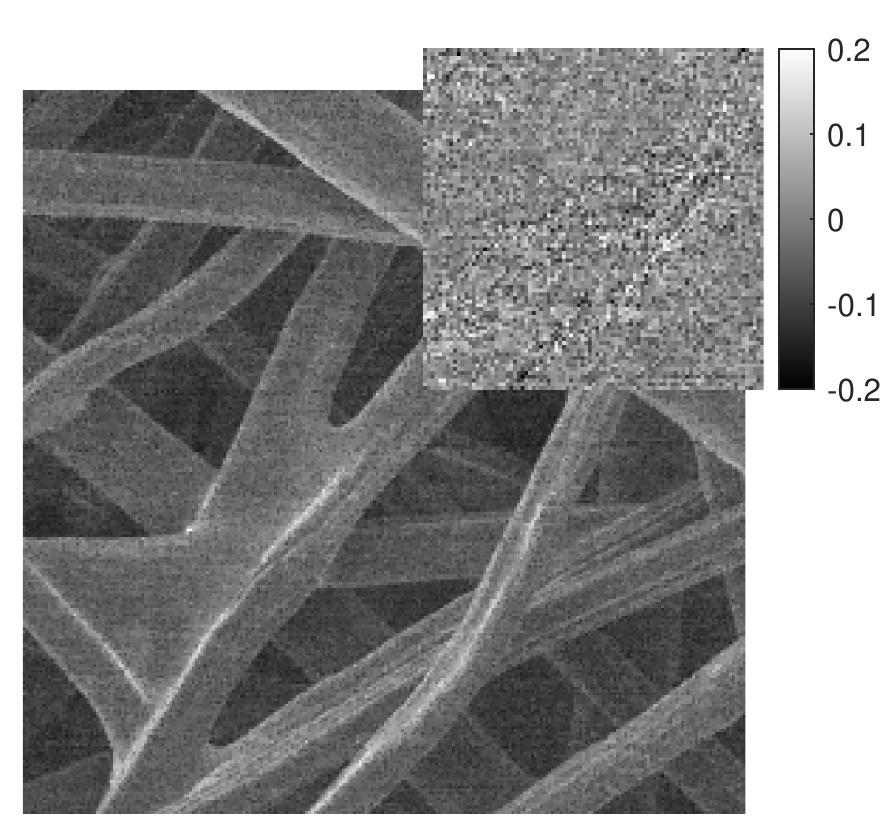}
     \caption{ $\etaCTVhatBold$, RMSE=5.54e-2}
     \label{fig:sticky_ctv}
    \end{subfigure}
    \begin{subfigure}{0.185\linewidth}
    \centering
     \includegraphics[width=1\linewidth]{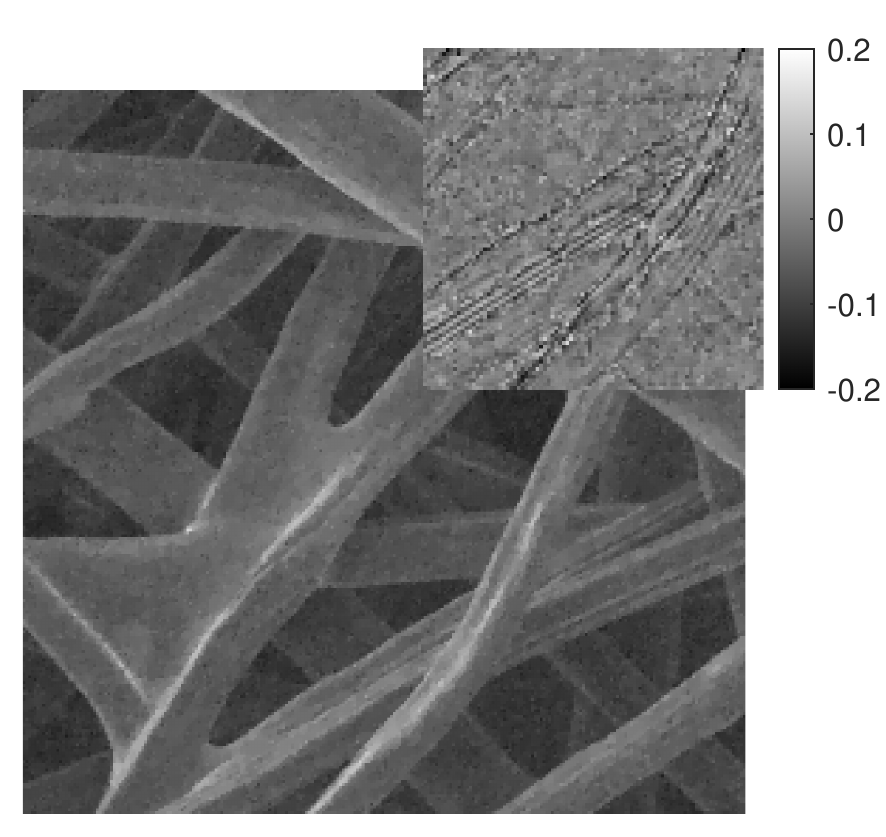}
     \caption{$\etaNCTVhat$, RMSE = 3.21e-2}
     \label{fig:sticky_nctv}
    \end{subfigure}
    \caption{ SEM example with ground truth $\eta \in [0.2,\,1]$, mean dose $\lambar = 200$ and nominal sub-acquisition dose $\Frac{\lamtilde}{n} = 0.1$.
    The actual dose $\lambda$ is a Gaussian autoregressive process with correlation coefficent of $0.999$ for neighboring pixels in a row and coefficient of variation $\Frac{\sigma_{\lambda}}{\bar{\lambda}} = 0.2$.
    All micrograph images are on the same scale shown in (a), chosen so that no more than 2\% of pixels are saturated in any given image.
    Inset images show error $\widehat{\etabf}-\etabf$ for a subset of the image taken from the bottom middle of the image.
    Tuning parameters are:
    $\betanc = 2000$, $\betanctv =  8$, $\betac = 100$, and $\betactv = 3e-4$.
    The non-causal estimator found $\widehat{\lambar} = 202.75$ and the non-causal estimator with TV regularization found $\widehat{\lambar} = 203.69$; the beam current empirical mean was $\frac{1}{p}\sum_{k=1}^p\lambda_k = 201.80$.}
    \label{fig:semExLowEta}
\end{figure*}

\Cref{fig:himEx,fig:semExLowEta} show estimated micrographs $\widehat{\etabf}$ using all the methods on each setting.
The inset images show error $\widehat{\etabf} - \etabf$ for portions of the micrographs.
In both HIM and SEM examples, stripe artifacts are more prominent in $\etaBaselineBold$ (\Cref{fig:bouquet_conv,fig:sticky_conv})
than in the TR reconstruction $\etaDTMLgivenBold$
(\Cref{fig:bouquet_trml,fig:sticky_trml}),
although they both assume the same knowledge of $\lambda$. In the higher $\eta$ FIB example, $\etaDTMLgivenBold$ even outperforms $\etaFTbold$, the frequency-domain filtered version of $\etaBaselineBold$.
This effect may be attributed to the natural robustness of TR methods to unknown beam current~\cite{Watkins2021a,Watkins2021b}.
Our oracle TR method $\etaOracleBold$
(\Cref{fig:bouquet_oracle,fig:sticky_oracle}),
which assumes perfect knowledge of the true dose at every pixel, exhibits no discernible striping. Without TV regularization, our causal joint estimates $\etaChatBold$
(\Cref{fig:bouquet_c,fig:sticky_c})
and non-causal joint estimates $\etaNChat$
(\Cref{fig:bouquet_nc,fig:sticky_nc})
exhibit lower RMSE, closer to our benchmark $\etaOracleBold$, with slightly more improvement seen with the non-causal version. Given TV regularization on $\etabf$, both algorithms meet or exceed the performance of $\etaOracleBold$, which does not employ any spatial regularization.
In fact, in both examples, $\etaNCTVhat$
(\Cref{fig:bouquet_nctv,fig:sticky_nctv})
outperforms the benchmark $\etaOracleBold$ approximately by a factor of 2\@.
\Cref{table:combinedTable} summarizes the RMSE results of all $\etabf$ estimation methods for both HIM and SEM examples.

\begin{table}
\centering
\begin{tabular}{|p{2cm}|cc|cr@{.}l|}
\hline
& \multicolumn{2}{c|}{\textbf{HIM Example}} & \multicolumn{3}{c|}{\textbf{SEM Example}}   \\
\textbf{Method} & RMSE($\widehat{\etabf}$) &RMSE($\widehat{\lambf}$)& RMSE($\widehat{\etabf}$) & \multicolumn{2}{c|}{RMSE($\widehat{\lambf}$)} \\ \hline 
Baseline                         & 1.1129 &  --    & 9.72e-2 & \multicolumn{2}{c|}{--} \\ 
FDF~\cite{Barlow2016}            & 0.9817 &  --    & 6.93e-2 & \multicolumn{2}{c|}{--} \\ 
DT$|\lambda$                     & 0.4974 &  --    & 5.54e-2 & \multicolumn{2}{c|}{--} \\
DT$|\lamtilde$~\cite{PENG2020}   & 0.5097 &  --    & 8.87e-2 & \multicolumn{2}{c|}{--} \\
Linear filter~\cite{watkins2021} & 0.4985 & 1.0020 & 6.63e-2 & 17&8849\\
Causal                           & 0.4984 & 0.9416 & 5.83e-2 &  9&4937\\
Non-causal                       & 0.4979 & 0.6765 & 5.68e-2 &  6&6761\\
Causal with TV                   & 0.4361 & 1.0215 & 5.54e-2 & 11&7524\\
Non-causal with TV               & 0.2298 & 0.6681 & 3.21e-2 &  5&8292\\
\hline
\end{tabular}
\caption{RMSE results by method for the HIM example in \Cref{fig:himEx} and the SEM example in \Cref{fig:semExLowEta}. For the frequency-domain filtering method, filter parameters were $w = 1$ and $h = 5$ for the HIM example and $w = 1$ and $h = 1$ for the SEM example. Our new joint estimation methods without TV regularization approach the performance of oracle estimator $\etaOracleBold$. When TV regularization is added, our causal and non-causal estimators outperform $\etaOracleBold$.
}
\label{table:combinedTable}
\end{table}

\begin{figure}
    \centering
    \begin{subfigure}{0.49\linewidth}
        \centering
        \includegraphics[width=1\linewidth]{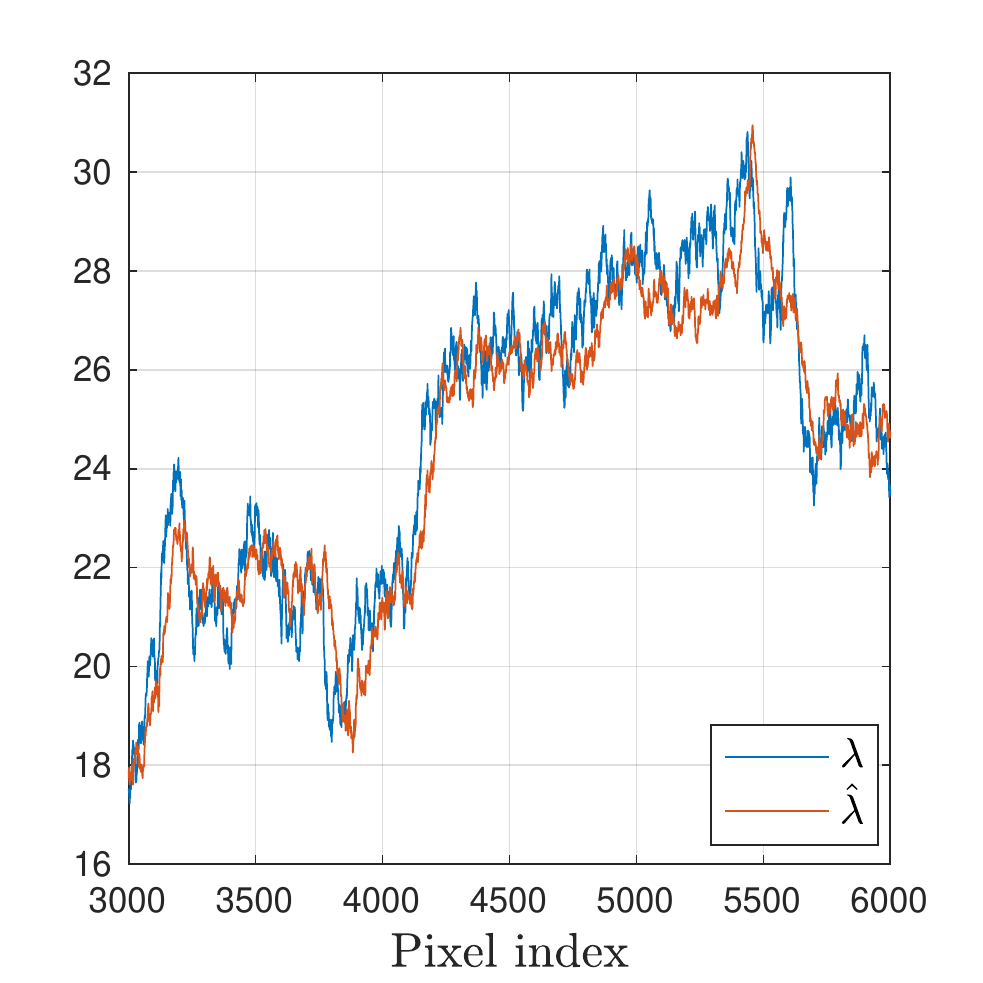}
        \caption{$\lamChatBold$, RMSE = 0.9416}
    \end{subfigure}
    \begin{subfigure}{0.49\linewidth}
        \centering
        \includegraphics[width=1\linewidth]{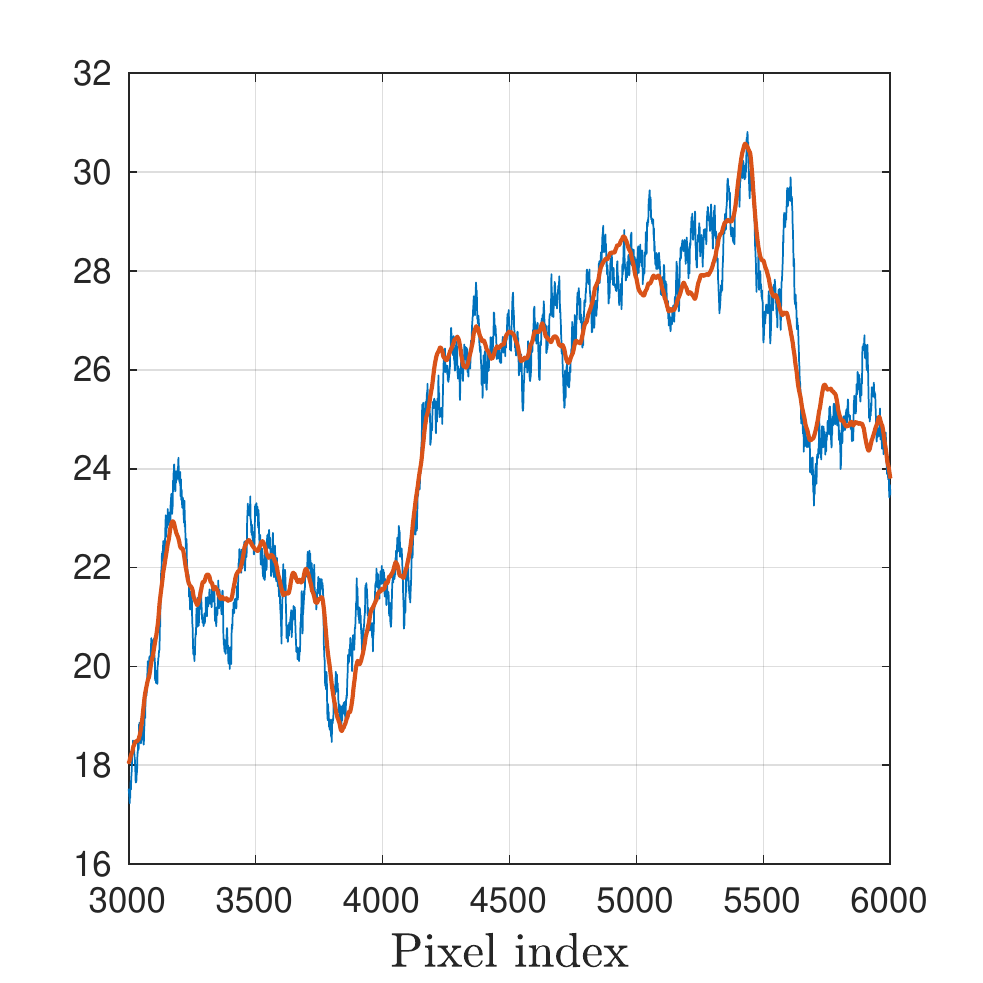}
        \caption{$\lamNChat$, RMSE = 0.6988}
    \end{subfigure}
    \caption{Beam current estimates for a representative subset of pixels for the HIM example in \Cref{fig:himEx}.}
    \label{fig:him_beam_current_est}
\end{figure}

\begin{figure}
    \centering
    \begin{subfigure}{0.49\linewidth}
        \centering
        \includegraphics[width=1\linewidth]{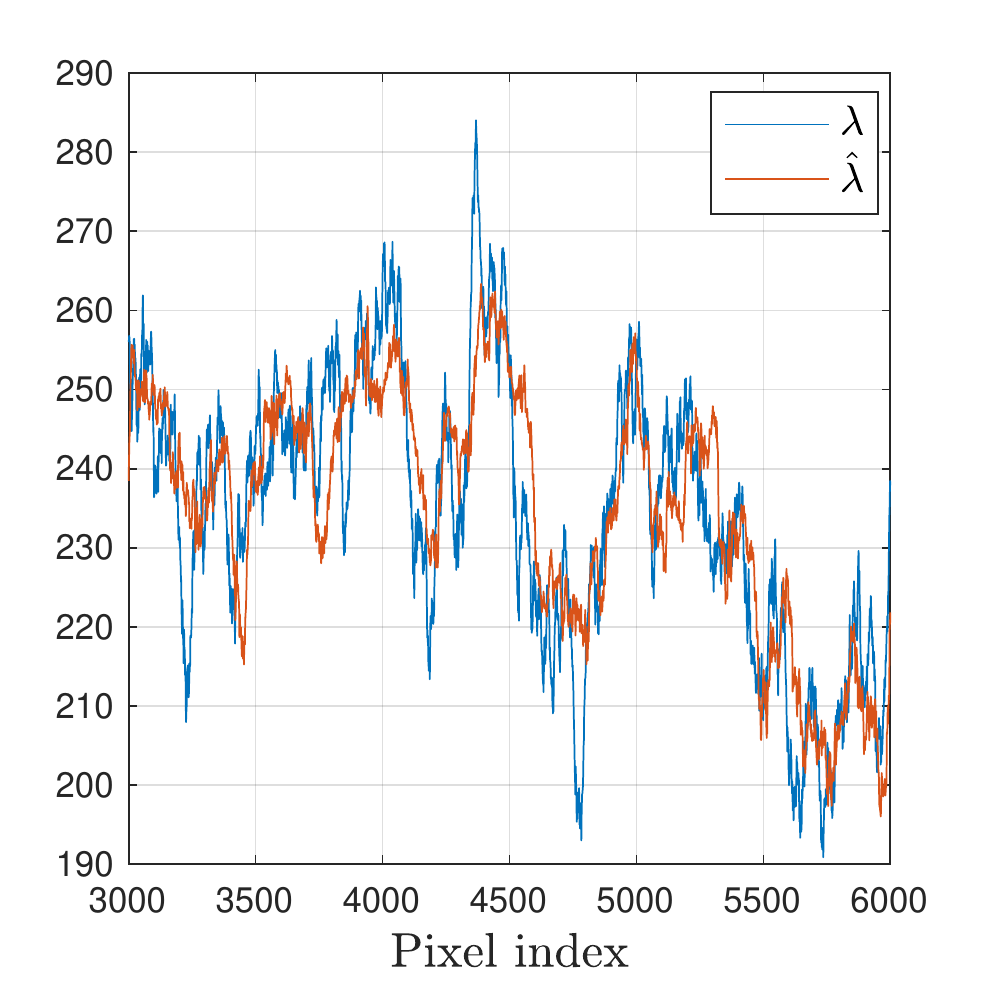}
        \caption{$\lamChatBold$, RMSE = 9.4937}
    \end{subfigure}
    \begin{subfigure}{0.49\linewidth}
        \centering
        \includegraphics[width=1\linewidth]{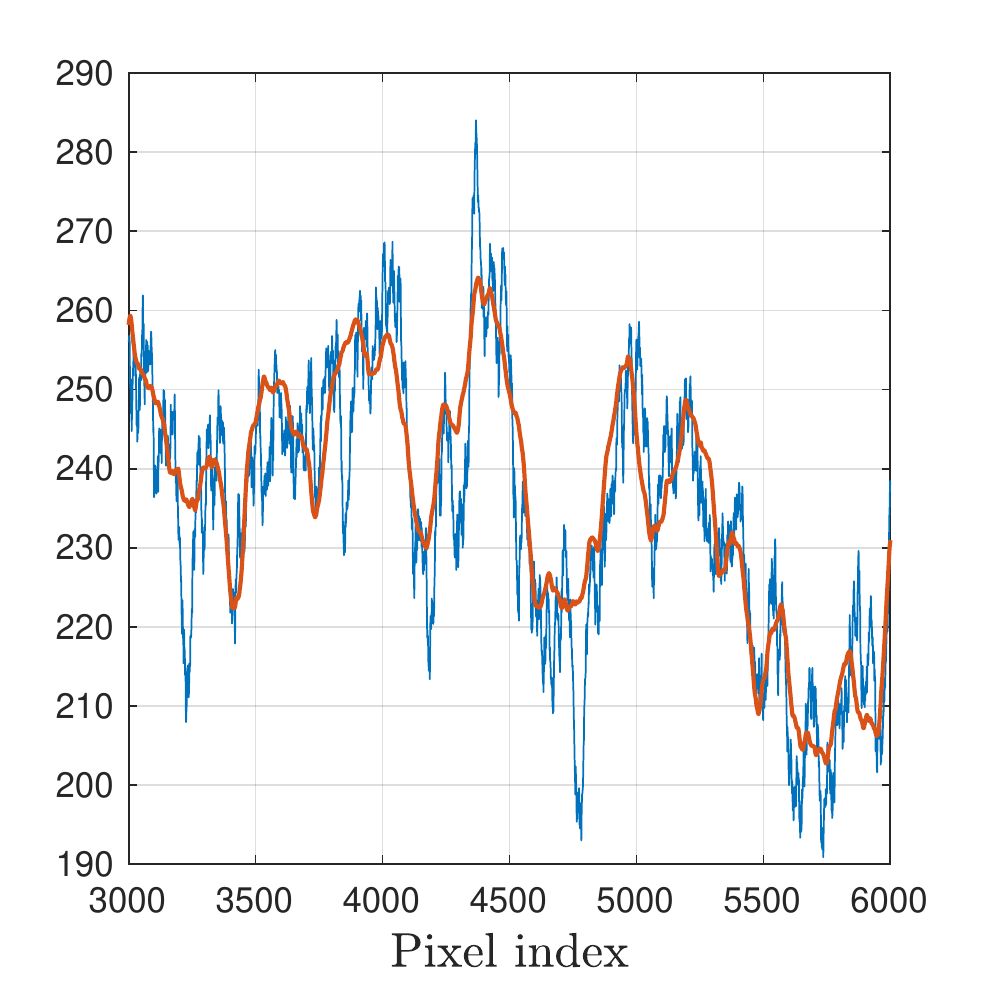}
        \caption{$\lamNChat$, RMSE =6.6761 }
    \end{subfigure}
    \caption{Beam current estimates for a representative subset of pixels for the SEM example in \Cref{fig:semExLowEta}.
    }
    \label{fig:sem2_beam_current_est}
\end{figure}

Beam current estimates $\lamChatBold$ and $\lamNChat$ are shown in \Cref{fig:him_beam_current_est} for the HIM example and in \Cref{fig:sem2_beam_current_est} for the SEM example.  
Both estimates closely match the true beam current.
The causal estimate $\lamChatBold$  has higher RMSE with a slight lag and more higher frequency noise.
All of our joint estimators outperform the initial joint estimator $\etaLFbold$ introduced in~\cite{watkins2021}.
The RMSE results of different $\lambf$ estimates, for both HIM and SEM examples, are summarized in \Cref{table:combinedTable}.

\begin{figure}
    \centering
     \begin{subfigure}{0.4\linewidth}
         \centering
         \includegraphics[width=1\linewidth]{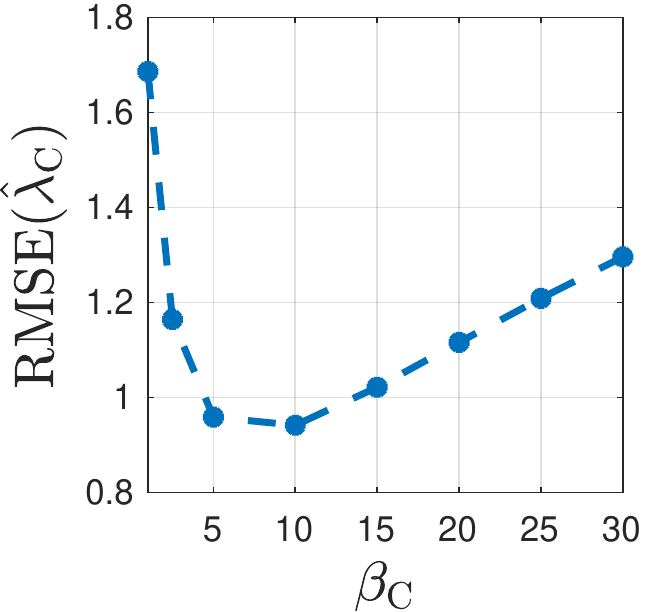}
         \caption{RMSE($\lamChatBold$) vs.\ $\betac$}
         \label{fig:mseLam_vs_beta_C}
     \end{subfigure}
     \begin{subfigure}{0.4\linewidth}
         \centering
         \includegraphics[width=1\linewidth]{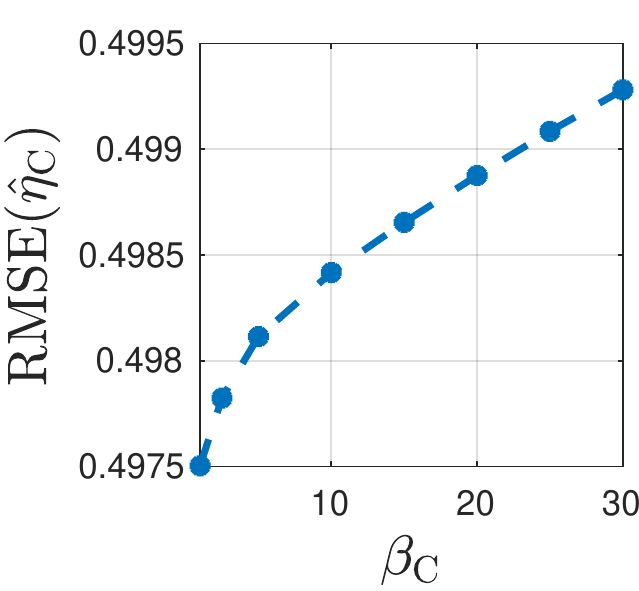}
         \caption{RMSE($\etaChatBold$) vs.\ $\betac$}
         \label{fig:mseEta_vs_beta_C}
     \end{subfigure}\\
         \begin{subfigure}{0.4\linewidth}
         \centering
         \includegraphics[width=1\linewidth]{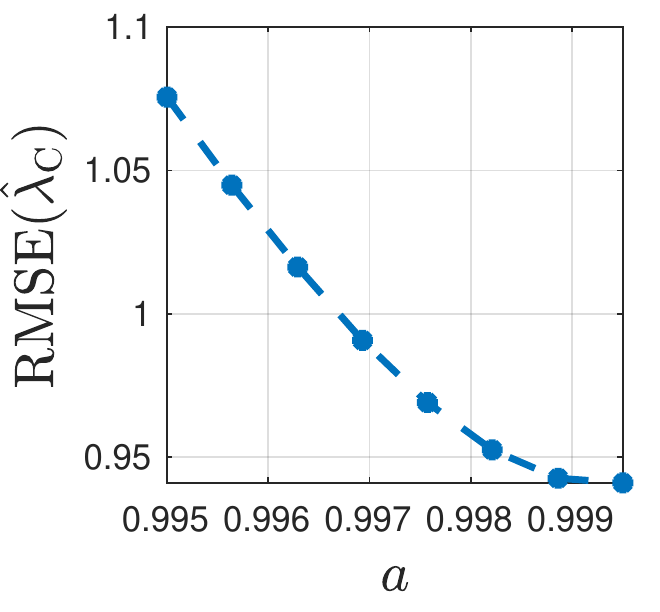}
         \caption{RMSE($\lamChatBold$) vs.\ $a$}
         \label{fig:mseLam_vs_a_C}
     \end{subfigure}
     \begin{subfigure}{0.4\linewidth}
         \centering
         \includegraphics[width=1\linewidth]{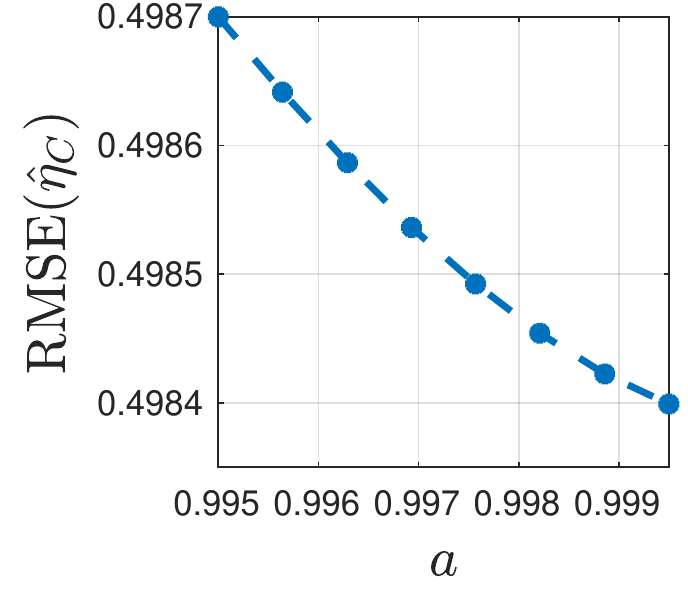}
         \caption{RMSE($\etaChatBold$) vs.\ $a$}
         \label{fig:mseEta_vs_a_C}
     \end{subfigure}
    \caption{Performance of causal joint estimation algorithm on data from \Cref{fig:himEx} for different values of $a$ and $\betac$. For (a) and (b) correlation is fixed at $a = 0.999$; for (c) and (d), tuning parameter is fixed at $\beta_{\rm C} = 10$. }
    \label{fig:robustness_demonstration_c}
\end{figure}

\begin{figure}
    \centering
        \begin{subfigure}{0.4\linewidth}
        \centering
        \includegraphics[width=1\linewidth]{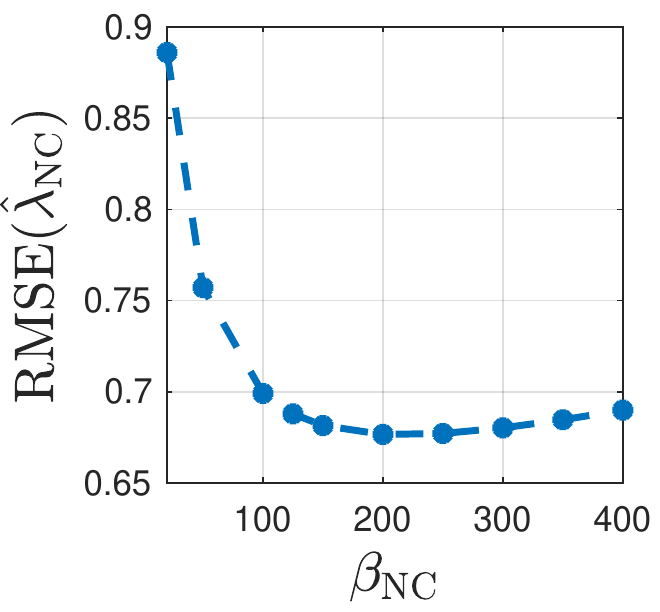}
        \caption{RMSE($\lamNChat$) vs.\ $\betanc$}
        \label{fig:mseLam_vs_beta_NC}
    \end{subfigure}
    \begin{subfigure}{0.4\linewidth}
        \centering
        \includegraphics[width=1\linewidth]{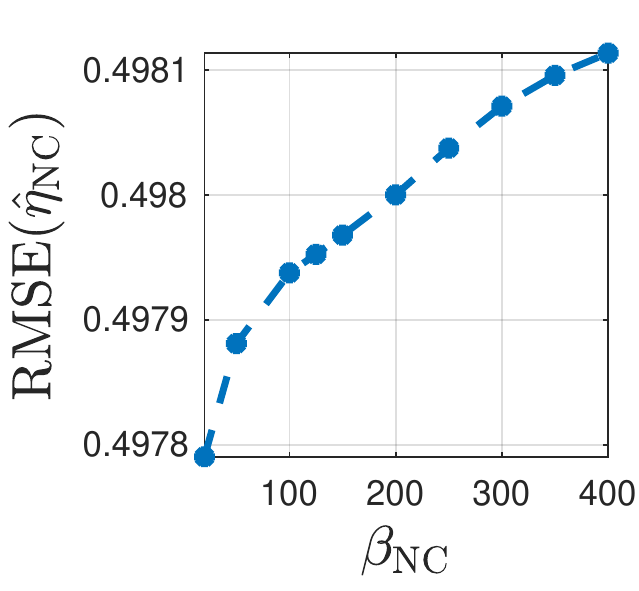}
        \caption{RMSE($\etaNChat$) vs.\ $\betanc$}
        \label{fig:mseEta_vs_beta_NC}
    \end{subfigure}\\
        \begin{subfigure}{0.4\linewidth}
        \centering
        \includegraphics[width=1\linewidth]{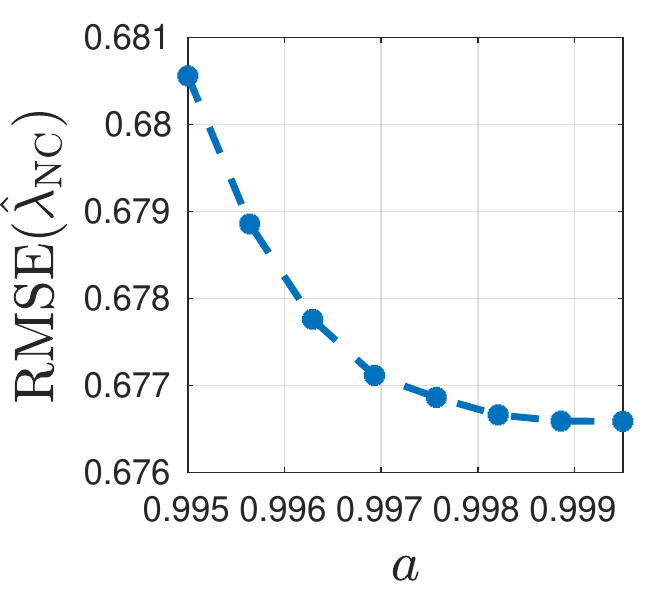}
        \caption{RMSE($\lamNChat$) vs.\ $a$}
        \label{fig:mseLam_vs_a_NC}
    \end{subfigure}
    \begin{subfigure}{0.4\linewidth}
        \centering
        \includegraphics[width=1\linewidth]{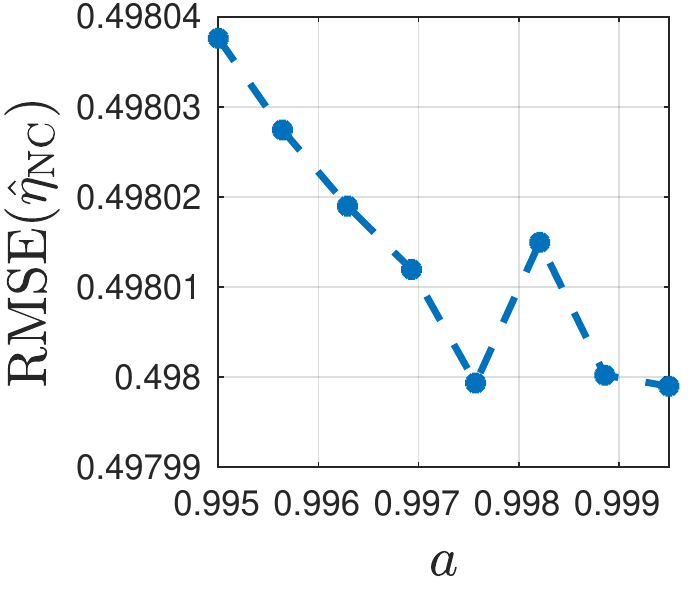}
        \caption{RMSE($\etaNChat$) vs.\ $a$}
    \end{subfigure}
    
    \caption{Performance of non-causal joint estimation algorithm on data from \Cref{fig:himEx} for different values of $a$ and $\betanc$. For (a) and (b), correlation is fixed at $a = 0.999$; for (c) and (d), tuning parameter is fixed $\beta_{\rm NC} = 200$. }
    \label{fig:robustness_demonstration_nc}
\end{figure}

As previously noted, all of our joint reconstruction algorithms include a tuning parameter ($\betac$ or $\betanc$ to control regularization on $\lambda$) and assume knowledge of correlation $a$.
In \Cref{fig:robustness_demonstration_c}, we show that performance of our causal estimation algorithm is not heavily dependent on ideal choices of either of these two parameters.
Using the data from \Cref{fig:himEx}, \Cref{fig:mseLam_vs_beta_C,fig:mseEta_vs_beta_C} form the joint estimate $(\etaChatBold,\,\lamChatBold)$ for different values of $\betac$, keeping the correlation fixed at $a = 0.999$.
\Cref{fig:mseLam_vs_a_C,fig:mseEta_vs_a_C} fix $\betac = 10$ and vary the assumed value of $a$.
We note that different values of $\betac$ and $a$ have a negligible effect on the RMSE of $\etaChatBold$.
Although the value of $\betac$ and $a$ does effect the RMSE of $\lamChatBold$, the RMSE remains relatively small across a wide range of values, with larger values of $\betac$ and $a$ (i.e., promoting more smoothing of $\lambf$) as safer choices when the true parameter value is not known.
In \Cref{fig:robustness_demonstration_nc}, we show similar trends for our non-causal algorithms.

\section{Exploiting a Discrete Markov Beam Current}
\label{sec:neon_beam_algorithms} 
In this section, we  demonstrate joint estimation when beam current flips back and forth between two values, as a simple model for a neon beam microscope~\cite{NotteNeon2010}.
Although the neon beam microscope could provide a number of functional advantages over the helium ion microscope~\cite{rahmanNeon2012}, it has been less widely adopted because of difficulties maintaining a stable beam current.

\subsection{A Discrete Markov Chain Model for Beam Current}
The beam current in a neon beam microscope may be modeled using a two-state hidden Markov model (HMM).
We assume that the nature of the beam current variation has been well characterized so that the states ${\bf s} \in \{s_1,\,s_2\}$ and transition probabilities
$q(s,r) = \mathrm{P}(\lambf_{k+1} = s \smid \lambf_k = r)$ are known.
The mean beam current under this model is denoted $\lambar$.
Based on this model, we propose causal and non-causal joint estimation algorithms for $\etabf$ and $\lambf$.

\subsection{Causal Estimation}
\Cref{alg:neonBeamRecursive} describes our causal joint estimator
$\big(\etaHmmBold,\,\lamHmmBold\big)$.
At each pixel, $\etaDTMLgivenBold_k (\lamtilde = \lambar)$ is used to form an initial estimate of $\etabf_k$.
As shown in~\cite{Watkins2021a} and~\cite{Watkins2021b},
as well as in our own results in \Cref{sec:performance_eval},
$\etaDTMLgiven$ is actually quite close to the true $\eta$ value, even when the beam current is imperfectly known.
Thus, we initially assume that $\etabf_k \approx \etaDTMLgivenBold_k (\lamtilde = \lambar)$ and use the Forward algorithm~\cite{baum1970maximization,baum1972inequality}, as described in \Cref{alg:neonBeamLamdaUpdate}, to compute the belief state $F_k(s)$ of $\lambf_k$ given the measurements from that pixel and all previous pixels:
\begin{equation*}
    F_k(s) := \mathrm{P}\big(\lambf_k = s \,|\, \ybf_{1:k}\big).
\end{equation*}
We pick $\lamHmmBold_k$ to be the state that maximizes $F_k(s)$.
The estimate $\etaHmmBold_k$ is produced by recomputing $\etaDTMLgivenBold_k$, using $\lamtilde = \lamHmmBold_k$ (\Cref{alg:neonBeamRecursive}, Line 4).
Note that \Cref{alg:neonBeamLamdaUpdate} operates recursively, requiring knowledge of $F_{k-1}(s)$ to compute $F_{k}(s)$.
At the first pixel,
\begin{align*}
    \mathrm{P}\big(&\yvec_1 = \ybf_1 | \lambf_1 = s\big)
    = \mathrm{P}_{\yvec_k}\big(\ybf_1\smid \lambf_k = s, \etaDTMLgivenBold_k(\lamtilde = \lambar)\big),
\end{align*}
where $\mathrm{P}_{\yvec_k}(\cdot \, ;\,\cdot,\cdot)$ is the PMF in \eqref{eq:DTTR-distribution} and we have assumed that
$\etabf_k \approx \etaDTMLgivenBold_k(\lamtilde = \lambar)$.
It follows from the law of total probability that 
\begin{align*}
    \mathrm{P}(&\yvec_1 = \ybf_1) = \sum_{s\in {\bf s}} \mathrm{P}\big(\yvec_1 = \ybf_1 \smid \lambf_1 = s\big)\mathrm{P}(\lambf_1 = s),
\end{align*}
where $\mathrm{P}(\lambf_1 = s)$ is the stationary distribution of the hidden Markov chain.
Applying Bayes's theorem, we find the initial belief state:
\begin{align*}
    F_1(s) &= \mathrm{P}(\lambf_1 = s\,|\,\yvec_1 = \ybf_1)\nonumber\\
    &= \frac{\mathrm{P}\big(\yvec_1 = \ybf_1\,|\,\lambf_1 = s\big)\mathrm{P}(\lambf_1 = s)}{ \mathrm{P}(\yvec_1 = \ybf_1)}.
\end{align*}

\begin{algorithm}
\caption{Causal joint estimation when beam current is modeled as a two-state hidden Markov chain
}
\label{alg:neonBeamRecursive}
\begin{algorithmic}[1]
\renewcommand{\algorithmicrequire}{\textbf{Input:}}
\REQUIRE $\ybf$, \, {\bf s}, \,$q(s,r)\, \forall \,s,r\in{\bf s}$ , $F_1(s) \forall \,s\in\bf{s} $
    \FOR{k = [2, 3, \dots p]}
    \STATE compute $F_k(s)$ using \Cref{alg:neonBeamLamdaUpdate} 
    
    with $\etabf_k = \etaDTMLgiven_k (\lamtilde = \lambar) $ and $\yvec_k = \ybf_k$
    \STATE $\lamHmmBold_k = \argmax_{s\in{\bf s}} F_k(s)$
    \STATE $\etaHmmBold_k = \etaDTMLgiven_k\big(\lamtilde = \lamHmmBold_k\big)$ 
    \ENDFOR
    \RETURN $\etaHmmBold$, $\lamHmmBold$
\end{algorithmic}
\end{algorithm}

\begin{algorithm*}
\caption{Forward Algorithm for computing belief state $F_k(s)$ at $k$th pixel given $\etabf_k$}
\label{alg:neonBeamLamdaUpdate}
\begin{algorithmic}[1]
\renewcommand{\algorithmicrequire}{\textbf{Input:}}
\REQUIRE $\yvec_k = \ybf_k\in\mathbb{R}^n$, $\etabf_k$,\, {\bf s}, \,$q(s,r)\, \forall \,s,r\in{\bf s}$ , $F_{k-1}(s) \forall \,s\in\bf{s} $
    \STATE Compute $P\big(\lambf_{k} = s\,|\,\text{all past measurements}) = \sum_{s'\in{\bf s}}q(s',s)F_{k-1}(s')$ $\forall \,s\in{\bf s}$
    \STATE  Compute $\mathrm{P}_{\yvec_k}\big(\ybf_k\smid \etabf_k, \lambf_k = s\big)$ $\forall\,s\in{\bf s}$
    \STATE $P\big(\yvec_k=\ybf_k\,|\,\text{all past measurements}\big)= \sum_{s\in{\bf s}}\mathrm{P}_{\yvec_k}\big(\ybf_k\smid \etabf_k, \lambf_k = s\big)P\big(\lambf_{k} = s\,|\,\text{all past measurements})$
    \STATE $F_k(s) = \Frac{\mathrm{P}_{\yvec_k}\big(\ybf_k\smid \etabf_k, \lambf_k = s\big)P\big(\lambf_{k} = s\,|\,\text{all past measurements}\big)}{P\big(\yvec_k=\ybf_k\,|\,\text{all past measurements}\big)}$
    \RETURN $F_k(s)$
\end{algorithmic}
\end{algorithm*}

\begin{figure*}
\centering
\begin{minipage}{.61\linewidth}
    \centering
        \begin{subfigure}{0.325\linewidth}
        \centering
        \includegraphics[width=1\linewidth]{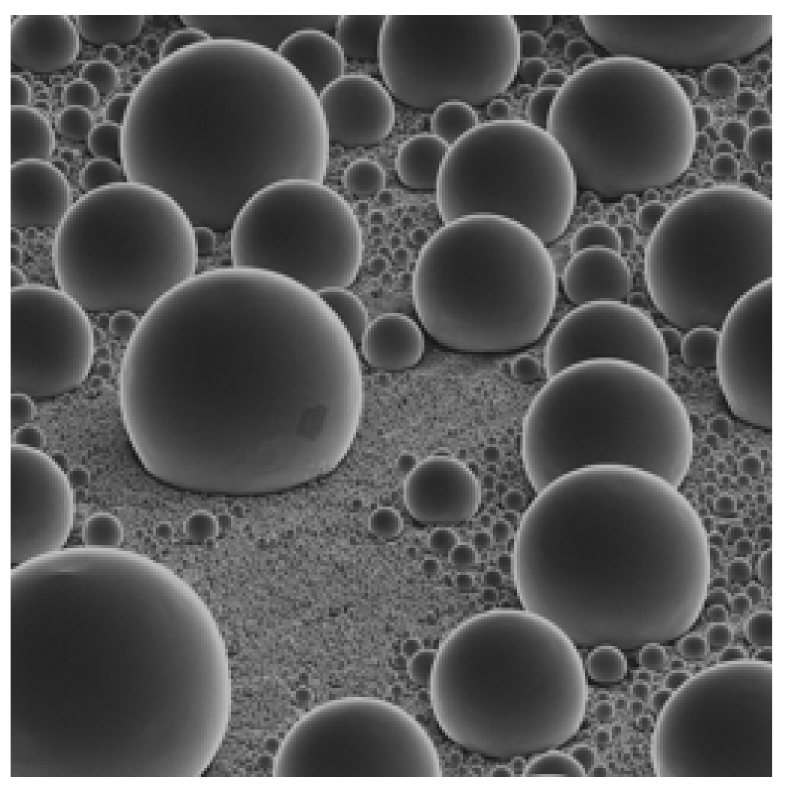}
        \caption{ground truth $\etabf$}
        \label{fig:neon_beam_gt}
    \end{subfigure}
    \begin{subfigure}{0.325\linewidth}
        \centering
        \includegraphics[width=1\linewidth]{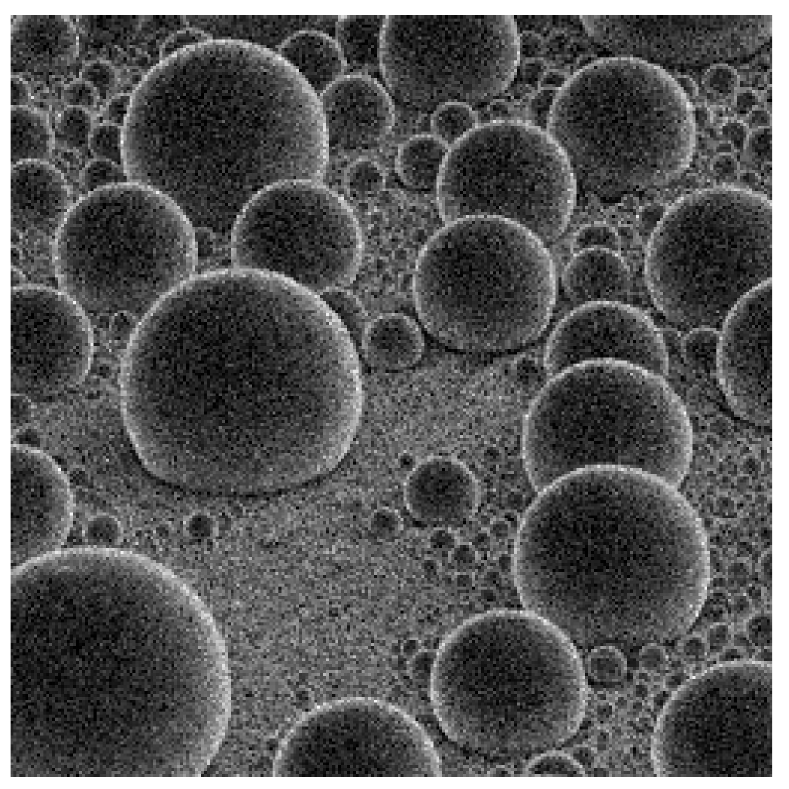}
        \caption{$\etaOracle$, RMSE = 0.5147}
        \label{fig:neon_beam_oracle}
    \end{subfigure}
    \begin{subfigure}{0.325\linewidth}
        \centering
        \includegraphics[width=1\linewidth]{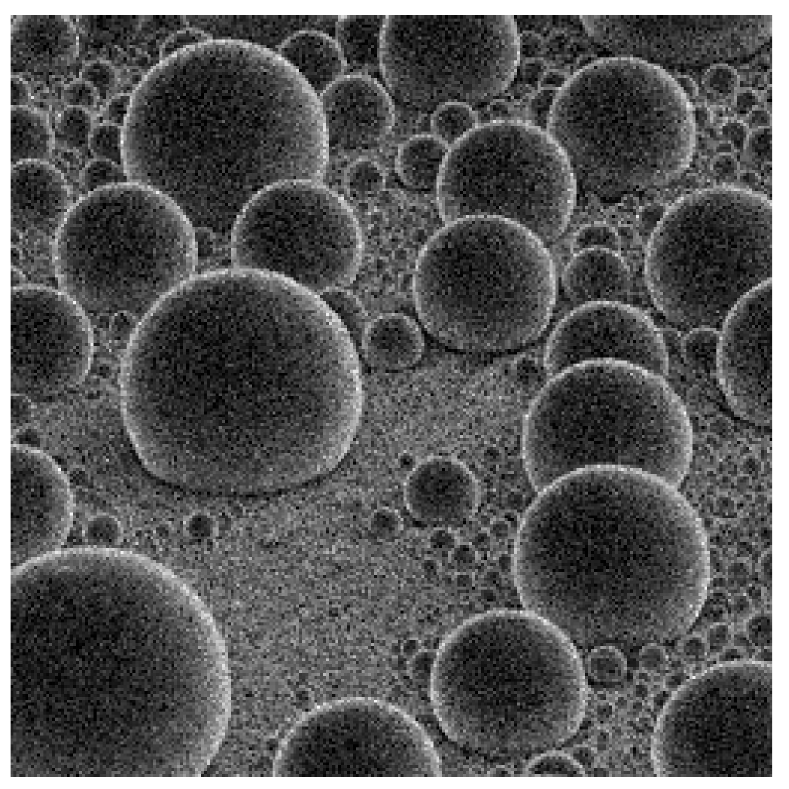}
        \caption{$\etaHmmBold$, RMSE = 0.5158}
        \label{fig:neon_beam_hmm}
    \end{subfigure}\\
    \begin{subfigure}{0.325\linewidth}
        \centering
        \includegraphics[width=1\linewidth]{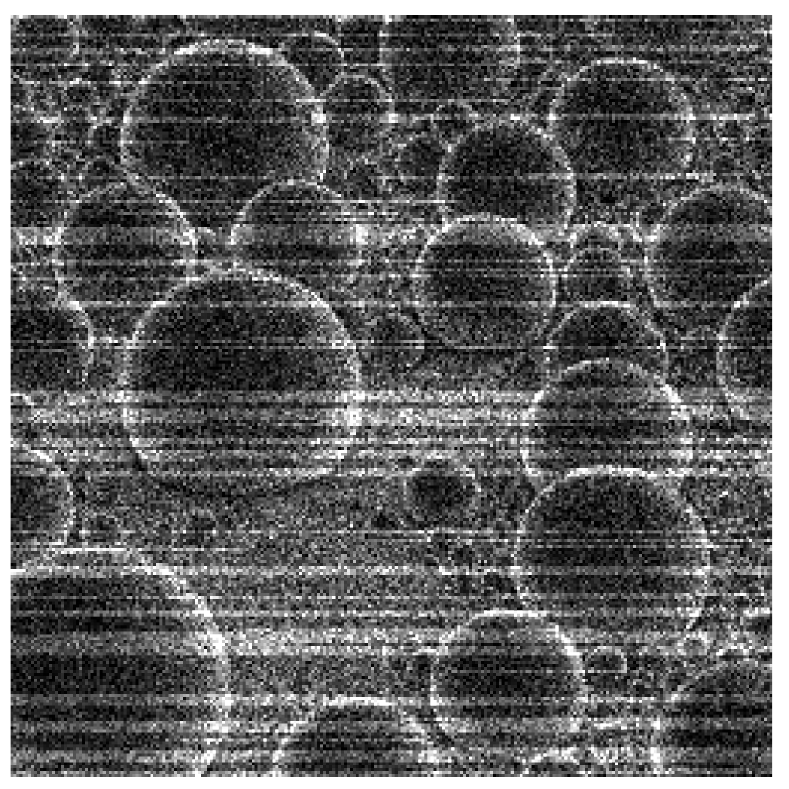}
        \caption{$\etaBaselineBold$, RMSE = 1.0689}
        \label{fig:neon_beam_conv}
    \end{subfigure}
    \begin{subfigure}{0.325\linewidth}
        \centering
        \includegraphics[width=1\linewidth]{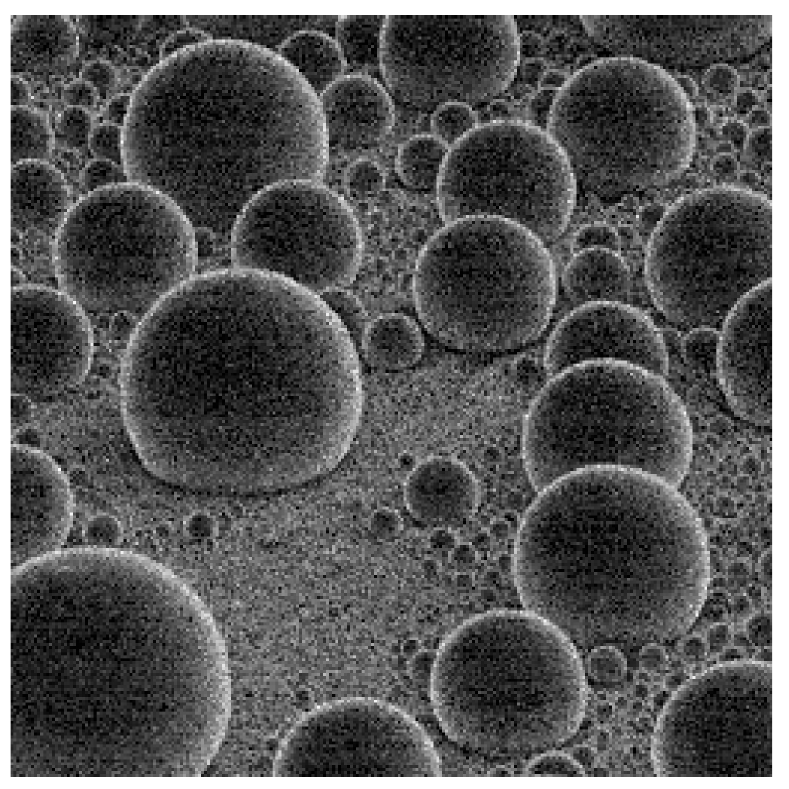}
        \caption{$\etaDTMLgivenBold$, RMSE = 0.5460}
        \label{fig:neon_beam_trml}
    \end{subfigure}
        \begin{subfigure}{0.325\linewidth}
        \centering
        \includegraphics[width=1\linewidth]{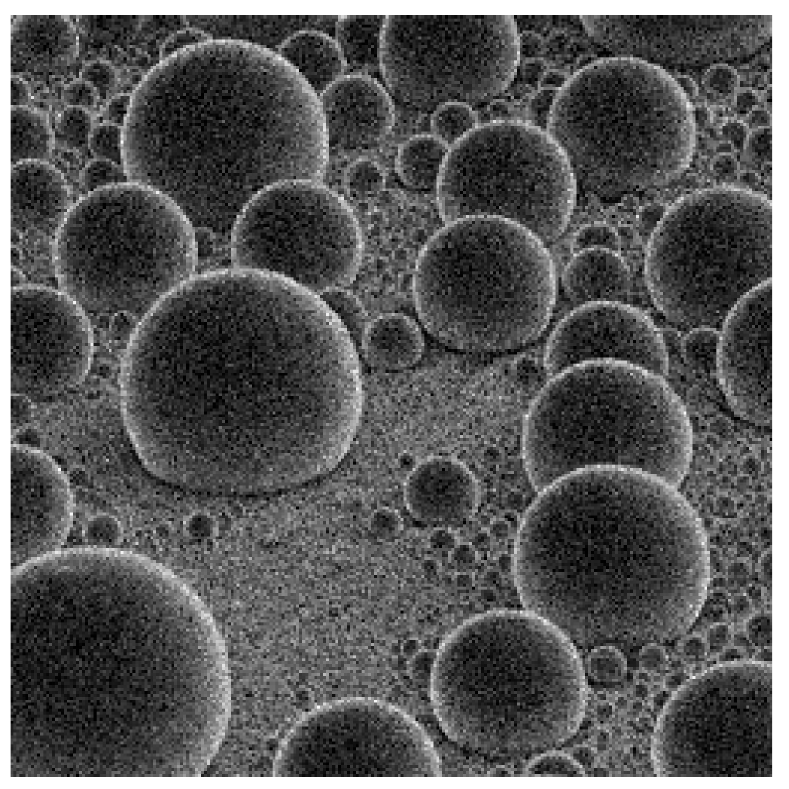}
        \caption{$\etaHmmNcBold$, RMSE = 0.5149}
        \label{fig:neon_beam_hmm_nc}
    \end{subfigure}
\end{minipage}
\begin{minipage}{.38\linewidth}
\centering
    \begin{subfigure}{1\linewidth}
        \centering
        \includegraphics[width=1\linewidth]{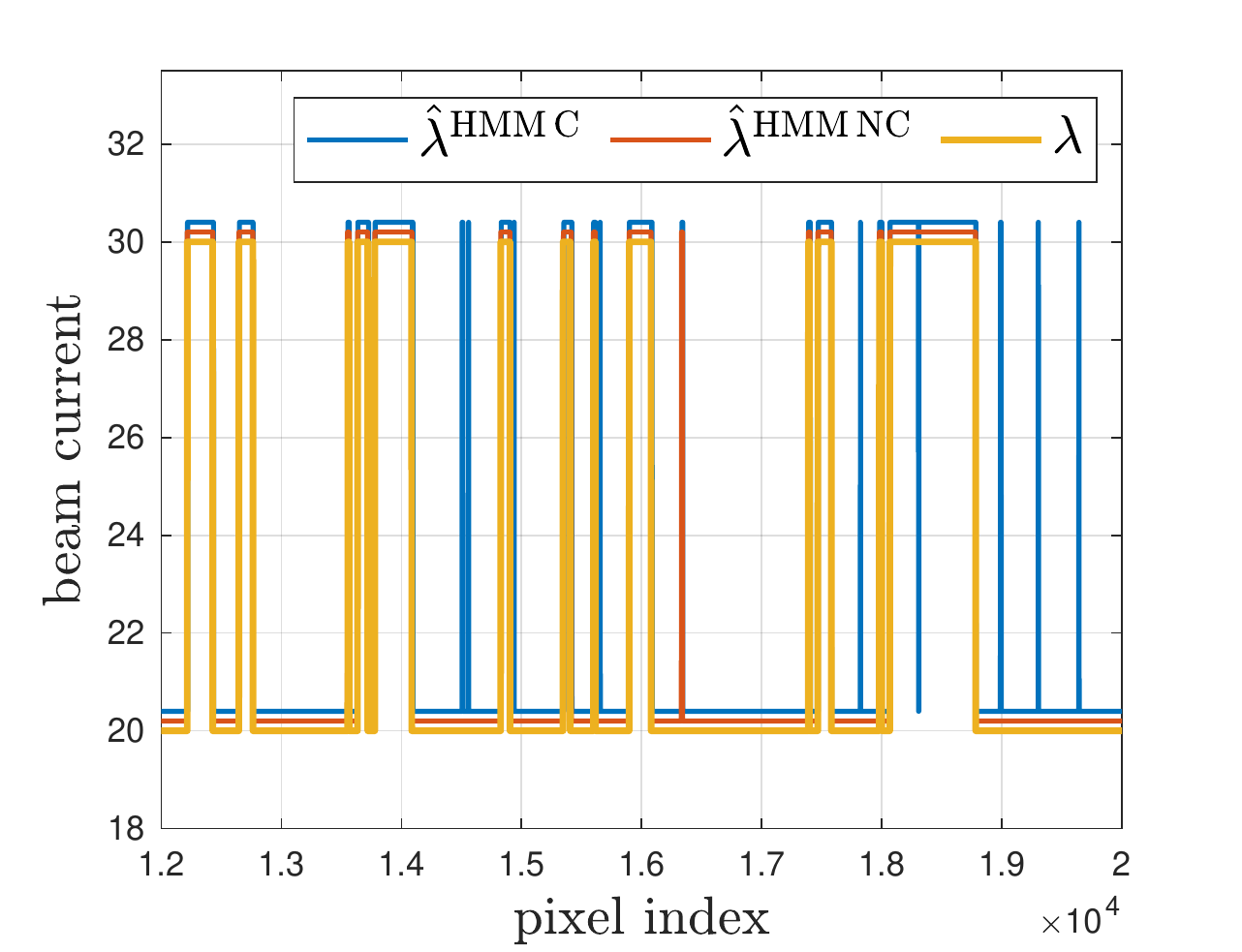}
        \caption{Ground truth $\lambf$ with estimates $\lamHmmBold$, and  $\lamHmmNcBold$ at a subset of pixels; $\lamHmmBold$ has 0.89\% error rate and $\lamHmmNcBold$ has 0.21\% error rate. To improve legibility, $\lamHmmNcBold$ is offset by 0.2, and $\lamHmmBold$ is offset by 0.4.}
        \label{fig:neon_beam_current}
    \end{subfigure}
\end{minipage}
    \caption{Results from a synthetic neon beam experiment with $\eta \in [2,\, 6]$, $n = 300$, and $\lambda$ modeled as a two-state hidden Markov chain with $\lambda \in \{20,\,30\}$.
    }
    \label{fig:neon_beam_results}
\end{figure*}

\subsection{Non-Causal Estimation}
\label{sec:nc_hmm}
Our non-causal joint estimate $\lamHmmNcBold_k$ selects the state with the greatest probability given the \emph{entire} measurement sequence:
\begin{equation}
    \lamHmmNcBold_k = \argmax_{s\in{\bf s}} \iP{\lambf_k = s \smid \ybf}.
    \label{eq:lamHat_hmm_nc}
\end{equation}
It uses the Forward-backward algorithm~\cite{baum1970maximization,baum1972inequality} to compute a quantity proportional to $\mathrm{P}(\lambf_k = s\,|\,\ybf)$ for each of the two possible states.
Note that $\mathrm{P}(\lambf_k = s\,|\,\ybf)$ may be factored as follows:
\begin{align}
\label{eq:forward_backward_terms}
    \mathrm{P}(&\lambf_k = s\,|\,\ybf) \propto \mathrm{P}(\lambf_k = s,\,\ybf)\nonumber\\
    &\eqlabel{a} \mathrm{P}\big(\lambf_k = s,\,\ybf_{1:k},\,\ybf_{k+1:p}\big)\nonumber\\
    &\eqlabel{b} \mathrm{P}\big(\ybf_{k+1:p}\,|\,\lambf_k = s,\,\ybf_{1:k}\big)\mathrm{P}\big(\lambf_k = s,\,\ybf_{1:k}\big)\nonumber\\
    &\eqlabel{c} \underbrace{\mathrm{P}\big(\ybf_{k+1:p}\,|\,\lambf_k = s\big)}_{\text{backward pass}}\underbrace{\mathrm{P}\big(\lambf_k = s,\,\ybf_{1:k}\big)}_{\text{forward pass}},
\end{align}
where (a) follows from splitting the components of $\ybf$;
(b) from the multiplication rule; and
(c) from the conditional independence of $\mathrm{P}\big(\ybf_{k+1:p}\,|\,\lambf_k = s\big)$ from $\ybf_{1:k}$.
The final two factors include a term readily available from the \emph{forward} pass of our our causal algorithm:
$\mathrm{P}\big(\lambf_k = s,\,\ybf_{1:k}\big) \propto F_k(s)$, and a second term computed in a new recursive \emph{backward} pass over the data.
This new term,
\begin{equation*}
    B_k(s) := \mathrm{P}\big(\ybf_{k+1:p}\,|\,\lambf_k = s\big),
\end{equation*}
may be computed using the following recursive formula moving backwards over the data sequence:
\begin{align*}
    B_k(s) &= \sum_{s'\in\bf{s}}B_{k+1}(s')\mathrm{P}_{\yvec_k}\big(\ybf_{k+1}\,|\,\lambf_{k+1} = s'\big)q(s,s').
\end{align*}
As stipulated by the Forward-backward algorithm, the last pixel is initialized with $B_n(s) = 1$ $\forall s \in \bf{s}$. Just as in the causal algorithm described in \Cref{alg:neonBeamRecursive}, our non-causal joint estimation algorithm forms an initial estimate of $\etabf_k$ at each pixel using $\etaDTMLgivenBold_k (\lamtilde = \lambar)$. Assuming $\etabf_k\approx\etaDTMLgivenBold_k (\lamtilde = \lambar)$, the estimate $\lamHmmNcBold_k$ is formed according to \eqref{eq:lamHat_hmm_nc}, requiring a forward and backward pass to compute the two terms in \eqref{eq:forward_backward_terms}. Then, $\etabf$ is estimated according to: $\etaHmmNcBold_k = \etaDTMLgivenBold_k (\lamtilde = \lamHmmNcBold_k) $.

\subsection{Simulated Microscopy Results}

Synthetic measurements were generated using an existing micrograph as the ground truth image.
The beam current time series was produced according to a two-state Markov chain model with $\lambda \in \{20,\,30\}$ using transition probabilities
\begin{align*}
\mathrm{P}(\lambda_t = 20 \smid \lambda_{t-1}=30) &= 0.003, \qquad \mbox{and} \\
\mathrm{P}(\lambda_t = 30 \smid \lambda_{t-1}=20) &= 0.002,
\end{align*}
resulting in $\lambar = 24$.
At each pixel, the dwell time was divided into $n = 300$ sub-acquisitions.

In \Cref{fig:neon_beam_results}, we compare the RMSE results for the following methods:
\begin{itemize}
    \item baseline: $\etaBaselineBold$ is the pixel-wise evaluation of \eqref{eq:eta-baseline} independently at each pixel using assumed dose $\lambar$.
     \item {DT$|\lambda$}: $\etaOracleBold$ is the pixel-wise ML estimate \eqref{eq:eta_DTML} computed with true beam current $\lambda$ (provided by an oracle).
    \item {DT$|\lamtilde$}: $\etaDTMLgivenBold$ is the pixel-wise ML estimate \eqref{eq:eta_DTML} computed using assumed dose $\lamtilde = \lambar$.
    \item HMM causal: $(\etaHmmBold, \lamHmmBold)$, computed using \Cref{alg:neonBeamLamdaUpdate}.
    \item HMM non-causal: $(\etaHmmNcBold, \lamHmmNcBold)$, computed according to \Cref{sec:nc_hmm}.
\end{itemize}
The baseline estimate $\etaBaselineBold$ in \Cref{fig:neon_beam_conv} exhibits prominent stripe artifacts.
\Cref{fig:neon_beam_trml} shows $\etaDTMLgivenBold$ with stripe artifacts greatly reduced but still visible.
\Cref{fig:neon_beam_hmm,fig:neon_beam_hmm_nc} show results for our causal and non-causal HMM joint estimation algorithms.
In both cases, RMSE is further reduced over $\etaDTMLgivenBold$, approaching the performance of $\etaOracleBold$ in \Cref{fig:neon_beam_oracle}, with $\etaHmmNcBold$ slightly outperforming $\etaHmmBold$. 

In \Cref{fig:neon_beam_current}, we plot the true beam current time series $\lambf$ with estimates $\lamHmmBold$ and $\lamHmmNcBold$. Note that both estimates match the true beam current at the vast majority of pixels, with an error percentage of 0.89\% for the causal algorithm and 0.21\% for the non-causal algorithm. Although performance is very good with only the causal forward pass, approximately four times fewer errors occur in the $\lambf$ estimate when all data is considered.

\section{Conclusion}
\label{sec:conclusion}
In this work, we explore the estimation of \emph{two} properties at each pixel of a particle beam micrograph:
mean SE yield $\eta$ and beam current $\lambda$.
Using the \CRB{} at a single pixel, we show the feasibility of joint estimation given time-resolved measurements.
Specifically, we show that at high $\eta$, joint estimation is only slightly more challenging than estimating $\eta$ when $\lambda$ is given.
We demonstrate that when the dose is sufficiently high, joint estimation is possible at even a single pixel.
To perform joint estimation at moderate doses, we exploit the fact that beam current does not vary arbitrarily.
The algorithms of \Cref{sec:multi_pixel_estimation} are motivated by electron and helium ion beams, where current is smoothly varying.
Algorithms in \Cref{sec:neon_beam_algorithms} are designed for neon beam microscopes, where beam current is known to jump among known values.
Through tests performed on synthetic microscopy data, we show that our $\eta$ estimators outperform existing methods and our novel $\lambda$ estimators closely match the ground truth.
This innovation not only prevents artifacts that arise when the beam current is not perfectly known, but also provides the operator with new and useful information that could transform microscopy.
Knowledge of the beam current could save costly instrument maintenance time, improve micrographs and milling outcomes, and even  further the proliferation of powerful new instruments like the neon beam microscope, where maintaining a stable beam current is a key challenge.

\appendix

In gradient algorithms, we require methods to evaluate or approximate derivatives of
the logarithm of the PMF $\mathrm{P}_{Y}(y \sMid \eta, \lambda)$ in \eqref{eq:neyman}.
The $y = 0$ case is mathematically simple and also important because many sub-acquisitions result in no detected SEs.
For $y > 0$, elegant expressions can be given using Touchard polynomials.
With discrete-time data with sufficiently large $n$, the relevant sub-acquisition dose is small,
so we also derive approximations that hold for small $\lambda$.

\subsection*{$y=0$ Case}
By substitution and simplification,
\begin{equation}
\log \mathrm{P}_{Y}(0 \sMid \eta, \lambda)
  = -\lambda(1-e^{-\eta}).
\end{equation}
The derivatives of this for optimization over $\lambda$ or $\eta$ are
\begin{subequations}
\begin{eqnarray}
\frac{d}{d\lambda} \left[ -\lambda(1-e^{-\eta}) \right]
  & = &  -(1 - e^{-\eta}),
         \label{eq:first-derivative-y0-lambda} \\ 
\frac{d^2}{d\lambda^2} \left[ -\lambda(1-e^{-\eta}) \right]
  & = &  0, 
         \label{eq:second-derivative-y0-lambda} \\ 
\frac{d}{d\eta} \left[ -\lambda(1-e^{-\eta}) \right]
  & = &  -\lambda e^{-\eta}, 
         \label{eq:first-derivative-y0-eta} \\ 
\frac{d^2}{d\eta^2} \left[ -\lambda(1-e^{-\eta}) \right]
  & = &  \lambda e^{-\eta}.
         \label{eq:second-derivative-y0-eta}
\end{eqnarray}
\end{subequations}

\subsection*{Touchard Polynomials}

The \emph{Touchard polynomials} are defined by
\begin{equation}
  T_n(x) = \sum_{k=0}^n S(n,k)x^k,
  \label{eq:Touchard}
\end{equation}
where $S(n,k)$ is a \emph{Stirling number of the second kind}, i.e.,
the number of partitions of a set of size $n$ into $k$ disjoint non-empty subsets.
Stirling numbers of the second kind can be used to write
\[
  m^y \ = \ \sum_{k=0}^y S(y,k) \frac{m!}{(m-k)!},
\]
where we regard $1/(m-k)! = 0$ if $k > m$~\cite{StackExchange_SangchulLee}.
Then
\begin{eqnarray}
 \lefteqn{ \sum_{m=0}^\infty \frac{m^y}{m!}x^m
   =   \sum_{m=0}^\infty \left( \sum_{k=0}^y S(y,k) \frac{m!}{(m-k)!} \right) \frac{x^m}{m!} } \nonumber \\
 & = & \sum_{k=0}^y S(y,k) \sum_{m=0}^\infty \frac{x^m}{(m-k)!} 
 \ = \ \sum_{k=0}^y S(y,k) x^k e^x \nonumber \\
 & = & T_y(x) \, e^x.
 \label{eq:series-to-Touchard}
\end{eqnarray}
Now we have
\begin{equation}
 \frac{d}{dx} \sum_{m=0}^\infty \frac{m^y}{m!}x^m
    \eqlabel{a}  \sum_{m=0}^\infty \frac{m^{y+1}}{m!}x^{m-1} 
    \eqlabel{b}  \frac{1}{x} T_{y+1}(x) e^x,
   \label{eq:diff-Touchard}
\end{equation}
where
(a) follows from term-by-term differentiation; and
(b) from \eqref{eq:series-to-Touchard}.
It is the derivative of the log that will be useful in what follows:
\begin{equation}
 \frac{d}{dx} \log \sum_{m=0}^\infty \frac{m^y}{m!}x^m
   =  \frac{T_{y+1}(x)}{x T_y(x)},
  \label{eq:diff-to-Touchard-ratio}
\end{equation}
which now follows from the chain rule and substitution of \eqref{eq:series-to-Touchard} and \eqref{eq:diff-Touchard}.

From \eqref{eq:Touchard},
a good approximation to $T_n(x)$ for small $x$ can be obtained by truncating to $k \in \{0,\,1,\,2\}$.
While Stirling numbers of the second kind $S(y,k)$ are not easy to work with in general,
for any $y \geq 1$, we have
$S(y,0) = 0$,
$S(y,1) = 1$,
and
$S(y,2) = 2^{y-1}-1$.
Therefore, for any $y \geq 1$,
\begin{subequations}
\label{eqs:Touchard-small-approx2}
  \begin{eqnarray}
    T_y(x)  & \approx & x + (2^{y-1}-1)x^2, \\
    T_y'(x) & \approx & 1 + (2^y-2)x,
    \end{eqnarray}
\end{subequations}
for small $x$.

\subsection*{Derivatives of log likelihood with respect to $\lambda$}
We have
\begin{eqnarray}
  \lefteqn{
   \frac{d}{d\lambda} \log\!\left[ \frac{e^{-\lambda}\eta^y}{y!}
         \sum_{m=0}^\infty
            \frac{(\lambda e^{-\eta})^m m^y}{m!} \right] } \nonumber \\
    & \!\!=\!\! & -1 + 
       \frac{d}{d\lambda} \log\!\left[ \sum_{m=0}^\infty
            \frac{(\lambda e^{-\eta})^m m^y}{m!} \right] \nonumber \\
    & \!\!\eqlabel{a}\!\! & -1 + e^{-\eta} \frac{T_{y+1}(\lambda e^{-\eta})}
                              {\lambda e^{-\eta} T_y(\lambda e^{-\eta})} 
     =  -1 + \frac{T_{y+1}(\lambda e^{-\eta})}
                    {\lambda T_y(\lambda e^{-\eta})}, \quad
                    \label{eq:loglike-first-derivative}
\end{eqnarray}
where (a) follows from the chain rule and \eqref{eq:diff-to-Touchard-ratio}.
It follows that
\begin{eqnarray}
  \lefteqn{
   \frac{d^2}{d\lambda^2} \log\!\left[ \frac{e^{-\lambda}\eta^y}{y!}
         \sum_{m=0}^\infty
            \frac{(\lambda e^{-\eta})^m m^y}{m!} \right] } \nonumber \\
    & = & \frac{e^{-\eta}T_{y+1}'(\lambda e^{-\eta})}{\lambda T_y(\lambda e^{-\eta})}
                  - \frac{T_{y+1}(\lambda e^{-\eta})}{\lambda^2 T_y(\lambda e^{-\eta})}
                  \nonumber \\
    & & \,        - \, \frac{\lambda e^{-\eta} T_{y+1}(\lambda e^{-\eta}) T_y'(\lambda e^{-\eta})}
                         {\left( \lambda T_y(\lambda e^{-\eta})\right)^2}.
                    \label{eq:loglike-second-derivative}
\end{eqnarray}
Substituting the second-order approximation \eqref{eqs:Touchard-small-approx2}
in \eqref{eq:loglike-first-derivative} and \eqref{eq:loglike-second-derivative}
gives the approximations
\begin{subequations}
\label{eqs:loglike-derivatives-approx2}
\begin{align}
   \frac{d}{d\lambda} \big[ \sim \big]
     &\approx  -1 + \frac{1}{\lambda} \cdot
                      \frac{1 + (2^y-1)\lambda e^{-\eta}}
                           {1 + (2^{y-1}-1)\lambda e^{-\eta}} ,
                    \label{eq:loglike-first-derivative-approx2} \\
   \frac{d^2}{d\lambda^2} \big[ \sim \big]
    &\approx  - \frac{1}{\lambda^2}
                    \left[
                     1 + \frac{(2^{y-1}-1)2^{y-1}(\lambda e^{-\eta})^2}
                              {\left(1 + (2^{y-1}-1)\lambda e^{-\eta}\right)^2}
                    \right].
                    \label{eq:loglike-second-derivative-approx2}
\end{align}
\end{subequations}

\subsection*{Derivatives of log likelihood with respect to $\eta$}
We have
\begin{eqnarray}
  \lefteqn{
   \frac{d}{d\eta} \log\!\left[ \frac{e^{-\lambda}\eta^y}{y!}
         \sum_{m=0}^\infty
            \frac{(\lambda e^{-\eta})^m m^y}{m!} \right] } \nonumber \\
    & = & \frac{y}{\eta} + 
       \frac{d}{d\eta} \log\!\left[ \sum_{m=0}^\infty
            \frac{(\lambda e^{-\eta})^m m^y}{m!} \right] \nonumber \\
    & \eqlabel{a} & \frac{y}{\eta} - \lambda e^{-\eta} \frac{T_{y+1}(\lambda e^{-\eta})}
                              {\lambda e^{-\eta} T_y(\lambda e^{-\eta})} 
    \ = \ \frac{y}{\eta} - \frac{T_{y+1}(\lambda e^{-\eta})}
                                {T_y(\lambda e^{-\eta})}, \qquad
                    \label{eq:loglike-first-derivative-eta}
\end{eqnarray}
where (a) follows from the chain rule and \eqref{eq:diff-to-Touchard-ratio}.
It follows that
\begin{eqnarray}
  \lefteqn{
   \frac{d^2}{d\eta^2} \log\!\left[ \frac{e^{-\lambda}\eta^y}{y!}
         \sum_{m=0}^\infty
            \frac{(\lambda e^{-\eta})^m m^y}{m!} \right] } \nonumber \\
    & = & -\frac{y}{\eta^2}
                  + \lambda e^{-\eta} \frac{T_{y+1}'(\lambda e^{-\eta})}{T_y(\lambda e^{-\eta})} \nonumber \\
    & & \,     - \, \lambda e^{-\eta} \frac{T_{y+1}(\lambda e^{-\eta}) T_y'(\lambda e^{-\eta})}
                            {(T_y(\lambda e^{-\eta}))^2}
                    \nonumber \\
    & = & -\frac{y}{\eta^2} + \lambda e^{-\eta} \nonumber \\
    & &        \times
                    \frac{T_{y+1}'(\lambda e^{-\eta}) T_y(\lambda e^{-\eta})
                        - T_{y+1}(\lambda e^{-\eta}) T_y'(\lambda e^{-\eta})}
                            {(T_y(\lambda e^{-\eta}))^2}.
                            \qquad
                    \label{eq:loglike-second-derivative-eta}
\end{eqnarray}
Substituting the second-order approximation \eqref{eqs:Touchard-small-approx2}
in \eqref{eq:loglike-first-derivative-eta} and \eqref{eq:loglike-second-derivative-eta}
gives the approximations
\begin{subequations}
\begin{align}
   \frac{d}{d\eta} \big[ \sim \big]
    &\approx \frac{y}{\eta} - \frac{1 + (2^y-1)\lambda e^{-\eta}}
                                {1 + (2^{y-1}-1)\lambda e^{-\eta}}, 
                    \label{eq:loglike-first-derivative-eta-approx2} \\
   \frac{d^2}{d\eta^2} \big[ \sim \big]
    & \approx -\frac{y}{\eta^2}
                       + \frac{2^{y-1} \lambda e^{-\eta}}
                              {\left(1 + (2^{y-1}-1)\lambda e^{-\eta}\right)^2}.
                    \label{eq:loglike-second-derivative-eta-approx2}
\end{align}
\end{subequations}

\section*{Acknowledgment}
The authors thank Sangchul Lee and Claude Leibovici
for answers to a Stack Exchange query~\cite{StackExchange_SangchulLee,StackExchange_ClaudeLeibovici}
that were instrumental to the derivations in
the appendix.

\ifCLASSOPTIONcaptionsoff
  \newpage
\fi



%
\bibliographystyle{IEEEtran}

\begin{thebibliography}{29}
\providecommand{\url}[1]{#1}
\csname url@samestyle\endcsname
\providecommand{\newblock}{\relax}
\providecommand{\bibinfo}[2]{#2}
\providecommand{\BIBentrySTDinterwordspacing}{\spaceskip=0pt\relax}
\providecommand{\BIBentryALTinterwordstretchfactor}{4}
\providecommand{\BIBentryALTinterwordspacing}{\spaceskip=\fontdimen2\font plus
\BIBentryALTinterwordstretchfactor\fontdimen3\font minus
  \fontdimen4\font\relax}
\providecommand{\BIBforeignlanguage}[2]{{%
\expandafter\ifx\csname l@#1\endcsname\relax
\typeout{** WARNING: IEEEtran.bst: No hyphenation pattern has been}%
\typeout{** loaded for the language `#1'. Using the pattern for}%
\typeout{** the default language instead.}%
\else
\language=\csname l@#1\endcsname
\fi
#2}}
\providecommand{\BIBdecl}{\relax}
\BIBdecl

\bibitem{WardNE:06}
B.~W. Ward, J.~A. Notte, and N.~P. Economou, ``Helium ion microscope: A new
  tool for nanoscale microscopy and metrology,'' \emph{J. Vacuum Sci. Technol.
  B}, vol.~24, no.~6, pp. 2871--2874, Nov. 2006.

\bibitem{Notte2016}
J.~Notte and J.~Huang, \emph{The Helium Ion Microscope}.\hskip 1em plus 0.5em
  minus 0.4em\relax Cham: Springer International Publishing, 2016, pp. 3--30.

\bibitem{morgan2006introduction}
J.~Morgan, J.~Notte, R.~Hill, and B.~Ward, ``An introduction to the helium ion
  microscope,'' \emph{Microscopy Today}, vol.~14, no.~4, pp. 24--31, 2006.

\bibitem{Rahman2013}
F.~F. Rahman, J.~A. Notte, R.~H. Livengood, and S.~Tan, ``Observation of
  synchronized atomic motions in the field ion microscope,''
  \emph{Ultramicroscopy}, vol. 126, pp. 10--18, 2013.

\bibitem{Barlow2016}
A.~J. Barlow, J.~F. Portoles, N.~Sano, and P.~J. Cumpson, ``Removing beam
  current artifacts in helium ion microscopy: A comparison of image processing
  techniques,'' \emph{Microsc. Microanal.}, vol.~22, no.~5, pp. 939--947, Oct.
  2016.

\bibitem{NotteNeon2010}
J.~Notte, F.~Rahman, S.~McVey, S.~Tan, and R.~Livengood, ``The neon gas field
  ion source - stability and lifetime,'' \emph{Microsc. Microanal.}, vol.~16,
  no.~S2, p. 28–29, 2010.

\bibitem{rahmanNeon2012}
F.~H.~M. Rahman, S.~McVey, L.~Farkas, J.~A. Notte, S.~Tan, and R.~H. Livengood,
  ``The prospects of a subnanometer focused neon ion beam,'' \emph{Scanning},
  vol.~34, no.~2, pp. 129--134, 2012.

\bibitem{Barlow2018}
A.~J. Barlow, N.~Sano, B.~J. Murdoch, J.~F. Portoles, P.~J. Pigram, and P.~J.
  Cumpson, ``Observing the evolution of regular nanostructured indium phosphide
  after gas cluster ion beam etching,'' \emph{Applied Surface Science}, vol.
  459, pp. 678--685, 2018.

\bibitem{schwartz2019}
J.~Schwartz, Y.~Jiang, Y.~Wang, A.~Aiello, P.~Bhattacharya, H.~Yuan, Z.~Mi,
  N.~Bassim, and R.~Hovden, ``Removing stripes, scratches, and curtaining with
  nonrecoverable compressed sensing,'' \emph{Microsc. Microanal.}, vol.~25,
  no.~3, p. 705–710, 2019.

\bibitem{Khalilian2019}
A.~Khalilian-Gourtani, M.~Tepper, V.~Minden, and D.~B. Chklovskii, ``Strip the
  stripes: Artifact detection and removal for scanning electron microscopy
  imaging,'' in \emph{Proc. IEEE Int. Conf. Acoust., Speech and Signal
  Process.}, 2019, pp. 1060--1064.

\bibitem{liu2018}
S.~Liu, L.~Sun, J.~Gao, and K.~Li, ``A fast curtain-removal method for {3D
  FIB-SEM} images of heterogeneous minerals,'' \emph{J. Microscopy}, vol. 272,
  no.~1, pp. 3--11, 2018.

\bibitem{ott2019}
T.~Ott, D.~Rold\'an, C.~Redenbach, K.~Schladitz, M.~Godehardt, and S.~H\"ohn,
  ``Three-dimensional structural comparison of tantalum glancing angle
  deposition thin films by {FIB-SEM},'' \emph{J. Sensors Sensor Syst.}, vol.~8,
  no.~2, pp. 305--315, 2019.

\bibitem{PENG2020}
M.~Peng, J.~Murray-Bruce, K.~K. Berggren, and V.~K. Goyal, ``Source shot noise
  mitigation in focused ion beam microscopy by time-resolved measurement,''
  \emph{Ultramicroscopy}, vol. 211, p. 112948, 2020.

\bibitem{peng2020analysis}
M.~Peng, J.~Murray-Bruce, and V.~K. Goyal, ``Time-resolved focused ion beam
  microscopy: Modeling, estimation methods, and analyses,'' \emph{IEEE Trans.
  Computational Imaging}, vol.~7, pp. 547--561, 2021.

\bibitem{Watkins2021a}
L.~Watkins, S.~W. Seidel, M.~Peng, A.~Agarwal, C.~C. Yu, and V.~K. Goyal,
  ``Robustness of time-resolved measurement to unknown and variable beam
  current in particle beam microscopy,'' in \emph{Proc. IEEE Int. Conf. Image
  Process.}, Anchorage, AK, Sep. 2021, pp. 3487--3491.

\bibitem{Watkins2021b}
------, ``Prevention beats removal: Avoiding stripe artifacts from current
  variation in particle beam microscopy through time-resolved sensing,''
  \emph{Microsc. Microanal.}, vol.~27, no.~S1, pp. 422--425, Aug. 2021.

\bibitem{yamada1991electron}
S.~Yamada, T.~Ito, K.~Gouhara, and Y.~Uchikawa, ``Electron-count imaging in
  {SEM},'' \emph{Scanning}, vol.~13, no.~2, pp. 165--171, 1991.

\bibitem{jin2008applications}
L.~Jin, A.-C. Milazzo, S.~Kleinfelder, S.~Li, P.~Leblanc, F.~Duttweiler, J.~C.
  Bouwer, S.~T. Peltier, M.~H. Ellisman, and N.-H. Xuong, ``Applications of
  direct detection device in transmission electron microscopy,'' \emph{J.
  Structural Biology}, vol. 161, no.~3, pp. 352--358, 2008.

\bibitem{mcmullan2009detective}
G.~McMullan, S.~Chen, R.~Henderson, and A.~R. Faruqi, ``Detective quantum
  efficiency of electron area detectors in electron microscopy,''
  \emph{Ultramicroscopy}, vol. 109, no.~9, pp. 1126--1143, 2009.

\bibitem{Agarwal:2021arXiv}
A.~Agarwal, J.~Simonaitis, V.~K. Goyal, and K.~K. Berggren, ``Secondary
  electron count imaging in {SEM},'' arXiv:2111.01862, Nov. 2021.

\bibitem{sim2004effect}
K.~S. Sim, J.~T.~L. Thong, and J.~C.~H. Phang, ``Effect of shot noise and
  secondary emission noise in scanning electron microscope images,''
  \emph{Scanning}, vol.~26, no.~1, pp. 36--40, 2004.

\bibitem{beckTV}
A.~Beck and M.~Teboulle, ``Fast gradient-based algorithms for constrained total
  variation image denoising and deblurring problems,'' \emph{IEEE Trans. Image
  Process.}, vol.~18, no.~11, pp. 2419--2434, 2009.

\bibitem{notte2007introduction}
J.~Notte, B.~Ward, N.~Economou, R.~Hill, R.~Percival, L.~Farkas, and S.~McVey,
  ``An introduction to the helium ion microscope,'' in \emph{AIP Conference
  Proceedings}, vol. 931.\hskip 1em plus 0.5em minus 0.4em\relax AIP, 2007, pp.
  489--496.

\bibitem{LinJ:05}
Y.~Lin and D.~C. Joy, ``A new examination of secondary electron yield data,''
  \emph{Surface Interface Anal.}, vol.~37, no.~11, pp. 895--900, 2005.

\bibitem{watkins2021}
L.~Watkins, ``Mitigating current variation in particle beam microscopy,''
  arXiv:2106.04686, Jun. 2021, {B.S.} honors thesis in electrical and computer
  engineering, Boston University.

\bibitem{baum1970maximization}
L.~E. Baum, T.~Petrie, G.~Soules, and N.~Weiss, ``A maximization technique
  occurring in the statistical analysis of probabilistic functions of {M}arkov
  chains,'' \emph{Annals Math. Stat.}, vol.~41, no.~1, pp. 164--171, 1970.

\bibitem{baum1972inequality}
L.~E. Baum, ``An inequality and associated maximization technique in
  statistical estimation for probabilistic functions of {M}arkov processes,''
  \emph{Inequalities}, vol.~3, no.~1, pp. 1--8, 1972.

\bibitem{StackExchange_SangchulLee}
S.~Lee, Mathematics Stack Exchange, Jul. 2020,
  https://math.stackexchange.com/q/3744576.

\bibitem{StackExchange_ClaudeLeibovici}
C.~Leibovici, Mathematics Stack Exchange, Jul. 2020,
  https://math.stackexchange.com/q/3744557.

\end{thebibliography}
\newcommand{\SortNoop}[1]{}




\end{document}